\documentclass[apsprl,twocolumn,superscriptaddress,longbibliography]{revtex4-2}

\usepackage{physics}

\usepackage{graphicx}
\usepackage{amsmath}
\usepackage{ulem}
\usepackage[usenamesdvipsnames]{xcolor}
\usepackage[colorlinks=truelinkcolor=blue,urlcolor=blue,citecolor=blue]{hyperref}
\usepackage{braket}
\usepackage{amssymb}
\usepackage{commath}
\usepackage{csquotes}
\usepackage{dcolumn}
\usepackage{bm}
\usepackage{epstopdf}
\usepackage{lipsum}
\usepackage{subfigure}
\usepackage{siunitx}
\usepackage{pifont}

\begin{document}

\title{Experimental realization of fragmented models in tilted Fermi-Hubbard chains}

\author{Thomas~Kohlert}
\affiliation{Fakult\"at f\"ur Physik, Ludwig-Maximilians-Universit\"at M\"unchen, 
	Munich, Germany}
\affiliation{Max-Planck-Institut f\"ur Quantenoptik, 
85748 Garching, Germany}
\affiliation{Munich Center for Quantum Science and Technology (MCQST), 
80799 München, Germany}

\author{Sebastian~Scherg}
\affiliation{Fakult\"at f\"ur Physik, Ludwig-Maximilians-Universit\"at M\"unchen, 
	Munich, Germany}
\affiliation{Max-Planck-Institut f\"ur Quantenoptik, 
85748 Garching, Germany}
\affiliation{Munich Center for Quantum Science and Technology (MCQST), 
80799 München, Germany}

\author{Pablo~Sala}
\affiliation{Munich Center for Quantum Science and Technology (MCQST), 
80799 München, Germany}
\affiliation{Department of Physics and Institute for Advanced Study, Technical University of Munich, 85748 Garching, Germany}

\author{Frank~Pollmann}
\affiliation{Munich Center for Quantum Science and Technology (MCQST),  
80799 München, Germany}
\affiliation{Department of Physics and Institute for Advanced Study, Technical University of Munich, 85748 Garching, Germany}

\author{Bharath~Hebbe~Madhusudhana}
\affiliation{Fakult\"at f\"ur Physik, Ludwig-Maximilians-Universit\"at M\"unchen, 
	Munich, Germany}
\affiliation{Max-Planck-Institut f\"ur Quantenoptik,  
85748 Garching, Germany}
\affiliation{Munich Center for Quantum Science and Technology (MCQST), 
80799 München, Germany}

\author{Immanuel~Bloch}
\affiliation{Fakult\"at f\"ur Physik, Ludwig-Maximilians-Universit\"at M\"unchen, 
	Munich, Germany}
\affiliation{Max-Planck-Institut f\"ur Quantenoptik, 
85748 Garching, Germany}
\affiliation{Munich Center for Quantum Science and Technology (MCQST),  
80799 München, Germany}

\author{Monika~Aidelsburger}
\affiliation{Fakult\"at f\"ur Physik, Ludwig-Maximilians-Universit\"at M\"unchen, 
	Munich, Germany}
\affiliation{Munich Center for Quantum Science and Technology (MCQST),  
80799 München, Germany}

\date{\today}

\begin{abstract}
Quantum many-body systems may defy thermalization even without disorder. Intriguingly, non-ergodicity may be caused by a fragmentation of the many-body Hilbert-space into dynamically disconnected subspaces. The tilted one-dimensional Fermi-Hubbard model was proposed as a platform to realize fragmented models perturbatively in the limit of large tilt. Here, we demonstrate the validity of this effective description for the transient dynamics using ultracold fermions. The effective analytic model allows for a detailed understanding of the emergent microscopic processes, which in our case exhibit a pronounced doublon-number dependence. We study this experimentally by tuning the doublon fraction in the initial state.
\end{abstract}

\maketitle

Quantum many-body systems out of equilibrium are typically characterized according to their long-time behavior of local observables. While generic quantum systems reach thermal equilibrium as predicted by the Eigenstate Thermalization Hypothesis (ETH)~\cite{Deutsch1991,Srednicki1994,Rigol2008}, well-known exceptions to this paradigm are integrable~\cite{Calabrese2011,Essler2016} and many-body localized (MBL) systems~\cite{Schreiber2015,Gornyi2005,Basko2006,NandkishoreMBL,abanin_colloquium_2019}. Their non-ergodicity is based on an extensive set of conserved quantities~\cite{Serbyn13,Huse14}. Recently, a new class of models with intermediate behavior, summarized as weak ergodicity breaking~\cite{Scars_review2020}, renewed the interest in questions of quantum thermalization. A key signature of weak ergodicity-breaking is the strong dependence of the dynamics on the initial conditions, which discriminates it from both fully thermal and strongly ergodicity-breaking systems. This is due to a special structure of the many-body Hilbert space, which exhibits (approximately) disconnected subspaces that are not characterized by the global symmetries of the model. 

\begin{figure}[t!]
	\includegraphics{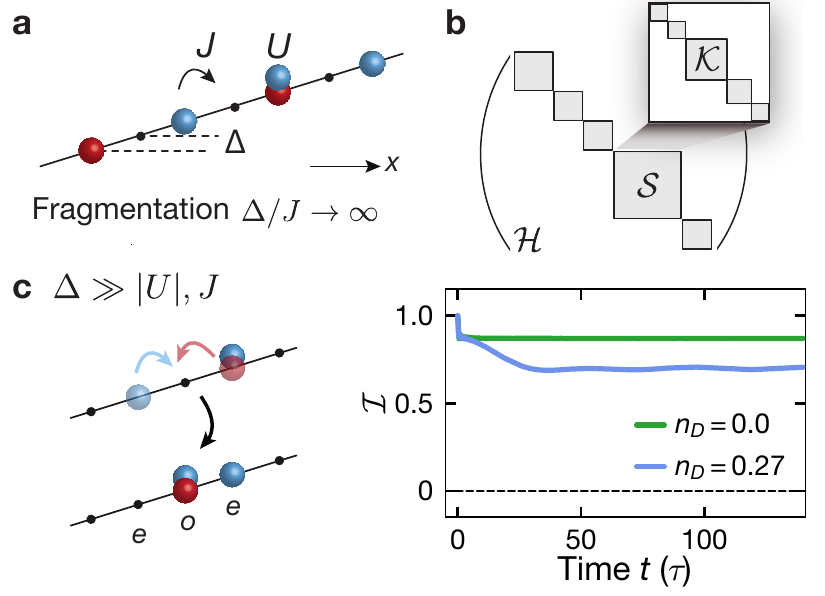}
	\caption{\textbf{Hilbert-space fragmentation in the tilted 1D Fermi Hubbard model.}
	\textbf{a} Schematic of the tilted Fermi-Hubbard model ($\uparrow$-atoms red, $\downarrow$-atoms blue) with linear potential (``tilt") of strength $\Delta$, tunneling $J$ and Hubbard interaction $U$.
	\textbf{b} Schematic illustration of Hilbert-space fragmentation. The symmetry sectors $\mathcal{S}$ of the total Hilbert space $\mathcal{H}$ decouple into (approximately) disconnected fragments $\mathcal{K}$.	  
	\textbf{c} Doublon-number dependent relaxation dynamics $\mathcal{I}(t)$ starting from an initial period-two charge-density wave with $\mathcal{I}(0)=1$ and doublon fraction $n_D$. The imbalance $\mathcal{I}$ is a measure for the relative occupation of even \textit{e} and odd \textit{o} lattice sites; $\tau$ is one tunneling time. The schematic illustrates the dominant correlated tunneling process of the effective Hamiltonian for $\Delta\gg|U|,J$ [Eq.~(\ref{eq:Heffdip})], which is only resonant for $n_D>0$. The solid lines are TEBD simulations (cumulative average) for $U=2.7J$ and $\Delta=8J$ for a lattice with $L=101\,(n_D=0)$ and $L=52\,(n_D=0.27)$ sites according to the effective Hamiltonian~(\ref{eq:Heffdip}). The dashed line indicates the steady-state value of $\mathcal{I}$ for a thermal system at infinite temperature (within $\mathcal{S}$).}
	\label{fig:fig1}
\end{figure}

For instance, quantum many-body scars~\cite{Turner2018Nature, Turner2018, Chen2018, ChengJu2019, Schecter2019, Pai2019, Zhao2020}, as recently observed with Rydberg atoms~\cite{Bernien2017,Bluvstein2020}, are the result of an atypical set of eigenstates, which are embedded in an otherwise thermal spectrum. Another prominent example are fractonic systems, such as one dimensional (1D) setups with conserved U(1) charge and its associated dipole moment~\cite{Sala2020, Khemani2020, Moudgalya2019,NandkishoreRC_2019, Taylor_Stark2020}. In these models the Hilbert-space fragments into exponentially many disconnected subspaces, a phenomenon known as \textit{Hilbert-space fragmentation}. In fact, similar phenomena also appear in classical stochastic dynamics in the realm of kinetically constrained spin systems, where it is known as reducibility~\cite{Ritort2003, Pancotti2020}. The major difference is that dipole and higher-moment conserving systems~\cite{Pancotti2020} provide a provable and systematic way of constructing such models without explicit kinetic constraints.

Remarkably, the tilted 1D Fermi-Hubbard model (Fig.~\ref{fig:fig1}a) was proposed to host a variety of distinct fragmented models, which are derived as effective descriptions in the limit of strong tilts $\Delta/J \rightarrow \infty$~\cite{Moudgalya2019,Khemani2020,Sala2020,Scherg2020}. In this regime, emergent conservation laws, such as dipole-moment conservation, result in a fragmentation of the Hilbert space into exponentially many fragments $\mathcal{K}$, such that even states belonging to the same symmetry sector $\mathcal{S}$, defined e.g. by the dipole moment, may become dynamically disconnected (Fig.~\ref{fig:fig1}b), when considering finite order in perturbation theory. In this work we experimentally study the properties of the underlying effective Hamiltonian that governs the transient dynamics. We verify the effective description by studying relaxation dynamics of a period-two charge density wave up to about 140 tunneling times for various doublon (doubly-occupied site) fractions $n_D$ (Fig.~\ref{fig:fig1}c). The doublon-number dependence is directly related to the underlying microscopic processes in the emergent fragmented models. 

Our experimental setup consists of a degenerate Fermi gas of about $50(5)\times 10^3$ ${}^{40}\mathrm{K}$ atoms at an average temperature of $0.12(2)T_F$, where $T_F$ is the Fermi temperature. The gas is prepared in an equal mixture of two magnetic hyperfine states, $\ket{\downarrow} = \ket{m_F=-9/2}$ and $\ket{\uparrow} = \ket{m_F=-7/2}$, in the $F=9/2$ ground-state hyperfine manifold. The fermions are loaded into a 3D optical lattice created by three pairs of retro-reflected laser beams. The lattice along the primary axis has a wavelength $\lambda_p = \SI{532}{\nano\meter}$ and the orthogonal lattices operate at $\lambda_\perp = \SI{738}{\nano\meter}$. We work at a primary lattice depth of $12E_{r,p}$, where one tunneling time $\tau=\hbar/J=\SI{0.75}{\milli\second}$. The orthogonal lattices are set to $55E_{r,\perp}$; here $E_{r,i} = h^2/(2m\lambda_i^2)$ is the recoil energy, with $i \in \{p,\perp\}$, $m$ the atomic mass, $\lambda_i$ the respective lattice wavelength and $h$ the Planck constant. This creates a 2D array of 1D chains, where the central chain has a length $L \approx 290$ and coupling to neighboring chains is suppressed by a factor of $\sim10^{-3}$, such that the system can be considered 1D on our experimental time scales. A magnetic field generated via a single coil induces a linear potential gradient (``tilt") along the primary lattice axis. Since the spins are encoded in different magnetic hyperfine states, the tilt $\Delta_\sigma$, $\sigma \in \{\uparrow,\downarrow\}$, is slightly state-dependent, $\Delta_\uparrow\simeq 0.9\Delta_\downarrow$~\cite{suppmat}. The dynamics of each 1D chain is described by the tilted 1D Fermi-Hubbard model (Fig.~\ref{fig:fig1}a)
\begin{align}
\begin{split}
\hat{H} = &-J \sum_{i,\sigma} (\hat{c}_{i+1,\sigma}^\dagger \hat{c}_{i,\sigma} + \mathrm{h.c.}) + U \sum_{i} \hat{n}_{i,\uparrow} \hat{n}_{i,\downarrow} \\
&+ \sum_{i,\sigma} \Delta_\sigma i\hat{n}_{i,\sigma},
\end{split}
\label{eq:Hamiltonian}
\end{align}
where $\hat{c}_{i,\sigma}$ ($\hat{c}_{i,\sigma}^\dagger$) denotes the fermionic annihilation (creation) operator for spin $\sigma$ on site $i$ and $\hat{n}_i = \hat{c}_i^\dagger \hat{c}_i$.

In order to study dynamics, we use a bichromatic superlattice to prepare a period-two charge-density wave (CDW), where only even sites are occupied~\cite{suppmat}. After a short dephasing time in the deep 3D lattice there are no residual coherences and the initial state can be described by an incoherent mixture of localized product states with random spin configurations at zero net magnetization. The dynamics is initiated by quenching the primary lattice to the desired value. After initiating the dynamics, we probe the relaxation by measuring the relative atom number on even ($N_e$) and odd ($N_o$) lattice sites, given by the ensemble-averaged imbalance $\mathcal{I} = (N_e-N_o)/(N_e+N_o)$, which we directly extract using a bandmapping technique~\cite{Sebby2006,Foelling07}. Moreover, using near-resonant light pulses to remove doubly-occupied sites before detection, we have access to singlon (singly-occupied site) and doublon-resolved imbalances, $\mathcal{I}_S$ and $\mathcal{I}_D$~\cite{suppmat}. In this work we restrict our observation times to $140\tau$, since for longer times light-assisted collisions significantly reduce the doublon fraction~\cite{suppmat}.

\begin{figure}[t]
	\includegraphics[width=3.3in]{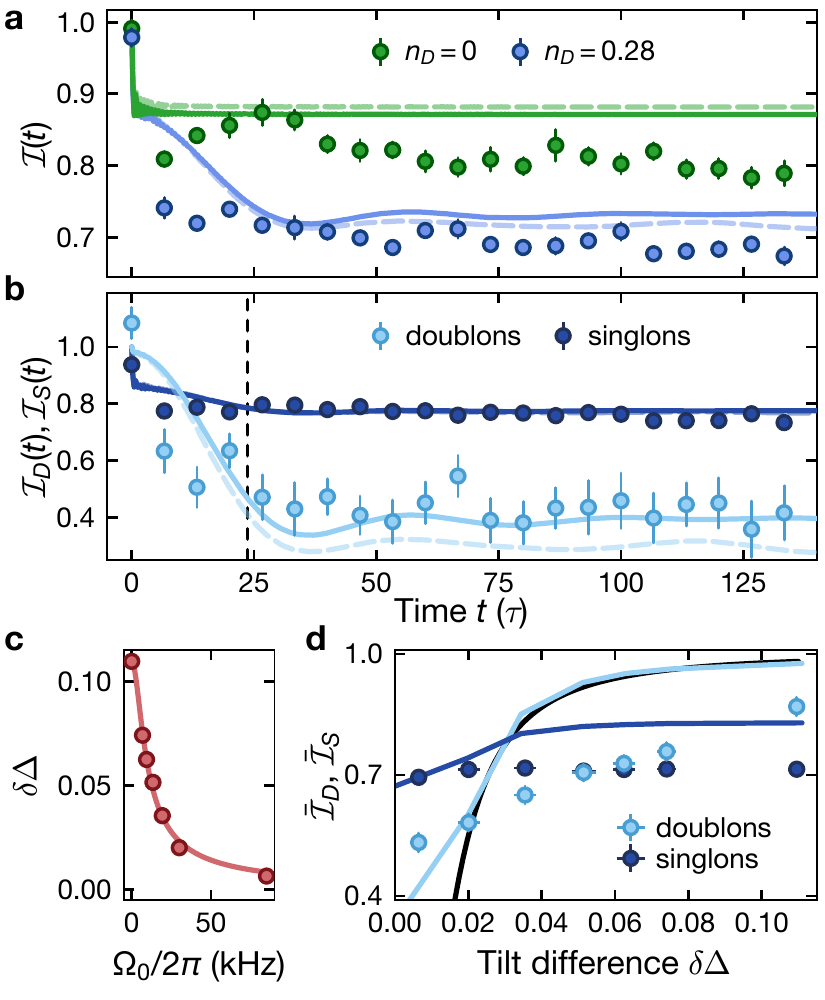}
	\caption{\textbf{Effective Hamiltonian dynamics for $\Delta \gg |U|,\, J$.} 
\textbf{a} Imbalance time trace for singlon ($n_D\simeq0$) and mixed $[n_D=0.28(2)]$ CDW initial states for $\Delta/J= 8.0(2)$, $U/J=2.7(2)$ and $\delta\Delta=0.6(2)\%$ [resonant Rabi frequency $\Omega_0=85(1)\,$kHz]. 
\textbf{b} Singlon- and doublon-resolved imbalance for the mixed initial state $[n_D=0.28(2)]$ for the same parameters. The lines in (a) and (b) are time-averaged TEBD simulations with $L=101$ lattice sites with the exact (dashed, transparent lines) and the effective model (solid lines) including a hole fraction of $20\%$ (comparable to Ref.~\cite{Scherg2018}). The dashed vertical line shows the effective timescale $1/(2\pi J^{(3)})$. Experimental data points are averaged ten times and error bars are the standard error of the mean (SEM). 
\textbf{c} Relative tilt difference $\delta\Delta$ as a function of the resonant Rabi frequency $\Omega_0$ in the presence of the RF dressing field. The solid line is a fit of the analytic model defined in Eq.~(\ref{eqn:tilt_diff}).
\textbf{d} Steady-state imbalance averaged over ten data points between $67\tau$ and $80\tau$ as a function of the tilt difference between both spins for $\Delta/J= 8.0(2)$, $U/J=2.7(2)$ and $n_D=0.47(4)$. Solid lines of the same color as the data points are TEBD simulations with the exact Hamiltonian (\ref{eq:Hamiltonian}) on 101~lattice sites, $n_D=0.46$ and a hole fraction of $20\%$. The black line represents the imbalance of a Wannier-Stark localized doublon-hole pair (main text). See table \ref{parameter_table} in~\cite{suppmat} for numerical details. } 
	\label{fig:fig2}
\end{figure}

In the limit of large tilts $\Delta \gg |U|,\, J$ and for $\Delta\equiv\Delta_\downarrow=\Delta_\uparrow$ we can use a Schrieffer-Wolff (SW) transformation to expand Hamiltonian~(\ref{eq:Hamiltonian}) in powers of $1/\Delta$~\cite{suppmat}. Up to third-order it reads~\cite{suppmat}
\begin{equation}
		\hat{H}_{\mathrm{eff}} = J^{(3)} \left( \hat{T}_3  + 2 \hat{T}_{XY} +2 \hat{V}\right)+ \tilde{U} \sum_i \hat{n}_{i,\uparrow} \hat{n}_{i,\downarrow}, \\
	\label{eq:Heffdip}
\end{equation}
with the effective tunneling $J^{(3)} = J^2U/\Delta^2$, the renormalized on-site Hubbard interaction $\tilde{U} = U \left( 1-4J^2/\Delta^2 \right)$ and a nearest-neighbor interaction $\hat{V}=\sum_{i,\sigma} \hat{n}_{i,\sigma} \hat{n}_{i+1,\bar{\sigma}}$, where $\bar{\sigma}$ denotes the opposite spin of $\sigma$. The dynamics is governed by $\hat{T}_3 = \sum_{i,\sigma} (\hat{c}_{i,\sigma}^\dagger \hat{c}_{i+1,\sigma} \hat{c}_{i+1,\bar{\sigma}} \hat{c}_{i+2,\bar{\sigma}}^\dagger + \mathrm{h.c.})$ and an exchange term $\hat{T}_{XY} = \sum_i (\hat{c}_{i,\uparrow}^\dagger \hat{c}_{i,\downarrow} \hat{c}_{i+1,\downarrow}^\dagger \hat{c}_{i+1,\uparrow} + \mathrm{h.c.})$. This Hamiltonian is SU$(2)$ invariant and conserves charge ($\hat{Q} = \sum_i \hat{n}_i$) and dipole moment ($\hat{P} = \sum_i i\hat{n}_i$). Similar to other dipole-conserving models studied previously in spin chains~\cite{Sala2020}, random unitary circuits~\cite{Khemani2020,NandkishoreRC_2019} and spinless Hubbard systems~\cite{Moudgalya2019,Taylor_Stark2020}, it is strongly fragmented~\cite{Scherg2020}. 

The spin-dependent tilt $\Delta_\sigma$ in Hamiltonian (\ref{eq:Hamiltonian}) introduces additional constraints. In order to tune this spin-dependence in the experiment we employ the technique of radio-frequency (RF) dressing~\cite{Skedrov2021, suppmat}, where we use an additional RF field to couple the two spins. Thereby we realize dressed states that see a weighted average of $\Delta_{\uparrow}$ and $\Delta_{\downarrow}$.  The weights are determined by the resonant RF coupling strength $\Omega_0$ and the detuning from resonance. This allows us to adjust the relative tilt difference $\delta\Delta = (\tilde{\Delta}_\downarrow -  \tilde{\Delta}_\uparrow)/\Delta_\downarrow$ by tuning the RF coupling strength (Fig.~\ref{fig:fig2}c), where $\tilde{\Delta}_\sigma$ denotes the spin-dependent tilt seen by the dressed states. The values were calibrated using non-interacting Bloch oscillations~\cite{suppmat}. The maximum tilt difference is determined by the $m_F$ quantum numbers in the absence of RF dressing and the smallest value of $\delta\Delta=0.6(2)\%$ is reached for the largest coupling strength. The Hubbard on-site interaction $U$ is invariant under RF dressing~\cite{Zwierlein2003, suppmat}, which allows us to tune its magnitude via the Feshbach resonance at $202.1\,\mathrm{G}$.

We start by measuring relaxation dynamics for initial states with and without doublons (Fig.~\ref{fig:fig2}a), where $n_D=N_D/(N_S+N_D)$ and $N_D$ ($N_S$) is the number of atoms on doubly- (singly-)occupied lattice sites. The different loading sequences are described in~\cite{suppmat}. We observe similar time traces in both cases, i.e., after a fast drop at short times, a steady-state value develops for evolution times $t>30\tau$, signaling non-ergodic behavior~\cite{Scherg2020}. For initial states with doublons, however, the steady-state value is reduced, in agreement with TEBD simulations~\cite{suppmat} of the full [Eq.~(\ref{eq:Hamiltonian})] and the effective model [Eq.~(\ref{eq:Heffdip})], including the SW transformation for the initial state. Here, the numerical traces are time-averaged (Figs~\ref{fig:fig2}a,b) in order to mimic dephasing of the observed oscillations, which in the experiment is realized by averaging over an inhomogeneous distribution. The slow residual decay of the experimental singlon time trace is due to technical heating caused by the transverse lattice laser beams. Comparing the exact and effective numerical traces, we observe a small systematic offset for $t\gtrsim10\tau$, which is due to higher order terms. It decreases for larger tilts or smaller interactions.

The effective model in Eq.~(\ref{fig:fig2}) allows for a microscopic understanding of the doublon-dependent dynamics. This is best revealed in the singlon- and doublon-resolved imbalance traces (Fig.~\ref{fig:fig2}b). Starting from a period-two CDW with $\mathcal{I}(0)=1$, there is only one correlated tunneling process governed by $\hat{T}_3$, that can initiate the dynamics (schematics in Fig.~\ref{fig:fig1}c). For pure singlon initial states this process is energetically suppressed by the effective on-site interaction energy $\tilde{U}$, since in the large-tilt limit $J^{(3)}/\tilde{U} \simeq (J/\Delta)^2 \ll 1$. The presence of doublons, however, renders this process resonant, which is expected to relax the CDW on a time scale governed by the hopping rate $J^{(3)}$. Indeed we find that while the singlon imbalance $\mathcal{I}_S$ remains stable even for mixed initial states, there is a pronounced decrease of the doublon imbalance $\mathcal{I}_D$. This is due to $\hat{T}_3$ (Fig.~\ref{fig:fig1}c), which leads to a fast rearrangement of doublons between even and odd lattice sites on a timescale $1/(2\pi J^{(3)}) \simeq 24\tau$, while keeping the singlon configuration fixed. 

Experimentally, we further explore the effect of SU$(2)$ symmetry breaking on the observed singlon- and doublon-resolved steady-state values by tuning the tilt difference $\delta\Delta$ (Fig.~\ref{fig:fig2}d). We consider the regime $J>|\tilde{\Delta}_\uparrow-\tilde{\Delta}_\downarrow|$, where the perturbative description [Eq.~(\ref{eq:Heffdip})] remains valid. For $|\tilde{\Delta}_\uparrow-\tilde{\Delta}_\downarrow|>0$, the relevant process enabled by $\hat{T}_3$ is energetically detuned even in the presence of doublons and we expect relaxation to be strongly suppressed for $|\tilde{\Delta}_\uparrow-\tilde{\Delta}_\downarrow|>J^{(3)}$. This is supported by the observed dependence of the doublon dynamics on the tilt difference. The signal obtained from the singlons on the other hand shows no significant dependence. 
For $|\tilde{\Delta}_\uparrow-\tilde{\Delta}_\downarrow|>J^{(3)}$ there is an intuitive description of the doublon steady-state imbalance in terms of a Wannier-Stark localized doublon-hole pair~\cite{suppmat}. Tunneling of a doublon according to $\hat{T}_3$ occurs with amplitude $J^{(3)}$ and the effective tilt, which detunes this process, is given by $\tilde{\Delta}_\downarrow - \tilde{\Delta}_\uparrow$. In analogy to single-particle Wannier-Stark localization the steady-state value is then given by the analytic expression $\bar{\mathcal{I}}_D = \mathcal{J}_0^2 \left(4J^{(3)}/(\tilde{\Delta}_\downarrow - \tilde{\Delta}_\uparrow) \right)$~\cite{suppmat}, which agrees remarkably well with the exact Hamiltonian dynamics; here $\mathcal{J}_0(x)$ is the zero-order Bessel function of the first kind. While the general trend observed in the experiment is reproduced by TEBD simulations, there is a systematic deviation for large $\delta\Delta$, which we attribute to technical heating induced by the transverse lattice laser beams, which affects mostly large imbalance values and mixed initial states.

\begin{figure}[t]
	\includegraphics[width=3.3in]{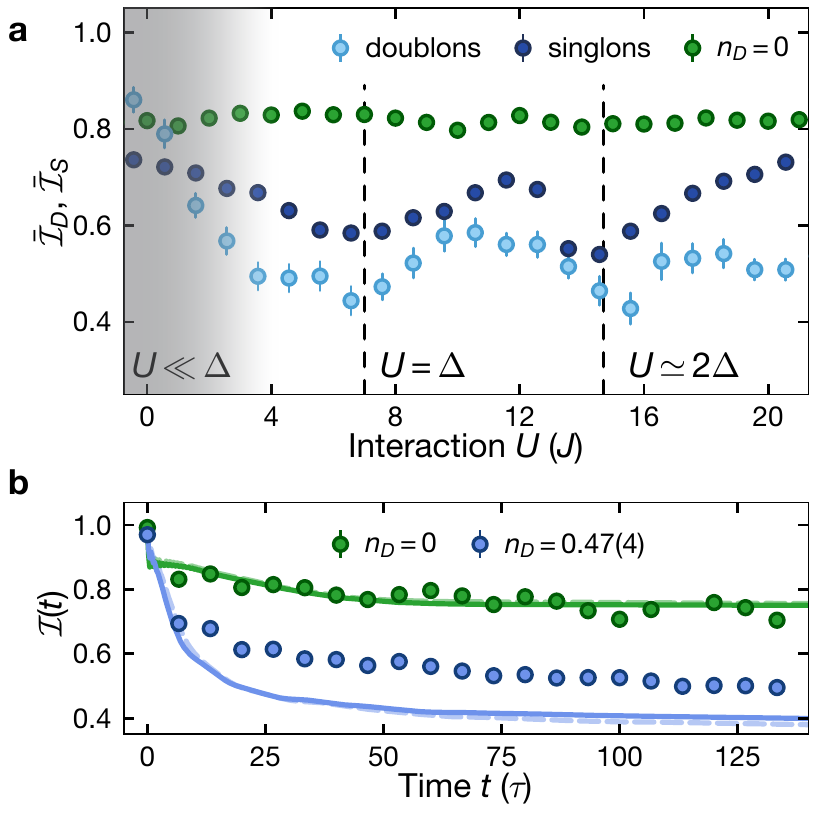}
	\caption{\textbf{Doublon-dependent dynamics for various interaction strengths and $\Delta/J = 8.0(2)$.} \textbf{a} Steady-state singlon and doublon imbalance for $n_D=0.47(4)$ and $\delta\Delta=0.6(2)\%$ averaged over $t\in[67\tau,80\tau]$. For comparison we also show the singlon initial state ($n_D\simeq 0$). The dashed vertical lines and the gray-shaded area highlight different regimes, where fragmented models have been found (main text). Data points contain eight averages over five points in time. Error bars are the SEM.
	 \textbf{b} Imbalance time traces at the resonance $U=14.7(2)J \simeq 2\Delta$ for a singlon initial state and with doublon fraction $n_D=0.47(4)$. The resonance is chosen as the local minimum of $\bar{\mathcal{I}}_S$ in (a) as indicated by the vertical dashed line. Error bars denote the SEM after ten averages. The lines represent time-averaged TEBD simulations of the effective (solid) and the exact Hamiltonian (dashed, transparent) for $L=51$, $n_D=0.46$ and a hole fraction of $20\%$ (See table \ref{parameter_table} in~\cite{suppmat} for numerical details).}
	\label{fig:fig3}
\end{figure}

The tilted 1D Fermi-Hubbard model exhibits rich non-ergodic phenomena depending on its microscopic parameters, including emergent Hilbert-space fragmentation and quantum scars. We study the doublon-dependent dynamics over a wide range of parameters by measuring the intermediate-time steady-state imbalance as a function of the Hubbard interaction strength for $\tilde{\Delta}_\uparrow\simeq\tilde{\Delta}_\downarrow$ (Fig.~\ref{fig:fig3}a).
We find that pure singlon initial states show no significant dependence, in agreement with previous experimental results~\cite{Scherg2020}. Initial states with doublons on the other hand, show a strong interaction as well as doublon-number dependent behavior, in particular near resonances between the tilt and interaction energy~\cite{Meinert2014}.
Intuitively, one may expect that away from the dipole-conserving regime ($\Delta \gg J, |U|$) the tilted Fermi-Hubbard model is ergodic due to the many resonances between the interaction $U$ and the tilt energy $\Delta$. Surprisingly, it was shown experimentally that non-ergodicity survives over large parameter ranges~\cite{Scherg2020}. Moreover, strongly-fragmented Hamiltonians have been derived, which govern the transient time dynamics even on resonance, as shown for $|U|\simeq2\Delta$ in Ref.~\cite{Scherg2020} and $|U|\simeq\Delta$ in Ref.~\cite{Desaules2021}. 

On the double-tilt resonance ($|U|\simeq2\Delta$) fragmentation is due to the conservation of dipole moment and doublon number: $\Delta \hat{P} + U\sum_i \hat{n}_{i,\uparrow} \hat{n}_{i,\downarrow}$. As before we compare the dynamics of initial states with and without doublons (Fig.~\ref{fig:fig3}b). We find a large reduction of the steady-state imbalance for $n_D>0$, similar to the results in (Fig.~\ref{fig:fig2}a). However, unlike in the previous regime, the singlon- and doublon-resolved imbalance time traces exhibit similar dynamics (see Fig.~\ref{fig:fig3supp} in~\cite{suppmat}), in agreement with numerical studies based on the exact [Eq.~\eqref{eq:Hamiltonian}] and effective Hamiltonian [Eq.~\eqref{eq:Heffdouble}]. This is a direct consequence of the microscopic processes, which directly affect the singlon imbalance in this regime [see Eq.~(\ref{eq:Hdoub_terms}) in~\cite{suppmat}].

\begin{figure}[t]
	\includegraphics[width=3.3in]{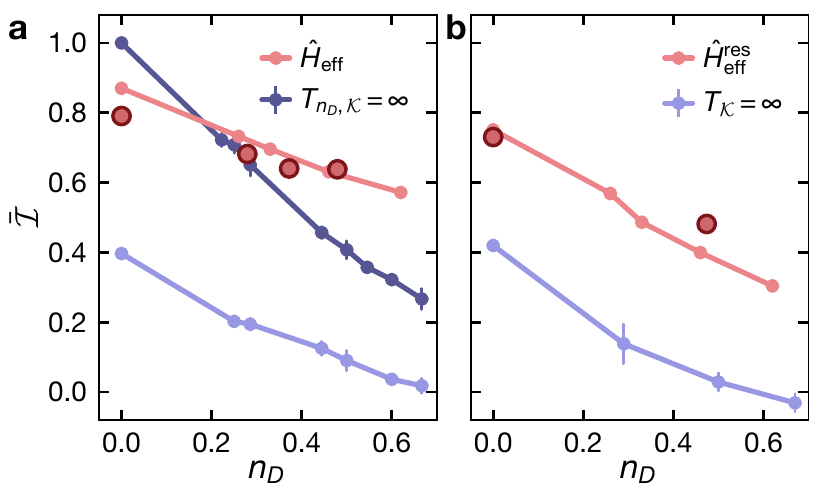}
	\caption{\textbf{Steady-state imbalance and infinite-temperature prediction.} The data points (big red dots) show the steady-state imbalance for $\delta\Delta=0.6(2)\%$ and $\Delta/J = 8.0(2)$, averaged between $120\tau$-$140\tau$ as a function of $n_D$ in the regime \textbf{a} $\Delta \gg J, |U|$ with $U/J = 2.7(2)$ and \textbf{b} $U=14.7(2)J \simeq 2\Delta$. Averaging has been done using four points in time and the error bars denote the respective SEM. Solid red lines are TEBD simulations (linear interpolation between numerical data points) with the effective Hamiltonians [Eq.~(\ref{eq:Heffdip}) in (a) and $\hat{H}_{\mathrm{eff}}^{\mathrm{res}}$ in Eq.~(\ref{eq:Heffdouble}) in (b)] with $L=51$ and a hole fraction of 20$\%$. The transparent blue lines (linear interpolation between numerical data points) show infinite-temperature predictions ($T_\mathcal{K}=\infty$) within the fragment $\mathcal{K}$ of the respective initial state calculated using exact diagonalization (ED) on 13, 15 and 17 lattice sites (without SW rotation of the initial state). The dark-blue line is the infinite-temperature prediction when imposing additional doublon-number conservation for Hamiltonian~(\ref{eq:Heffdip}). (See table \ref{parameter_table} in~\cite{suppmat} for numerical details).}
	\label{fig:fig4}
\end{figure}

The fragmented structure of the many-body Hilbert-space (Fig.~\ref{fig:fig1}b) naturally calls for a modified definition of ETH, where thermalization is defined not only with respect to a symmetry sector $\mathcal{S}$ but also to a particular fragment $\mathcal{K}$ (also known as Krylov-restricted thermalization). Within this modified framework the usual characteristics for identifying non-ergodic behavior within a fragment apply~\cite{Moudgalya2019,rakovszky_statistical_2020}. In this work we have studied the intermediate-time steady-state imbalance as a function of the fraction of doublons in the initial state in two regimes, where emergent fragmented models exist: the regime $\Delta\gg J, |U|$ (Fig.~\ref{fig:fig4}a) and the double-tilt resonance, where $|U|\simeq 2\Delta$ (Fig.~\ref{fig:fig4}b). Note, that initial states with different $n_D$ belong to different symmetry sectors $\mathcal{S}$. In both regimes the steady-state value $\bar{\mathcal{I}}$ decreases with increasing doublon fraction (for $n_D< 0.5$) in agreement with numerical simulations using the respective effective Hamiltonian [Eq.~(\ref{eq:Heffdip}) and $\hat{H}_{\mathrm{eff}}^{\mathrm{res}}$ in Eq.~(\ref{eq:Heffdouble})].
The system further relaxes to a finite value in contrast to ergodic systems thermalizing to infinite temperature within the symmetry sector $\mathcal{S}$, where $\mathcal{I}(\infty)=0$. Moreover, the steady-state value $\bar{\mathcal{I}}$ does not agree with the infinite-temperature prediction for thermalization within the corresponding fragments (blue theory
data without Schrieffer-Wolff rotation~\cite{suppmat}). Thus, in both regimes our experimental data
indicates that the system does not thermalize to an infinite-temperature state up to 140$\tau$. We emphasize that in general the dynamics within one fragment can be further constrained and exhibit non-ergodic behavior~\cite{Moudgalya2019}, so a priori we do not know, if the system would display thermal behavior even within this subspace. For instance, additional constraints have recently been identified in the resonant regime $|U|\simeq \Delta$~\cite{Desaules2021}. Indeed, if we impose an additional doublon-number conservation for the Hamiltonian in Eq.~(\ref{eq:Heffdip}) as suggested in Refs~\cite{Nieuwenburg2019,Scherg2020}, resulting in the effective Hamiltonian given in Eq.~\eqref{eq:T3_doublon} in~\cite{suppmat}, we find an infinite-temperature prediction~\cite{suppmat} that is much closer to our experimental data (Fig.~\ref{fig:fig4}a). 

In conclusion, we have shown that the dynamics of the tilted Fermi-Hubbard model in the large-tilt limit is well captured by effective perturbative Hamiltonians, which in certain parameter regimes are fragmented and give rise to non-ergodic behavior. We have studied non-equilibrium dynamics in two specific regimes, where this has been shown theoretically ($\Delta\gg J, |U|$ and the resonant regime $|U|\simeq 2\Delta$). At the same time we find no experimental evidence for ergodic behavior away from these limits, in agreement with previous experimental studies~\cite{Scherg2020}. We further demonstrate that the observed dependence of the dynamics on the number of doublons in the initial state is directly related to the microscopic processes of the effective Hamiltonian. It will be interesting to further systematically explore thermalization within individual fragments for the various different parameter regimes of the tilted Fermi-Hubbard model. Moreover, to reveal the fragmented nature of the spectrum more directly, one could further look at the thermalization of different initial states within the same symmetry sector $\mathcal{S}$. Moreover, at the resonance $|U|\simeq\Delta$ it was found that additional constraints result in scarring~\cite{Desaules2021}, which highlights the potential of this experimental platform for studying the interplay of both phenomena. Additionally, it is expected that higher-order terms in the perturbative expansion will generally lead to thermalization~\cite{Scherg2020}. However, due to the small amplitude of the higher-order terms even weak disorder or inhomogeneities can render these terms inefficient, which results in an interesting interplay between thermalization and localization connecting to the phenomenon of Stark-MBL~\cite{Schulz2019,Nieuwenburg2019}. The RF-dressing technique may further pave the way towards the implementation of effective spin models via precise control over the tilt difference. Intriguingly, by extending our system to 2D, we should be able to connect our studies to the emergence of hydrodynamic behavior~\cite{Bakr2020} and potentially realize higher-dimensional models with multipole moment conservation~\cite{Khemani2020,Nieuwenburg2019}.

\paragraph*{\textbf{Acknowledgments}}
M.~A. acknowledges fruitful discussions with K.~Shkedrov and Y.~Sagi. P.~S. acknowledges K. Hemery and J. Hauschild for their help implementing the MPS numerical simulations; and G. Tomasi, N. Pancotti, T. Rakovszky and C.Turner for helpful discussions. The MPS simulations were performed using the Tensor Network Python (TeNPy) package~\cite{tenpy}. This work was supported by the Deutsche Forschungsgemeinschaft (DFG, German Research Foundation) under Germany's Excellence Strategy -- EXC-2111 -- 39081486. The work at LMU was additionally supported by DIP and B.~H.~M. acknowledges support from the European Union (Marie Curie, Pasquans). Moreover, the work at TU was supported by the European Research Council (ERC) under the European Union's Horizon 2020 research and innovation program (grant agreement No. 771537).

\bibliography{Fragmentation}

\begin{thebibliography}{58}%
\makeatletter
\providecommand \@ifxundefined [1]{%
 \@ifx{#1\undefined}
}%
\providecommand \@ifnum [1]{%
 \ifnum #1\expandafter \@firstoftwo
 \else \expandafter \@secondoftwo
 \fi
}%
\providecommand \@ifx [1]{%
 \ifx #1\expandafter \@firstoftwo
 \else \expandafter \@secondoftwo
 \fi
}%
\providecommand \natexlab [1]{#1}%
\providecommand \enquote  [1]{``#1''}%
\providecommand \bibnamefont  [1]{#1}%
\providecommand \bibfnamefont [1]{#1}%
\providecommand \citenamefont [1]{#1}%
\providecommand \href@noop [0]{\@secondoftwo}%
\providecommand \href [0]{\begingroup \@sanitize@url \@href}%
\providecommand \@href[1]{\@@startlink{#1}\@@href}%
\providecommand \@@href[1]{\endgroup#1\@@endlink}%
\providecommand \@sanitize@url [0]{\catcode `\\12\catcode `\$12\catcode
  `\&12\catcode `\#12\catcode `\^12\catcode `\_12\catcode `\%12\relax}%
\providecommand \@@startlink[1]{}%
\providecommand \@@endlink[0]{}%
\providecommand \url  [0]{\begingroup\@sanitize@url \@url }%
\providecommand \@url [1]{\endgroup\@href {#1}{\urlprefix }}%
\providecommand \urlprefix  [0]{URL }%
\providecommand \Eprint [0]{\href }%
\providecommand \doibase [0]{https://doi.org/}%
\providecommand \selectlanguage [0]{\@gobble}%
\providecommand \bibinfo  [0]{\@secondoftwo}%
\providecommand \bibfield  [0]{\@secondoftwo}%
\providecommand \translation [1]{[#1]}%
\providecommand \BibitemOpen [0]{}%
\providecommand \bibitemStop [0]{}%
\providecommand \bibitemNoStop [0]{.\EOS\space}%
\providecommand \EOS [0]{\spacefactor3000\relax}%
\providecommand \BibitemShut  [1]{\csname bibitem#1\endcsname}%
\let\auto@bib@innerbib\@empty
\bibitem [{\citenamefont {Deutsch}(1991)}]{Deutsch1991}%
  \BibitemOpen
  \bibfield  {author} {\bibinfo {author} {\bibfnamefont {J.~M.}\ \bibnamefont
  {Deutsch}},\ }\bibfield  {title} {\bibinfo {title} {Quantum statistical
  mechanics in a closed system},\ }\href
  {https://doi.org/10.1103/PhysRevA.43.2046} {\bibfield  {journal} {\bibinfo
  {journal} {Phys. Rev. A}\ }\textbf {\bibinfo {volume} {43}},\ \bibinfo
  {pages} {2046} (\bibinfo {year} {1991})}\BibitemShut {NoStop}%
\bibitem [{\citenamefont {Srednicki}(1994)}]{Srednicki1994}%
  \BibitemOpen
  \bibfield  {author} {\bibinfo {author} {\bibfnamefont {M.}~\bibnamefont
  {Srednicki}},\ }\bibfield  {title} {\bibinfo {title} {Chaos and quantum
  thermalization},\ }\href {https://doi.org/10.1103/PhysRevE.50.888} {\bibfield
   {journal} {\bibinfo  {journal} {Phys. Rev. E}\ }\textbf {\bibinfo {volume}
  {50}},\ \bibinfo {pages} {888} (\bibinfo {year} {1994})}\BibitemShut
  {NoStop}%
\bibitem [{\citenamefont {Rigol}\ \emph {et~al.}(2008)\citenamefont {Rigol},
  \citenamefont {Dunjko},\ and\ \citenamefont {Olshanii}}]{Rigol2008}%
  \BibitemOpen
  \bibfield  {author} {\bibinfo {author} {\bibfnamefont {M.}~\bibnamefont
  {Rigol}}, \bibinfo {author} {\bibfnamefont {V.}~\bibnamefont {Dunjko}},\ and\
  \bibinfo {author} {\bibfnamefont {M.}~\bibnamefont {Olshanii}},\ }\bibfield
  {title} {\bibinfo {title} {Thermalization and its mechanism for generic
  isolated quantum systems},\ }\href {https://doi.org/10.1038/nature06838}
  {\bibfield  {journal} {\bibinfo  {journal} {Nature}\ }\textbf {\bibinfo
  {volume} {452}},\ \bibinfo {pages} {854} (\bibinfo {year}
  {2008})}\BibitemShut {NoStop}%
\bibitem [{\citenamefont {Calabrese}\ \emph {et~al.}(2011)\citenamefont
  {Calabrese}, \citenamefont {Essler},\ and\ \citenamefont
  {Fagotti}}]{Calabrese2011}%
  \BibitemOpen
  \bibfield  {author} {\bibinfo {author} {\bibfnamefont {P.}~\bibnamefont
  {Calabrese}}, \bibinfo {author} {\bibfnamefont {F.~H.~L.}\ \bibnamefont
  {Essler}},\ and\ \bibinfo {author} {\bibfnamefont {M.}~\bibnamefont
  {Fagotti}},\ }\bibfield  {title} {\bibinfo {title} {Quantum quench in the
  transverse-field {I}sing chain},\ }\href
  {https://doi.org/10.1103/PhysRevLett.106.227203} {\bibfield  {journal}
  {\bibinfo  {journal} {Phys. Rev. Lett.}\ }\textbf {\bibinfo {volume} {106}},\
  \bibinfo {pages} {227203} (\bibinfo {year} {2011})}\BibitemShut {NoStop}%
\bibitem [{\citenamefont {Essler}\ and\ \citenamefont
  {Fagotti}(2016)}]{Essler2016}%
  \BibitemOpen
  \bibfield  {author} {\bibinfo {author} {\bibfnamefont {F.~H.~L.}\
  \bibnamefont {Essler}}\ and\ \bibinfo {author} {\bibfnamefont
  {M.}~\bibnamefont {Fagotti}},\ }\bibfield  {title} {\bibinfo {title} {Quench
  dynamics and relaxation in isolated integrable quantum spin chains},\ }\href
  {https://doi.org/10.1088/1742-5468/2016/06/064002} {\bibfield  {journal}
  {\bibinfo  {journal} {Journal of Statistical Mechanics: Theory and
  Experiment}\ }\textbf {\bibinfo {volume} {2016}},\ \bibinfo {pages} {064002}
  (\bibinfo {year} {2016})}\BibitemShut {NoStop}%
\bibitem [{\citenamefont {Schreiber}\ \emph {et~al.}(2015)\citenamefont
  {Schreiber}, \citenamefont {Hodgman}, \citenamefont {Bordia}, \citenamefont
  {L\"uschen}, \citenamefont {Fischer}, \citenamefont {Vosk}, \citenamefont
  {Altman}, \citenamefont {Schneider},\ and\ \citenamefont
  {Bloch}}]{Schreiber2015}%
  \BibitemOpen
  \bibfield  {author} {\bibinfo {author} {\bibfnamefont {M.}~\bibnamefont
  {Schreiber}}, \bibinfo {author} {\bibfnamefont {S.~S.}\ \bibnamefont
  {Hodgman}}, \bibinfo {author} {\bibfnamefont {P.}~\bibnamefont {Bordia}},
  \bibinfo {author} {\bibfnamefont {H.~P.}\ \bibnamefont {L\"uschen}}, \bibinfo
  {author} {\bibfnamefont {M.~H.}\ \bibnamefont {Fischer}}, \bibinfo {author}
  {\bibfnamefont {R.}~\bibnamefont {Vosk}}, \bibinfo {author} {\bibfnamefont
  {E.}~\bibnamefont {Altman}}, \bibinfo {author} {\bibfnamefont
  {U.}~\bibnamefont {Schneider}},\ and\ \bibinfo {author} {\bibfnamefont
  {I.}~\bibnamefont {Bloch}},\ }\bibfield  {title} {\bibinfo {title}
  {Observation of many-body localization of interacting fermions in a
  quasirandom optical lattice},\ }\href
  {https://doi.org/10.1126/science.aaa7432} {\bibfield  {journal} {\bibinfo
  {journal} {Science}\ }\textbf {\bibinfo {volume} {349}},\ \bibinfo {pages}
  {842} (\bibinfo {year} {2015})}\BibitemShut {NoStop}%
\bibitem [{\citenamefont {Gornyi}\ \emph {et~al.}(2005)\citenamefont {Gornyi},
  \citenamefont {Mirlin},\ and\ \citenamefont {Polyakov}}]{Gornyi2005}%
  \BibitemOpen
  \bibfield  {author} {\bibinfo {author} {\bibfnamefont {I.~V.}\ \bibnamefont
  {Gornyi}}, \bibinfo {author} {\bibfnamefont {A.~D.}\ \bibnamefont {Mirlin}},\
  and\ \bibinfo {author} {\bibfnamefont {D.~G.}\ \bibnamefont {Polyakov}},\
  }\bibfield  {title} {\bibinfo {title} {Interacting electrons in disordered
  wires: Anderson localization and low-{$T$} transport},\ }\href
  {https://doi.org/10.1103/PhysRevLett.95.206603} {\bibfield  {journal}
  {\bibinfo  {journal} {Phys. Rev. Lett.}\ }\textbf {\bibinfo {volume} {95}},\
  \bibinfo {pages} {206603} (\bibinfo {year} {2005})}\BibitemShut {NoStop}%
\bibitem [{\citenamefont {Basko}\ \emph {et~al.}(2006)\citenamefont {Basko},
  \citenamefont {Aleiner},\ and\ \citenamefont {Altshuler}}]{Basko2006}%
  \BibitemOpen
  \bibfield  {author} {\bibinfo {author} {\bibfnamefont {D.}~\bibnamefont
  {Basko}}, \bibinfo {author} {\bibfnamefont {I.}~\bibnamefont {Aleiner}},\
  and\ \bibinfo {author} {\bibfnamefont {B.}~\bibnamefont {Altshuler}},\
  }\bibfield  {title} {\bibinfo {title} {Metal–insulator transition in a
  weakly interacting many-electron system with localized single-particle
  states},\ }\href {https://doi.org/https://doi.org/10.1016/j.aop.2005.11.014}
  {\bibfield  {journal} {\bibinfo  {journal} {Annals of Physics}\ }\textbf
  {\bibinfo {volume} {321}},\ \bibinfo {pages} {1126 } (\bibinfo {year}
  {2006})}\BibitemShut {NoStop}%
\bibitem [{\citenamefont {Nandkishore}\ and\ \citenamefont
  {Huse}(2015)}]{NandkishoreMBL}%
  \BibitemOpen
  \bibfield  {author} {\bibinfo {author} {\bibfnamefont {R.}~\bibnamefont
  {Nandkishore}}\ and\ \bibinfo {author} {\bibfnamefont {D.~A.}\ \bibnamefont
  {Huse}},\ }\bibfield  {title} {\bibinfo {title} {Many-body localization and
  thermalization in quantum statistical mechanics},\ }\href
  {https://doi.org/10.1146/annurev-conmatphys-031214-014726} {\bibfield
  {journal} {\bibinfo  {journal} {Annual Review of Condensed Matter Physics}\
  }\textbf {\bibinfo {volume} {6}},\ \bibinfo {pages} {15} (\bibinfo {year}
  {2015})}\BibitemShut {NoStop}%
\bibitem [{\citenamefont {Abanin}\ \emph {et~al.}(2019)\citenamefont {Abanin},
  \citenamefont {Altman}, \citenamefont {Bloch},\ and\ \citenamefont
  {Serbyn}}]{abanin_colloquium_2019}%
  \BibitemOpen
  \bibfield  {author} {\bibinfo {author} {\bibfnamefont {D.~A.}\ \bibnamefont
  {Abanin}}, \bibinfo {author} {\bibfnamefont {E.}~\bibnamefont {Altman}},
  \bibinfo {author} {\bibfnamefont {I.}~\bibnamefont {Bloch}},\ and\ \bibinfo
  {author} {\bibfnamefont {M.}~\bibnamefont {Serbyn}},\ }\bibfield  {title}
  {\bibinfo {title} {Colloquium: {Many}-body localization, thermalization, and
  entanglement},\ }\href {https://doi.org/10.1103/RevModPhys.91.021001}
  {\bibfield  {journal} {\bibinfo  {journal} {Rev. Mod. Phys.}\ }\textbf
  {\bibinfo {volume} {91}},\ \bibinfo {pages} {021001} (\bibinfo {year}
  {2019})}\BibitemShut {NoStop}%
\bibitem [{\citenamefont {Serbyn}\ \emph {et~al.}(2013)\citenamefont {Serbyn},
  \citenamefont {Papi\ifmmode~\acute{c}\else \'{c}\fi{}},\ and\ \citenamefont
  {Abanin}}]{Serbyn13}%
  \BibitemOpen
  \bibfield  {author} {\bibinfo {author} {\bibfnamefont {M.}~\bibnamefont
  {Serbyn}}, \bibinfo {author} {\bibfnamefont {Z.}~\bibnamefont
  {Papi\ifmmode~\acute{c}\else \'{c}\fi{}}},\ and\ \bibinfo {author}
  {\bibfnamefont {D.~A.}\ \bibnamefont {Abanin}},\ }\bibfield  {title}
  {\bibinfo {title} {Local conservation laws and the structure of the many-body
  localized states},\ }\href {https://doi.org/10.1103/PhysRevLett.111.127201}
  {\bibfield  {journal} {\bibinfo  {journal} {Phys. Rev. Lett.}\ }\textbf
  {\bibinfo {volume} {111}},\ \bibinfo {pages} {127201} (\bibinfo {year}
  {2013})}\BibitemShut {NoStop}%
\bibitem [{\citenamefont {Huse}\ \emph {et~al.}(2014)\citenamefont {Huse},
  \citenamefont {Nandkishore},\ and\ \citenamefont {Oganesyan}}]{Huse14}%
  \BibitemOpen
  \bibfield  {author} {\bibinfo {author} {\bibfnamefont {D.~A.}\ \bibnamefont
  {Huse}}, \bibinfo {author} {\bibfnamefont {R.}~\bibnamefont {Nandkishore}},\
  and\ \bibinfo {author} {\bibfnamefont {V.}~\bibnamefont {Oganesyan}},\
  }\bibfield  {title} {\bibinfo {title} {Phenomenology of fully
  many-body-localized systems},\ }\href
  {https://doi.org/10.1103/PhysRevB.90.174202} {\bibfield  {journal} {\bibinfo
  {journal} {Phys. Rev. B}\ }\textbf {\bibinfo {volume} {90}},\ \bibinfo
  {pages} {174202} (\bibinfo {year} {2014})}\BibitemShut {NoStop}%
\bibitem [{\citenamefont {Serbyn}\ \emph {et~al.}(2020)\citenamefont {Serbyn},
  \citenamefont {Abanin},\ and\ \citenamefont {Papi\'{c}}}]{Scars_review2020}%
  \BibitemOpen
  \bibfield  {author} {\bibinfo {author} {\bibfnamefont {M.}~\bibnamefont
  {Serbyn}}, \bibinfo {author} {\bibfnamefont {D.~A.}\ \bibnamefont {Abanin}},\
  and\ \bibinfo {author} {\bibfnamefont {Z.}~\bibnamefont {Papi\'{c}}},\
  }\bibfield  {title} {\bibinfo {title} {Quantum many-body scars and weak
  breaking of ergodicity},\ }\href {https://arxiv.org/pdf/2011.09486.pdf}
  {\bibfield  {journal} {\bibinfo  {journal} {arXiv:2011.09486}\ } (\bibinfo
  {year} {2020})}\BibitemShut {NoStop}%
\bibitem [{\citenamefont {Turner}\ \emph
  {et~al.}(2018{\natexlab{a}})\citenamefont {Turner}, \citenamefont
  {Michailidis}, \citenamefont {Abanin}, \citenamefont {Serbyn},\ and\
  \citenamefont {Papi\'{c}}}]{Turner2018Nature}%
  \BibitemOpen
  \bibfield  {author} {\bibinfo {author} {\bibfnamefont {C.~J.}\ \bibnamefont
  {Turner}}, \bibinfo {author} {\bibfnamefont {A.~A.}\ \bibnamefont
  {Michailidis}}, \bibinfo {author} {\bibfnamefont {D.~A.}\ \bibnamefont
  {Abanin}}, \bibinfo {author} {\bibfnamefont {M.}~\bibnamefont {Serbyn}},\
  and\ \bibinfo {author} {\bibfnamefont {Z.}~\bibnamefont {Papi\'{c}}},\
  }\bibfield  {title} {\bibinfo {title} {Weak ergodicity breaking from quantum
  many-body scars},\ }\href {https://doi.org/10.1038/s41567-018-0137-5}
  {\bibfield  {journal} {\bibinfo  {journal} {Nat. Phys.}\ }\textbf {\bibinfo
  {volume} {14}},\ \bibinfo {pages} {745} (\bibinfo {year}
  {2018}{\natexlab{a}})}\BibitemShut {NoStop}%
\bibitem [{\citenamefont {Turner}\ \emph
  {et~al.}(2018{\natexlab{b}})\citenamefont {Turner}, \citenamefont
  {Michailidis}, \citenamefont {Abanin}, \citenamefont {Serbyn},\ and\
  \citenamefont {Papi\'{c}}}]{Turner2018}%
  \BibitemOpen
  \bibfield  {author} {\bibinfo {author} {\bibfnamefont {C.~J.}\ \bibnamefont
  {Turner}}, \bibinfo {author} {\bibfnamefont {A.~A.}\ \bibnamefont
  {Michailidis}}, \bibinfo {author} {\bibfnamefont {D.~A.}\ \bibnamefont
  {Abanin}}, \bibinfo {author} {\bibfnamefont {M.}~\bibnamefont {Serbyn}},\
  and\ \bibinfo {author} {\bibfnamefont {Z.}~\bibnamefont {Papi\'{c}}},\
  }\bibfield  {title} {\bibinfo {title} {Quantum scarred eigenstates in a
  {R}ydberg atom chain: Entanglement, breakdown of thermalization, and
  stability to perturbations},\ }\href
  {https://doi.org/10.1103/PhysRevB.98.155134} {\bibfield  {journal} {\bibinfo
  {journal} {Phys. Rev. B}\ }\textbf {\bibinfo {volume} {98}},\ \bibinfo
  {pages} {155134} (\bibinfo {year} {2018}{\natexlab{b}})}\BibitemShut
  {NoStop}%
\bibitem [{\citenamefont {Chen}\ \emph {et~al.}(2018)\citenamefont {Chen},
  \citenamefont {Burnell},\ and\ \citenamefont {Chandran}}]{Chen2018}%
  \BibitemOpen
  \bibfield  {author} {\bibinfo {author} {\bibfnamefont {C.}~\bibnamefont
  {Chen}}, \bibinfo {author} {\bibfnamefont {F.}~\bibnamefont {Burnell}},\ and\
  \bibinfo {author} {\bibfnamefont {A.}~\bibnamefont {Chandran}},\ }\bibfield
  {title} {\bibinfo {title} {How does a locally constrained quantum system
  localize?},\ }\href {https://doi.org/10.1103/PhysRevLett.121.085701}
  {\bibfield  {journal} {\bibinfo  {journal} {Phys. Rev. Lett.}\ }\textbf
  {\bibinfo {volume} {121}},\ \bibinfo {pages} {085701} (\bibinfo {year}
  {2018})}\BibitemShut {NoStop}%
\bibitem [{\citenamefont {Lin}\ and\ \citenamefont
  {Motrunich}(2019)}]{ChengJu2019}%
  \BibitemOpen
  \bibfield  {author} {\bibinfo {author} {\bibfnamefont {C.-J.}\ \bibnamefont
  {Lin}}\ and\ \bibinfo {author} {\bibfnamefont {O.~I.}\ \bibnamefont
  {Motrunich}},\ }\bibfield  {title} {\bibinfo {title} {Exact quantum many-body
  scar states in the {R}ydberg-blockaded atom chain},\ }\href
  {https://doi.org/10.1103/PhysRevLett.122.173401} {\bibfield  {journal}
  {\bibinfo  {journal} {Phys. Rev. Lett.}\ }\textbf {\bibinfo {volume} {122}},\
  \bibinfo {pages} {173401} (\bibinfo {year} {2019})}\BibitemShut {NoStop}%
\bibitem [{\citenamefont {Schecter}\ and\ \citenamefont
  {Iadecola}(2019)}]{Schecter2019}%
  \BibitemOpen
  \bibfield  {author} {\bibinfo {author} {\bibfnamefont {M.}~\bibnamefont
  {Schecter}}\ and\ \bibinfo {author} {\bibfnamefont {T.}~\bibnamefont
  {Iadecola}},\ }\bibfield  {title} {\bibinfo {title} {Weak ergodicity breaking
  and quantum many-body scars in spin-1 {$XY$} magnets},\ }\href
  {https://doi.org/10.1103/PhysRevLett.123.147201} {\bibfield  {journal}
  {\bibinfo  {journal} {Phys. Rev. Lett.}\ }\textbf {\bibinfo {volume} {123}},\
  \bibinfo {pages} {147201} (\bibinfo {year} {2019})}\BibitemShut {NoStop}%
\bibitem [{\citenamefont {Pai}\ and\ \citenamefont {Pretko}(2019)}]{Pai2019}%
  \BibitemOpen
  \bibfield  {author} {\bibinfo {author} {\bibfnamefont {S.}~\bibnamefont
  {Pai}}\ and\ \bibinfo {author} {\bibfnamefont {M.}~\bibnamefont {Pretko}},\
  }\bibfield  {title} {\bibinfo {title} {Dynamical scar states in driven
  fracton systems},\ }\href {https://doi.org/10.1103/PhysRevLett.123.136401}
  {\bibfield  {journal} {\bibinfo  {journal} {Phys. Rev. Lett.}\ }\textbf
  {\bibinfo {volume} {123}},\ \bibinfo {pages} {136401} (\bibinfo {year}
  {2019})}\BibitemShut {NoStop}%
\bibitem [{\citenamefont {Zhao}\ \emph {et~al.}(2020)\citenamefont {Zhao},
  \citenamefont {Vovrosh}, \citenamefont {Mintert},\ and\ \citenamefont
  {Knolle}}]{Zhao2020}%
  \BibitemOpen
  \bibfield  {author} {\bibinfo {author} {\bibfnamefont {H.}~\bibnamefont
  {Zhao}}, \bibinfo {author} {\bibfnamefont {J.}~\bibnamefont {Vovrosh}},
  \bibinfo {author} {\bibfnamefont {F.}~\bibnamefont {Mintert}},\ and\ \bibinfo
  {author} {\bibfnamefont {J.}~\bibnamefont {Knolle}},\ }\bibfield  {title}
  {\bibinfo {title} {Quantum many-body scars in optical lattices},\ }\href
  {https://doi.org/10.1103/PhysRevLett.124.160604} {\bibfield  {journal}
  {\bibinfo  {journal} {Phys. Rev. Lett.}\ }\textbf {\bibinfo {volume} {124}},\
  \bibinfo {pages} {160604} (\bibinfo {year} {2020})}\BibitemShut {NoStop}%
\bibitem [{\citenamefont {Bernien}\ \emph {et~al.}(2017)\citenamefont
  {Bernien}, \citenamefont {Schwartz}, \citenamefont {Keesling}, \citenamefont
  {Levine}, \citenamefont {Omran}, \citenamefont {Pichler}, \citenamefont
  {Choi}, \citenamefont {Zibrov}, \citenamefont {Endres}, \citenamefont
  {Greiner}, \citenamefont {Vuleti\'{c}},\ and\ \citenamefont
  {Lukin}}]{Bernien2017}%
  \BibitemOpen
  \bibfield  {author} {\bibinfo {author} {\bibfnamefont {H.}~\bibnamefont
  {Bernien}}, \bibinfo {author} {\bibfnamefont {S.}~\bibnamefont {Schwartz}},
  \bibinfo {author} {\bibfnamefont {A.}~\bibnamefont {Keesling}}, \bibinfo
  {author} {\bibfnamefont {H.}~\bibnamefont {Levine}}, \bibinfo {author}
  {\bibfnamefont {A.}~\bibnamefont {Omran}}, \bibinfo {author} {\bibfnamefont
  {H.}~\bibnamefont {Pichler}}, \bibinfo {author} {\bibfnamefont
  {S.}~\bibnamefont {Choi}}, \bibinfo {author} {\bibfnamefont {A.~S.}\
  \bibnamefont {Zibrov}}, \bibinfo {author} {\bibfnamefont {M.}~\bibnamefont
  {Endres}}, \bibinfo {author} {\bibfnamefont {M.}~\bibnamefont {Greiner}},
  \bibinfo {author} {\bibfnamefont {V.}~\bibnamefont {Vuleti\'{c}}},\ and\
  \bibinfo {author} {\bibfnamefont {M.~D.}\ \bibnamefont {Lukin}},\ }\bibfield
  {title} {\bibinfo {title} {Probing many-body dynamics on a 51-atom quantum
  simulator},\ }\href {https://www.nature.com/articles/nature24622} {\bibfield
  {journal} {\bibinfo  {journal} {Nature}\ }\textbf {\bibinfo {volume} {551}},\
  \bibinfo {pages} {579} (\bibinfo {year} {2017})}\BibitemShut {NoStop}%
\bibitem [{\citenamefont {Bluvstein}\ \emph {et~al.}(2021)\citenamefont
  {Bluvstein}, \citenamefont {Omran}, \citenamefont {Levine}, \citenamefont
  {Keesling}, \citenamefont {Semeghini}, \citenamefont {Ebadi}, \citenamefont
  {Wang}, \citenamefont {Michailidis}, \citenamefont {Maskara}, \citenamefont
  {Ho}, \citenamefont {Choi}, \citenamefont {Serbyn}, \citenamefont {Greiner},
  \citenamefont {Vuleti{\'c}},\ and\ \citenamefont {Lukin}}]{Bluvstein2020}%
  \BibitemOpen
  \bibfield  {author} {\bibinfo {author} {\bibfnamefont {D.}~\bibnamefont
  {Bluvstein}}, \bibinfo {author} {\bibfnamefont {A.}~\bibnamefont {Omran}},
  \bibinfo {author} {\bibfnamefont {H.}~\bibnamefont {Levine}}, \bibinfo
  {author} {\bibfnamefont {A.}~\bibnamefont {Keesling}}, \bibinfo {author}
  {\bibfnamefont {G.}~\bibnamefont {Semeghini}}, \bibinfo {author}
  {\bibfnamefont {S.}~\bibnamefont {Ebadi}}, \bibinfo {author} {\bibfnamefont
  {T.~T.}\ \bibnamefont {Wang}}, \bibinfo {author} {\bibfnamefont {A.~A.}\
  \bibnamefont {Michailidis}}, \bibinfo {author} {\bibfnamefont
  {N.}~\bibnamefont {Maskara}}, \bibinfo {author} {\bibfnamefont {W.~W.}\
  \bibnamefont {Ho}}, \bibinfo {author} {\bibfnamefont {S.}~\bibnamefont
  {Choi}}, \bibinfo {author} {\bibfnamefont {M.}~\bibnamefont {Serbyn}},
  \bibinfo {author} {\bibfnamefont {M.}~\bibnamefont {Greiner}}, \bibinfo
  {author} {\bibfnamefont {V.}~\bibnamefont {Vuleti{\'c}}},\ and\ \bibinfo
  {author} {\bibfnamefont {M.~D.}\ \bibnamefont {Lukin}},\ }\bibfield  {title}
  {\bibinfo {title} {Controlling quantum many-body dynamics in driven {R}ydberg
  atom arrays},\ }\href {https://doi.org/10.1126/science.abg2530} {\bibfield
  {journal} {\bibinfo  {journal} {Science}\ }\textbf {\bibinfo {volume}
  {371}},\ \bibinfo {pages} {1355} (\bibinfo {year} {2021})}\BibitemShut
  {NoStop}%
\bibitem [{\citenamefont {Sala}\ \emph {et~al.}(2020)\citenamefont {Sala},
  \citenamefont {Rakovszky}, \citenamefont {Verresen}, \citenamefont {Knap},\
  and\ \citenamefont {Pollmann}}]{Sala2020}%
  \BibitemOpen
  \bibfield  {author} {\bibinfo {author} {\bibfnamefont {P.}~\bibnamefont
  {Sala}}, \bibinfo {author} {\bibfnamefont {T.}~\bibnamefont {Rakovszky}},
  \bibinfo {author} {\bibfnamefont {R.}~\bibnamefont {Verresen}}, \bibinfo
  {author} {\bibfnamefont {M.}~\bibnamefont {Knap}},\ and\ \bibinfo {author}
  {\bibfnamefont {F.}~\bibnamefont {Pollmann}},\ }\bibfield  {title} {\bibinfo
  {title} {Ergodicity breaking arising from {H}ilbert space fragmentation in
  dipole-conserving {H}amiltonians},\ }\href
  {https://doi.org/10.1103/PhysRevX.10.011047} {\bibfield  {journal} {\bibinfo
  {journal} {Phys. Rev. X}\ }\textbf {\bibinfo {volume} {10}},\ \bibinfo
  {pages} {011047} (\bibinfo {year} {2020})}\BibitemShut {NoStop}%
\bibitem [{\citenamefont {Khemani}\ \emph {et~al.}(2020)\citenamefont
  {Khemani}, \citenamefont {Hermele},\ and\ \citenamefont
  {Nandkishore}}]{Khemani2020}%
  \BibitemOpen
  \bibfield  {author} {\bibinfo {author} {\bibfnamefont {V.}~\bibnamefont
  {Khemani}}, \bibinfo {author} {\bibfnamefont {M.}~\bibnamefont {Hermele}},\
  and\ \bibinfo {author} {\bibfnamefont {R.}~\bibnamefont {Nandkishore}},\
  }\bibfield  {title} {\bibinfo {title} {Localization from {H}ilbert space
  shattering: From theory to physical realizations},\ }\href
  {https://doi.org/10.1103/PhysRevB.101.174204} {\bibfield  {journal} {\bibinfo
   {journal} {Phys. Rev. B}\ }\textbf {\bibinfo {volume} {101}},\ \bibinfo
  {pages} {174204} (\bibinfo {year} {2020})}\BibitemShut {NoStop}%
\bibitem [{\citenamefont {Moudgalya}\ \emph {et~al.}(2019)\citenamefont
  {Moudgalya}, \citenamefont {Prem}, \citenamefont {Nandkishore}, \citenamefont
  {Regnault},\ and\ \citenamefont {Bernevig}}]{Moudgalya2019}%
  \BibitemOpen
  \bibfield  {author} {\bibinfo {author} {\bibfnamefont {S.}~\bibnamefont
  {Moudgalya}}, \bibinfo {author} {\bibfnamefont {A.}~\bibnamefont {Prem}},
  \bibinfo {author} {\bibfnamefont {R.}~\bibnamefont {Nandkishore}}, \bibinfo
  {author} {\bibfnamefont {N.}~\bibnamefont {Regnault}},\ and\ \bibinfo
  {author} {\bibfnamefont {B.~A.}\ \bibnamefont {Bernevig}},\ }\bibfield
  {title} {\bibinfo {title} {Thermalization and its absence within {K}rylov
  subspaces of a constrained {H}amiltonian},\ }\href
  {https://arxiv.org/abs/1910.14048} {\bibfield  {journal} {\bibinfo  {journal}
  {arXiv:1910.14048}\ } (\bibinfo {year} {2019})}\BibitemShut {NoStop}%
\bibitem [{\citenamefont {Pai}\ \emph {et~al.}(2019)\citenamefont {Pai},
  \citenamefont {Pretko},\ and\ \citenamefont
  {Nandkishore}}]{NandkishoreRC_2019}%
  \BibitemOpen
  \bibfield  {author} {\bibinfo {author} {\bibfnamefont {S.}~\bibnamefont
  {Pai}}, \bibinfo {author} {\bibfnamefont {M.}~\bibnamefont {Pretko}},\ and\
  \bibinfo {author} {\bibfnamefont {R.~M.}\ \bibnamefont {Nandkishore}},\
  }\bibfield  {title} {\bibinfo {title} {Localization in fractonic random
  circuits},\ }\href {https://doi.org/10.1103/PhysRevX.9.021003} {\bibfield
  {journal} {\bibinfo  {journal} {Phys. Rev. X}\ }\textbf {\bibinfo {volume}
  {9}},\ \bibinfo {pages} {021003} (\bibinfo {year} {2019})}\BibitemShut
  {NoStop}%
\bibitem [{\citenamefont {Taylor}\ \emph {et~al.}(2020)\citenamefont {Taylor},
  \citenamefont {Schulz}, \citenamefont {Pollmann},\ and\ \citenamefont
  {Moessner}}]{Taylor_Stark2020}%
  \BibitemOpen
  \bibfield  {author} {\bibinfo {author} {\bibfnamefont {S.~R.}\ \bibnamefont
  {Taylor}}, \bibinfo {author} {\bibfnamefont {M.}~\bibnamefont {Schulz}},
  \bibinfo {author} {\bibfnamefont {F.}~\bibnamefont {Pollmann}},\ and\
  \bibinfo {author} {\bibfnamefont {R.}~\bibnamefont {Moessner}},\ }\bibfield
  {title} {\bibinfo {title} {Experimental probes of {S}tark many-body
  localization},\ }\href {https://doi.org/10.1103/PhysRevB.102.054206}
  {\bibfield  {journal} {\bibinfo  {journal} {Phys. Rev. B}\ }\textbf {\bibinfo
  {volume} {102}},\ \bibinfo {pages} {054206} (\bibinfo {year}
  {2020})}\BibitemShut {NoStop}%
\bibitem [{\citenamefont {Ritort}\ and\ \citenamefont
  {Sollich}(2003)}]{Ritort2003}%
  \BibitemOpen
  \bibfield  {author} {\bibinfo {author} {\bibfnamefont {F.}~\bibnamefont
  {Ritort}}\ and\ \bibinfo {author} {\bibfnamefont {P.}~\bibnamefont
  {Sollich}},\ }\bibfield  {title} {\bibinfo {title} {Glassy dynamics of
  kinetically constrained models},\ }\href
  {https://doi.org/10.1080/0001873031000093582} {\bibfield  {journal} {\bibinfo
   {journal} {Advances in Physics}\ }\textbf {\bibinfo {volume} {52}},\
  \bibinfo {pages} {219} (\bibinfo {year} {2003})}\BibitemShut {NoStop}%
\bibitem [{\citenamefont {Pancotti}\ \emph {et~al.}(2020)\citenamefont
  {Pancotti}, \citenamefont {Giudice}, \citenamefont {Cirac}, \citenamefont
  {Garrahan},\ and\ \citenamefont {Ba\~nuls}}]{Pancotti2020}%
  \BibitemOpen
  \bibfield  {author} {\bibinfo {author} {\bibfnamefont {N.}~\bibnamefont
  {Pancotti}}, \bibinfo {author} {\bibfnamefont {G.}~\bibnamefont {Giudice}},
  \bibinfo {author} {\bibfnamefont {J.~I.}\ \bibnamefont {Cirac}}, \bibinfo
  {author} {\bibfnamefont {J.~P.}\ \bibnamefont {Garrahan}},\ and\ \bibinfo
  {author} {\bibfnamefont {M.~C.}\ \bibnamefont {Ba\~nuls}},\ }\bibfield
  {title} {\bibinfo {title} {Quantum {E}ast {M}odel: Localization, nonthermal
  eigenstates, and slow dynamics},\ }\href
  {https://doi.org/10.1103/PhysRevX.10.021051} {\bibfield  {journal} {\bibinfo
  {journal} {Phys. Rev. X}\ }\textbf {\bibinfo {volume} {10}},\ \bibinfo
  {pages} {021051} (\bibinfo {year} {2020})}\BibitemShut {NoStop}%
\bibitem [{\citenamefont {Scherg}\ \emph {et~al.}(2020)\citenamefont {Scherg},
  \citenamefont {Kohlert}, \citenamefont {Sala}, \citenamefont {Pollmann},
  \citenamefont {H.~M.}, \citenamefont {Bloch},\ and\ \citenamefont
  {Aidelsburger}}]{Scherg2020}%
  \BibitemOpen
  \bibfield  {author} {\bibinfo {author} {\bibfnamefont {S.}~\bibnamefont
  {Scherg}}, \bibinfo {author} {\bibfnamefont {T.}~\bibnamefont {Kohlert}},
  \bibinfo {author} {\bibfnamefont {P.}~\bibnamefont {Sala}}, \bibinfo {author}
  {\bibfnamefont {F.}~\bibnamefont {Pollmann}}, \bibinfo {author}
  {\bibfnamefont {B.}~\bibnamefont {H.~M.}}, \bibinfo {author} {\bibfnamefont
  {I.}~\bibnamefont {Bloch}},\ and\ \bibinfo {author} {\bibfnamefont
  {M.}~\bibnamefont {Aidelsburger}},\ }\bibfield  {title} {\bibinfo {title}
  {Observing non-ergodicity due to kinetic constraints in tilted
  {F}ermi-{H}ubbard chains},\ }\href {https://arxiv.org/abs/2010.12965}
  {\bibfield  {journal} {\bibinfo  {journal} {arXiv:2010.12965}\ } (\bibinfo
  {year} {2020})}\BibitemShut {NoStop}%
\bibitem [{sup()}]{suppmat}%
  \BibitemOpen
  \href@noop {} {}\bibinfo {note} {See Supplementary Material, which includes
  Refs.~\cite{Schneider2008, Bordia2016, Feldmann1992, Salomon1996,
  Dalibard1992, Carmichael1992, Fischer2016, Schneider2012, Bharath2021,
  LewensteinBook, Abanin_theory, Vidal2003, Else2017, BRAVYI20112793,
  Weiss1999}, not cited in the main text, for details on: Experimental
  sequence, calibration and data acquisition techniques, initial state
  preparation and characterization, RF dressing, experimental imperfections,
  supplementary experimental data as well as details on analytical and
  numerical techniques (details on effective Hamiltonians, ED and TEBD
  simulations, Krylov methods and a quantitative analysis of experimental
  imperfections via an approximate method).}\BibitemShut {Stop}%
\bibitem [{\citenamefont {Sebby-Strabley}\ \emph {et~al.}(2006)\citenamefont
  {Sebby-Strabley}, \citenamefont {Anderlini}, \citenamefont {Jessen},\ and\
  \citenamefont {Porto}}]{Sebby2006}%
  \BibitemOpen
  \bibfield  {author} {\bibinfo {author} {\bibfnamefont {J.}~\bibnamefont
  {Sebby-Strabley}}, \bibinfo {author} {\bibfnamefont {M.}~\bibnamefont
  {Anderlini}}, \bibinfo {author} {\bibfnamefont {P.~S.}\ \bibnamefont
  {Jessen}},\ and\ \bibinfo {author} {\bibfnamefont {J.~V.}\ \bibnamefont
  {Porto}},\ }\bibfield  {title} {\bibinfo {title} {Lattice of double wells for
  manipulating pairs of cold atoms},\ }\href
  {https://doi.org/10.1103/PhysRevA.73.033605} {\bibfield  {journal} {\bibinfo
  {journal} {Phys. Rev. A}\ }\textbf {\bibinfo {volume} {73}},\ \bibinfo
  {pages} {033605} (\bibinfo {year} {2006})}\BibitemShut {NoStop}%
\bibitem [{\citenamefont {F{\"o}lling}\ \emph {et~al.}(2007)\citenamefont
  {F{\"o}lling}, \citenamefont {Trotzky}, \citenamefont {Cheinet},
  \citenamefont {Feld}, \citenamefont {Saers}, \citenamefont {Widera},
  \citenamefont {M{\"u}ller},\ and\ \citenamefont {Bloch}}]{Foelling07}%
  \BibitemOpen
  \bibfield  {author} {\bibinfo {author} {\bibfnamefont {S.}~\bibnamefont
  {F{\"o}lling}}, \bibinfo {author} {\bibfnamefont {S.}~\bibnamefont
  {Trotzky}}, \bibinfo {author} {\bibfnamefont {P.}~\bibnamefont {Cheinet}},
  \bibinfo {author} {\bibfnamefont {M.}~\bibnamefont {Feld}}, \bibinfo {author}
  {\bibfnamefont {R.}~\bibnamefont {Saers}}, \bibinfo {author} {\bibfnamefont
  {A.}~\bibnamefont {Widera}}, \bibinfo {author} {\bibfnamefont
  {T.}~\bibnamefont {M{\"u}ller}},\ and\ \bibinfo {author} {\bibfnamefont
  {I.}~\bibnamefont {Bloch}},\ }\bibfield  {title} {\bibinfo {title} {Direct
  observation of second-order atom tunnelling},\ }\href
  {https://www.nature.com/articles/nature06112} {\bibfield  {journal} {\bibinfo
   {journal} {Nature}\ }\textbf {\bibinfo {volume} {448}},\ \bibinfo {pages}
  {1029} (\bibinfo {year} {2007})}\BibitemShut {NoStop}%
\bibitem [{\citenamefont {Scherg}\ \emph {et~al.}(2018)\citenamefont {Scherg},
  \citenamefont {Kohlert}, \citenamefont {Herbrych}, \citenamefont {Stolpp},
  \citenamefont {Bordia}, \citenamefont {Schneider}, \citenamefont
  {Heidrich-Meisner}, \citenamefont {Bloch},\ and\ \citenamefont
  {Aidelsburger}}]{Scherg2018}%
  \BibitemOpen
  \bibfield  {author} {\bibinfo {author} {\bibfnamefont {S.}~\bibnamefont
  {Scherg}}, \bibinfo {author} {\bibfnamefont {T.}~\bibnamefont {Kohlert}},
  \bibinfo {author} {\bibfnamefont {J.}~\bibnamefont {Herbrych}}, \bibinfo
  {author} {\bibfnamefont {J.}~\bibnamefont {Stolpp}}, \bibinfo {author}
  {\bibfnamefont {P.}~\bibnamefont {Bordia}}, \bibinfo {author} {\bibfnamefont
  {U.}~\bibnamefont {Schneider}}, \bibinfo {author} {\bibfnamefont
  {F.}~\bibnamefont {Heidrich-Meisner}}, \bibinfo {author} {\bibfnamefont
  {I.}~\bibnamefont {Bloch}},\ and\ \bibinfo {author} {\bibfnamefont
  {M.}~\bibnamefont {Aidelsburger}},\ }\bibfield  {title} {\bibinfo {title}
  {Nonequilibrium mass transport in the 1{D} {F}ermi-{H}ubbard model},\ }\href
  {https://doi.org/10.1103/PhysRevLett.121.130402} {\bibfield  {journal}
  {\bibinfo  {journal} {Phys. Rev. Lett.}\ }\textbf {\bibinfo {volume} {121}},\
  \bibinfo {pages} {130402} (\bibinfo {year} {2018})}\BibitemShut {NoStop}%
\bibitem [{\citenamefont {Skedrov}\ \emph {et~al.}(2021)\citenamefont
  {Skedrov}, \citenamefont {Menashes}, \citenamefont {Ness}, \citenamefont
  {Vainbaum},\ and\ \citenamefont {Sagi}}]{Skedrov2021}%
  \BibitemOpen
  \bibfield  {author} {\bibinfo {author} {\bibfnamefont {C.}~\bibnamefont
  {Skedrov}}, \bibinfo {author} {\bibfnamefont {M.}~\bibnamefont {Menashes}},
  \bibinfo {author} {\bibfnamefont {G.}~\bibnamefont {Ness}}, \bibinfo {author}
  {\bibfnamefont {A.}~\bibnamefont {Vainbaum}},\ and\ \bibinfo {author}
  {\bibfnamefont {Y.}~\bibnamefont {Sagi}},\ }\bibfield  {title} {\bibinfo
  {title} {Absence of heating in a uniform {F}ermi gas created by periodic
  driving},\ }\href {https://arxiv.org/pdf/2102.09506.pdf} {\bibfield
  {journal} {\bibinfo  {journal} {arXiv:2102.09506}\ } (\bibinfo {year}
  {2021})}\BibitemShut {NoStop}%
\bibitem [{\citenamefont {Zwierlein}\ \emph {et~al.}(2003)\citenamefont
  {Zwierlein}, \citenamefont {Hadzibabic}, \citenamefont {Gupta},\ and\
  \citenamefont {Ketterle}}]{Zwierlein2003}%
  \BibitemOpen
  \bibfield  {author} {\bibinfo {author} {\bibfnamefont {M.~W.}\ \bibnamefont
  {Zwierlein}}, \bibinfo {author} {\bibfnamefont {Z.}~\bibnamefont
  {Hadzibabic}}, \bibinfo {author} {\bibfnamefont {S.}~\bibnamefont {Gupta}},\
  and\ \bibinfo {author} {\bibfnamefont {W.}~\bibnamefont {Ketterle}},\
  }\bibfield  {title} {\bibinfo {title} {Spectroscopic insensitivity to cold
  collisions in a two-state mixture of fermions},\ }\href
  {https://doi.org/10.1103/PhysRevLett.91.250404} {\bibfield  {journal}
  {\bibinfo  {journal} {Phys. Rev. Lett.}\ }\textbf {\bibinfo {volume} {91}},\
  \bibinfo {pages} {250404} (\bibinfo {year} {2003})}\BibitemShut {NoStop}%
\bibitem [{\citenamefont {Meinert}\ \emph {et~al.}(2014)\citenamefont
  {Meinert}, \citenamefont {Mark}, \citenamefont {Kirilov}, \citenamefont
  {Lauber}, \citenamefont {Weinmann}, \citenamefont {Gr{\"o}bner},
  \citenamefont {Daley},\ and\ \citenamefont {N{\"a}gerl}}]{Meinert2014}%
  \BibitemOpen
  \bibfield  {author} {\bibinfo {author} {\bibfnamefont {F.}~\bibnamefont
  {Meinert}}, \bibinfo {author} {\bibfnamefont {M.~J.}\ \bibnamefont {Mark}},
  \bibinfo {author} {\bibfnamefont {E.}~\bibnamefont {Kirilov}}, \bibinfo
  {author} {\bibfnamefont {K.}~\bibnamefont {Lauber}}, \bibinfo {author}
  {\bibfnamefont {P.}~\bibnamefont {Weinmann}}, \bibinfo {author}
  {\bibfnamefont {M.}~\bibnamefont {Gr{\"o}bner}}, \bibinfo {author}
  {\bibfnamefont {A.~J.}\ \bibnamefont {Daley}},\ and\ \bibinfo {author}
  {\bibfnamefont {H.-C.}\ \bibnamefont {N{\"a}gerl}},\ }\bibfield  {title}
  {\bibinfo {title} {Observation of many-body dynamics in long-range tunneling
  after a quantum quench},\ }\href {https://doi.org/10.1126/science.1248402}
  {\bibfield  {journal} {\bibinfo  {journal} {Science}\ }\textbf {\bibinfo
  {volume} {344}},\ \bibinfo {pages} {1259} (\bibinfo {year}
  {2014})}\BibitemShut {NoStop}%
\bibitem [{\citenamefont {Desaules}\ \emph {et~al.}(2021)\citenamefont
  {Desaules}, \citenamefont {Hudomal}, \citenamefont {Turner},\ and\
  \citenamefont {Papi\'{c}}}]{Desaules2021}%
  \BibitemOpen
  \bibfield  {author} {\bibinfo {author} {\bibfnamefont {J.-Y.}\ \bibnamefont
  {Desaules}}, \bibinfo {author} {\bibfnamefont {A.}~\bibnamefont {Hudomal}},
  \bibinfo {author} {\bibfnamefont {C.~J.}\ \bibnamefont {Turner}},\ and\
  \bibinfo {author} {\bibfnamefont {Z.}~\bibnamefont {Papi\'{c}}},\ }\bibfield
  {title} {\bibinfo {title} {A proposal for realizing quantum scars in the
  tilted {1D} {F}ermi-{H}ubbard model},\ }\href
  {https://arxiv.org/pdf/2102.01675.pdf} {\bibfield  {journal} {\bibinfo
  {journal} {arXiv:2102.01675}\ } (\bibinfo {year} {2021})}\BibitemShut
  {NoStop}%
\bibitem [{\citenamefont {Rakovszky}\ \emph {et~al.}(2020)\citenamefont
  {Rakovszky}, \citenamefont {Sala}, \citenamefont {Verresen}, \citenamefont
  {Knap},\ and\ \citenamefont {Pollmann}}]{rakovszky_statistical_2020}%
  \BibitemOpen
  \bibfield  {author} {\bibinfo {author} {\bibfnamefont {T.}~\bibnamefont
  {Rakovszky}}, \bibinfo {author} {\bibfnamefont {P.}~\bibnamefont {Sala}},
  \bibinfo {author} {\bibfnamefont {R.}~\bibnamefont {Verresen}}, \bibinfo
  {author} {\bibfnamefont {M.}~\bibnamefont {Knap}},\ and\ \bibinfo {author}
  {\bibfnamefont {F.}~\bibnamefont {Pollmann}},\ }\bibfield  {title} {\bibinfo
  {title} {Statistical localization: {From} strong fragmentation to strong edge
  modes},\ }\href {https://doi.org/10.1103/PhysRevB.101.125126} {\bibfield
  {journal} {\bibinfo  {journal} {Phys. Rev. B}\ }\textbf {\bibinfo {volume}
  {101}},\ \bibinfo {pages} {125126} (\bibinfo {year} {2020})}\BibitemShut
  {NoStop}%
\bibitem [{\citenamefont {van Nieuwenburg}\ \emph {et~al.}(2019)\citenamefont
  {van Nieuwenburg}, \citenamefont {Baum},\ and\ \citenamefont
  {Refael}}]{Nieuwenburg2019}%
  \BibitemOpen
  \bibfield  {author} {\bibinfo {author} {\bibfnamefont {E.}~\bibnamefont {van
  Nieuwenburg}}, \bibinfo {author} {\bibfnamefont {Y.}~\bibnamefont {Baum}},\
  and\ \bibinfo {author} {\bibfnamefont {G.}~\bibnamefont {Refael}},\
  }\bibfield  {title} {\bibinfo {title} {From {B}loch oscillations to many-body
  localization in clean interacting systems},\ }\href
  {https://doi.org/10.1073/pnas.1819316116} {\bibfield  {journal} {\bibinfo
  {journal} {Proceedings of the National Academy of Sciences}\ }\textbf
  {\bibinfo {volume} {116}},\ \bibinfo {pages} {9269} (\bibinfo {year}
  {2019})}\BibitemShut {NoStop}%
\bibitem [{\citenamefont {Schulz}\ \emph {et~al.}(2019)\citenamefont {Schulz},
  \citenamefont {Hooley}, \citenamefont {Moessner},\ and\ \citenamefont
  {Pollmann}}]{Schulz2019}%
  \BibitemOpen
  \bibfield  {author} {\bibinfo {author} {\bibfnamefont {M.}~\bibnamefont
  {Schulz}}, \bibinfo {author} {\bibfnamefont {C.~A.}\ \bibnamefont {Hooley}},
  \bibinfo {author} {\bibfnamefont {R.}~\bibnamefont {Moessner}},\ and\
  \bibinfo {author} {\bibfnamefont {F.}~\bibnamefont {Pollmann}},\ }\bibfield
  {title} {\bibinfo {title} {Stark many-body localization},\ }\href
  {https://doi.org/10.1103/PhysRevLett.122.040606} {\bibfield  {journal}
  {\bibinfo  {journal} {Phys. Rev. Lett.}\ }\textbf {\bibinfo {volume} {122}},\
  \bibinfo {pages} {040606} (\bibinfo {year} {2019})}\BibitemShut {NoStop}%
\bibitem [{\citenamefont {Guardado-Sanchez}\ \emph {et~al.}(2020)\citenamefont
  {Guardado-Sanchez}, \citenamefont {Morningstar}, \citenamefont {Spar},
  \citenamefont {Brown}, \citenamefont {Huse},\ and\ \citenamefont
  {Bakr}}]{Bakr2020}%
  \BibitemOpen
  \bibfield  {author} {\bibinfo {author} {\bibfnamefont {E.}~\bibnamefont
  {Guardado-Sanchez}}, \bibinfo {author} {\bibfnamefont {A.}~\bibnamefont
  {Morningstar}}, \bibinfo {author} {\bibfnamefont {B.~M.}\ \bibnamefont
  {Spar}}, \bibinfo {author} {\bibfnamefont {P.~T.}\ \bibnamefont {Brown}},
  \bibinfo {author} {\bibfnamefont {D.~A.}\ \bibnamefont {Huse}},\ and\
  \bibinfo {author} {\bibfnamefont {W.~S.}\ \bibnamefont {Bakr}},\ }\bibfield
  {title} {\bibinfo {title} {Subdiffusion and heat transport in a tilted
  two-dimensional {F}ermi-{H}ubbard system},\ }\href
  {https://doi.org/10.1103/PhysRevX.10.011042} {\bibfield  {journal} {\bibinfo
  {journal} {Phys. Rev. X}\ }\textbf {\bibinfo {volume} {10}},\ \bibinfo
  {pages} {011042} (\bibinfo {year} {2020})}\BibitemShut {NoStop}%
\bibitem [{\citenamefont {Hauschild}\ and\ \citenamefont
  {Pollmann}(2018)}]{tenpy}%
  \BibitemOpen
  \bibfield  {author} {\bibinfo {author} {\bibfnamefont {J.}~\bibnamefont
  {Hauschild}}\ and\ \bibinfo {author} {\bibfnamefont {F.}~\bibnamefont
  {Pollmann}},\ }\bibfield  {title} {\bibinfo {title} {Efficient numerical
  simulations with tensor networks: {T}ensor {N}etwork {P}ython ({TeNPy})},\
  }\href {https://doi.org/10.21468/SciPostPhysLectNotes.5} {\bibfield
  {journal} {\bibinfo  {journal} {SciPost Phys. Lect. Notes}\ ,\ \bibinfo
  {pages} {5}} (\bibinfo {year} {2018})},\ \bibinfo {note} {code available from
  \url{https://github.com/tenpy/tenpy}},\ \Eprint
  {https://arxiv.org/abs/1805.00055} {arXiv:1805.00055} \BibitemShut {NoStop}%
\bibitem [{\citenamefont {Schneider}\ \emph {et~al.}(2008)\citenamefont
  {Schneider}, \citenamefont {Hackerm{\"u}ller}, \citenamefont {Will},
  \citenamefont {Best}, \citenamefont {Bloch}, \citenamefont {Costi},
  \citenamefont {Helmes}, \citenamefont {Rasch},\ and\ \citenamefont
  {Rosch}}]{Schneider2008}%
  \BibitemOpen
  \bibfield  {author} {\bibinfo {author} {\bibfnamefont {U.}~\bibnamefont
  {Schneider}}, \bibinfo {author} {\bibfnamefont {L.}~\bibnamefont
  {Hackerm{\"u}ller}}, \bibinfo {author} {\bibfnamefont {S.}~\bibnamefont
  {Will}}, \bibinfo {author} {\bibfnamefont {T.}~\bibnamefont {Best}}, \bibinfo
  {author} {\bibfnamefont {I.}~\bibnamefont {Bloch}}, \bibinfo {author}
  {\bibfnamefont {T.~A.}\ \bibnamefont {Costi}}, \bibinfo {author}
  {\bibfnamefont {R.~W.}\ \bibnamefont {Helmes}}, \bibinfo {author}
  {\bibfnamefont {D.}~\bibnamefont {Rasch}},\ and\ \bibinfo {author}
  {\bibfnamefont {A.}~\bibnamefont {Rosch}},\ }\bibfield  {title} {\bibinfo
  {title} {Metallic and insulating phases of repulsively interacting fermions
  in a 3{D} optical lattice},\ }\href {https://doi.org/10.1126/science.1165449}
  {\bibfield  {journal} {\bibinfo  {journal} {Science}\ }\textbf {\bibinfo
  {volume} {322}},\ \bibinfo {pages} {1520} (\bibinfo {year}
  {2008})}\BibitemShut {NoStop}%
\bibitem [{\citenamefont {Bordia}\ \emph {et~al.}(2016)\citenamefont {Bordia},
  \citenamefont {L\"uschen}, \citenamefont {Hodgman}, \citenamefont
  {Schreiber}, \citenamefont {Bloch},\ and\ \citenamefont
  {Schneider}}]{Bordia2016}%
  \BibitemOpen
  \bibfield  {author} {\bibinfo {author} {\bibfnamefont {P.}~\bibnamefont
  {Bordia}}, \bibinfo {author} {\bibfnamefont {H.~P.}\ \bibnamefont
  {L\"uschen}}, \bibinfo {author} {\bibfnamefont {S.~S.}\ \bibnamefont
  {Hodgman}}, \bibinfo {author} {\bibfnamefont {M.}~\bibnamefont {Schreiber}},
  \bibinfo {author} {\bibfnamefont {I.}~\bibnamefont {Bloch}},\ and\ \bibinfo
  {author} {\bibfnamefont {U.}~\bibnamefont {Schneider}},\ }\bibfield  {title}
  {\bibinfo {title} {Coupling identical one-dimensional many-body localized
  systems},\ }\href {https://doi.org/10.1103/PhysRevLett.116.140401} {\bibfield
   {journal} {\bibinfo  {journal} {Phys. Rev. Lett.}\ }\textbf {\bibinfo
  {volume} {116}},\ \bibinfo {pages} {140401} (\bibinfo {year}
  {2016})}\BibitemShut {NoStop}%
\bibitem [{\citenamefont {Feldmann}\ \emph {et~al.}(1992)\citenamefont
  {Feldmann}, \citenamefont {Leo}, \citenamefont {Shah}, \citenamefont
  {Miller}, \citenamefont {Cunningham}, \citenamefont {Meier}, \citenamefont
  {von Plessen}, \citenamefont {Schulze}, \citenamefont {Thomas},\ and\
  \citenamefont {Schmitt-Rink}}]{Feldmann1992}%
  \BibitemOpen
  \bibfield  {author} {\bibinfo {author} {\bibfnamefont {J.}~\bibnamefont
  {Feldmann}}, \bibinfo {author} {\bibfnamefont {K.}~\bibnamefont {Leo}},
  \bibinfo {author} {\bibfnamefont {J.}~\bibnamefont {Shah}}, \bibinfo {author}
  {\bibfnamefont {D.~A.~B.}\ \bibnamefont {Miller}}, \bibinfo {author}
  {\bibfnamefont {J.~E.}\ \bibnamefont {Cunningham}}, \bibinfo {author}
  {\bibfnamefont {T.}~\bibnamefont {Meier}}, \bibinfo {author} {\bibfnamefont
  {G.}~\bibnamefont {von Plessen}}, \bibinfo {author} {\bibfnamefont
  {A.}~\bibnamefont {Schulze}}, \bibinfo {author} {\bibfnamefont
  {P.}~\bibnamefont {Thomas}},\ and\ \bibinfo {author} {\bibfnamefont
  {S.}~\bibnamefont {Schmitt-Rink}},\ }\bibfield  {title} {\bibinfo {title}
  {Optical investigation of {B}loch oscillations in a semiconductor
  superlattice},\ }\href {https://doi.org/10.1103/PhysRevB.46.7252} {\bibfield
  {journal} {\bibinfo  {journal} {Phys. Rev. B}\ }\textbf {\bibinfo {volume}
  {46}},\ \bibinfo {pages} {7252} (\bibinfo {year} {1992})}\BibitemShut
  {NoStop}%
\bibitem [{\citenamefont {Ben~Dahan}\ \emph {et~al.}(1996)\citenamefont
  {Ben~Dahan}, \citenamefont {Peik}, \citenamefont {Reichel}, \citenamefont
  {Castin},\ and\ \citenamefont {Salomon}}]{Salomon1996}%
  \BibitemOpen
  \bibfield  {author} {\bibinfo {author} {\bibfnamefont {M.}~\bibnamefont
  {Ben~Dahan}}, \bibinfo {author} {\bibfnamefont {E.}~\bibnamefont {Peik}},
  \bibinfo {author} {\bibfnamefont {J.}~\bibnamefont {Reichel}}, \bibinfo
  {author} {\bibfnamefont {Y.}~\bibnamefont {Castin}},\ and\ \bibinfo {author}
  {\bibfnamefont {C.}~\bibnamefont {Salomon}},\ }\bibfield  {title} {\bibinfo
  {title} {Bloch oscillations of atoms in an optical potential},\ }\href
  {https://doi.org/10.1103/PhysRevLett.76.4508} {\bibfield  {journal} {\bibinfo
   {journal} {Phys. Rev. Lett.}\ }\textbf {\bibinfo {volume} {76}},\ \bibinfo
  {pages} {4508} (\bibinfo {year} {1996})}\BibitemShut {NoStop}%
\bibitem [{\citenamefont {Dalibard}\ \emph {et~al.}(1992)\citenamefont
  {Dalibard}, \citenamefont {Castin},\ and\ \citenamefont
  {M\o{}lmer}}]{Dalibard1992}%
  \BibitemOpen
  \bibfield  {author} {\bibinfo {author} {\bibfnamefont {J.}~\bibnamefont
  {Dalibard}}, \bibinfo {author} {\bibfnamefont {Y.}~\bibnamefont {Castin}},\
  and\ \bibinfo {author} {\bibfnamefont {K.}~\bibnamefont {M\o{}lmer}},\
  }\bibfield  {title} {\bibinfo {title} {Wave-function approach to dissipative
  processes in quantum optics},\ }\href
  {https://doi.org/10.1103/PhysRevLett.68.580} {\bibfield  {journal} {\bibinfo
  {journal} {Phys. Rev. Lett.}\ }\textbf {\bibinfo {volume} {68}},\ \bibinfo
  {pages} {580} (\bibinfo {year} {1992})}\BibitemShut {NoStop}%
\bibitem [{\citenamefont {Carmichael}(1993)}]{Carmichael1992}%
  \BibitemOpen
  \bibfield  {author} {\bibinfo {author} {\bibfnamefont {H.~J.}\ \bibnamefont
  {Carmichael}},\ }\bibfield  {title} {\bibinfo {title} {Quantum trajectory
  theory for cascaded open systems},\ }\href
  {https://doi.org/10.1103/PhysRevLett.70.2273} {\bibfield  {journal} {\bibinfo
   {journal} {Phys. Rev. Lett.}\ }\textbf {\bibinfo {volume} {70}},\ \bibinfo
  {pages} {2273} (\bibinfo {year} {1993})}\BibitemShut {NoStop}%
\bibitem [{\citenamefont {Fischer}\ \emph {et~al.}(2016)\citenamefont
  {Fischer}, \citenamefont {Maksymenko},\ and\ \citenamefont
  {Altman}}]{Fischer2016}%
  \BibitemOpen
  \bibfield  {author} {\bibinfo {author} {\bibfnamefont {M.~H.}\ \bibnamefont
  {Fischer}}, \bibinfo {author} {\bibfnamefont {M.}~\bibnamefont
  {Maksymenko}},\ and\ \bibinfo {author} {\bibfnamefont {E.}~\bibnamefont
  {Altman}},\ }\bibfield  {title} {\bibinfo {title} {Dynamics of a
  many-body-localized system coupled to a bath},\ }\href
  {https://doi.org/10.1103/PhysRevLett.116.160401} {\bibfield  {journal}
  {\bibinfo  {journal} {Phys. Rev. Lett.}\ }\textbf {\bibinfo {volume} {116}},\
  \bibinfo {pages} {160401} (\bibinfo {year} {2016})}\BibitemShut {NoStop}%
\bibitem [{\citenamefont {Schneider}\ \emph {et~al.}(2012)\citenamefont
  {Schneider}, \citenamefont {Hackermüller}, \citenamefont {Ronzheimer},
  \citenamefont {Will}, \citenamefont {Braun}, \citenamefont {Best},
  \citenamefont {Bloch}, \citenamefont {Demler}, \citenamefont {Mandt},
  \citenamefont {Rasch},\ and\ \citenamefont {Rosch}}]{Schneider2012}%
  \BibitemOpen
  \bibfield  {author} {\bibinfo {author} {\bibfnamefont {U.}~\bibnamefont
  {Schneider}}, \bibinfo {author} {\bibfnamefont {L.}~\bibnamefont
  {Hackermüller}}, \bibinfo {author} {\bibfnamefont {J.~P.}\ \bibnamefont
  {Ronzheimer}}, \bibinfo {author} {\bibfnamefont {S.}~\bibnamefont {Will}},
  \bibinfo {author} {\bibfnamefont {S.}~\bibnamefont {Braun}}, \bibinfo
  {author} {\bibfnamefont {T.}~\bibnamefont {Best}}, \bibinfo {author}
  {\bibfnamefont {I.}~\bibnamefont {Bloch}}, \bibinfo {author} {\bibfnamefont
  {E.}~\bibnamefont {Demler}}, \bibinfo {author} {\bibfnamefont
  {S.}~\bibnamefont {Mandt}}, \bibinfo {author} {\bibfnamefont
  {D.}~\bibnamefont {Rasch}},\ and\ \bibinfo {author} {\bibfnamefont
  {A.}~\bibnamefont {Rosch}},\ }\bibfield  {title} {\bibinfo {title} {Fermionic
  transport and out-of-equilibrium dynamics in a homogeneous {H}ubbard model
  with ultracold atoms},\ }\href {https://doi.org/10.1038/nphys2205} {\bibfield
   {journal} {\bibinfo  {journal} {Nature Physics}\ }\textbf {\bibinfo {volume}
  {8}},\ \bibinfo {pages} {213} (\bibinfo {year} {2012})}\BibitemShut {NoStop}%
\bibitem [{\citenamefont {H.~M.}\ \emph {et~al.}(2021)\citenamefont {H.~M.},
  \citenamefont {Scherg}, \citenamefont {Kohlert}, \citenamefont {Bloch},\ and\
  \citenamefont {Aidelsburger}}]{Bharath2021}%
  \BibitemOpen
  \bibfield  {author} {\bibinfo {author} {\bibfnamefont {B.}~\bibnamefont
  {H.~M.}}, \bibinfo {author} {\bibfnamefont {S.}~\bibnamefont {Scherg}},
  \bibinfo {author} {\bibfnamefont {T.}~\bibnamefont {Kohlert}}, \bibinfo
  {author} {\bibfnamefont {I.}~\bibnamefont {Bloch}},\ and\ \bibinfo {author}
  {\bibfnamefont {M.}~\bibnamefont {Aidelsburger}},\ }\bibfield  {title}
  {\bibinfo {title} {Benchmarking a novel effiecient numerical method for
  localized 1{D} {F}ermi-{H}ubbard systems on a quantum simulator},\ }\href
  {https://arxiv.org/pdf/2105.06372.pdf} {\bibfield  {journal} {\bibinfo
  {journal} {arXiv:2105.06372}\ } (\bibinfo {year} {2021})}\BibitemShut
  {NoStop}%
\bibitem [{\citenamefont {Lewenstein}\ \emph {et~al.}(2012)\citenamefont
  {Lewenstein}, \citenamefont {Sanpera},\ and\ \citenamefont
  {Ahufinger}}]{LewensteinBook}%
  \BibitemOpen
  \bibfield  {author} {\bibinfo {author} {\bibfnamefont {M.}~\bibnamefont
  {Lewenstein}}, \bibinfo {author} {\bibfnamefont {A.}~\bibnamefont
  {Sanpera}},\ and\ \bibinfo {author} {\bibfnamefont {V.}~\bibnamefont
  {Ahufinger}},\ }\href@noop {} {\emph {\bibinfo {title} {Ultracold Atoms in
  Optical Lattices}}}\ (\bibinfo  {publisher} {Oxford University Press},\
  \bibinfo {year} {2012})\BibitemShut {NoStop}%
\bibitem [{\citenamefont {Abanin}\ \emph {et~al.}(2017)\citenamefont {Abanin},
  \citenamefont {De~Roeck}, \citenamefont {Ho},\ and\ \citenamefont
  {Huveneers}}]{Abanin_theory}%
  \BibitemOpen
  \bibfield  {author} {\bibinfo {author} {\bibfnamefont {D.}~\bibnamefont
  {Abanin}}, \bibinfo {author} {\bibfnamefont {W.}~\bibnamefont {De~Roeck}},
  \bibinfo {author} {\bibfnamefont {W.~W.}\ \bibnamefont {Ho}},\ and\ \bibinfo
  {author} {\bibfnamefont {F.}~\bibnamefont {Huveneers}},\ }\bibfield  {title}
  {\bibinfo {title} {A rigorous theory of many-body prethermalization for
  periodically driven and closed quantum systems},\ }\href
  {https://doi.org/10.1007/s00220-017-2930-x} {\bibfield  {journal} {\bibinfo
  {journal} {Communications in Mathematical Physics}\ }\textbf {\bibinfo
  {volume} {354}},\ \bibinfo {pages} {809} (\bibinfo {year}
  {2017})}\BibitemShut {NoStop}%
\bibitem [{\citenamefont {Vidal}(2003)}]{Vidal2003}%
  \BibitemOpen
  \bibfield  {author} {\bibinfo {author} {\bibfnamefont {G.}~\bibnamefont
  {Vidal}},\ }\bibfield  {title} {\bibinfo {title} {Efficient classical
  simulation of slightly entangled quantum computations},\ }\href
  {https://doi.org/10.1103/PhysRevLett.91.147902} {\bibfield  {journal}
  {\bibinfo  {journal} {Phys. Rev. Lett.}\ }\textbf {\bibinfo {volume} {91}},\
  \bibinfo {pages} {147902} (\bibinfo {year} {2003})}\BibitemShut {NoStop}%
\bibitem [{\citenamefont {Else}\ \emph {et~al.}(2017)\citenamefont {Else},
  \citenamefont {Bauer},\ and\ \citenamefont {Nayak}}]{Else2017}%
  \BibitemOpen
  \bibfield  {author} {\bibinfo {author} {\bibfnamefont {D.~V.}\ \bibnamefont
  {Else}}, \bibinfo {author} {\bibfnamefont {B.}~\bibnamefont {Bauer}},\ and\
  \bibinfo {author} {\bibfnamefont {C.}~\bibnamefont {Nayak}},\ }\bibfield
  {title} {\bibinfo {title} {Prethermal phases of matter protected by
  time-translation symmetry},\ }\href
  {https://doi.org/10.1103/PhysRevX.7.011026} {\bibfield  {journal} {\bibinfo
  {journal} {Phys. Rev. X}\ }\textbf {\bibinfo {volume} {7}},\ \bibinfo {pages}
  {011026} (\bibinfo {year} {2017})}\BibitemShut {NoStop}%
\bibitem [{\citenamefont {DePue}\ \emph {et~al.}(1999)\citenamefont {DePue},
  \citenamefont {McCormick}, \citenamefont {Winoto}, \citenamefont {Oliver},\
  and\ \citenamefont {Weiss}}]{Weiss1999}%
  \BibitemOpen
  \bibfield  {author} {\bibinfo {author} {\bibfnamefont {M.~T.}\ \bibnamefont
  {DePue}}, \bibinfo {author} {\bibfnamefont {C.}~\bibnamefont {McCormick}},
  \bibinfo {author} {\bibfnamefont {S.~L.}\ \bibnamefont {Winoto}}, \bibinfo
  {author} {\bibfnamefont {S.}~\bibnamefont {Oliver}},\ and\ \bibinfo {author}
  {\bibfnamefont {D.~S.}\ \bibnamefont {Weiss}},\ }\bibfield  {title} {\bibinfo
  {title} {Unity occupation of sites in a 3d optical lattice},\ }\href
  {https://doi.org/10.1103/PhysRevLett.82.2262} {\bibfield  {journal} {\bibinfo
   {journal} {Phys. Rev. Lett.}\ }\textbf {\bibinfo {volume} {82}},\ \bibinfo
  {pages} {2262} (\bibinfo {year} {1999})}\BibitemShut {NoStop}%
\bibitem [{\citenamefont {Bravyi}\ \emph {et~al.}(2011)\citenamefont {Bravyi},
  \citenamefont {DiVincenzo},\ and\ \citenamefont {Loss}}]{BRAVYI20112793}%
  \BibitemOpen
  \bibfield  {author} {\bibinfo {author} {\bibfnamefont {S.}~\bibnamefont
  {Bravyi}}, \bibinfo {author} {\bibfnamefont {D.~P.}\ \bibnamefont
  {DiVincenzo}},\ and\ \bibinfo {author} {\bibfnamefont {D.}~\bibnamefont
  {Loss}},\ }\bibfield  {title} {\bibinfo {title} {Schrieffer–wolff
  transformation for quantum many-body systems},\ }\href
  {https://doi.org/https://doi.org/10.1016/j.aop.2011.06.004} {\bibfield
  {journal} {\bibinfo  {journal} {Annals of Physics}\ }\textbf {\bibinfo
  {volume} {326}},\ \bibinfo {pages} {2793} (\bibinfo {year}
  {2011})}\BibitemShut {NoStop}%
\end{thebibliography}%
\cleardoublepage

\setcounter{figure}{0}
\setcounter{equation}{0}
\setcounter{page}{1}

\renewcommand{\thepage}{S\arabic{page}} 
\renewcommand{\thesection}{S\arabic{section}} 
\renewcommand{\thetable}{S\arabic{table}}  
\renewcommand{\thefigure}{S\arabic{figure}} 
\renewcommand{\theequation}{S\arabic{equation}} 
\newcommand{\up}{\uparrow}
\newcommand{\dn}{\downarrow}

\renewcommand{\thesection}{\Roman{section}}
\renewcommand{\thesubsection}{S\arabic{subsection}}
\def\tocname{Table of contents}

\onecolumngrid
\addtocontents{toc}{\string\tocdepth@munge}
\section*{\Large{Supporting material}}
\addtocontents{toc}{\string\tocdepth@restore}
\twocolumngrid
\setlength{\intextsep}{0.8cm}
\setlength{\textfloatsep}{0.8cm}

This supporting material provides a complete overview of the experimental setup, the measurement and data acquisition techniques, as well as analytical and numerical methods employed in this work.
\tableofcontents

\section{Experimental techniques}
This section gives a detailed overview of the experimental sequence and techniques, initial state preparation and data acquisition as well as systematic effects induced by experimental imperfections.

\subsection{General sequence}
\label{sec:general}
The fermionic gas of ${}^{40}\mathrm{K}$ atoms is cooled to quantum degeneracy in a three-dimensional crossed-beam dipole trap, where we end up with an equal mixture of the spin components $\ket{\uparrow} = \ket{F=9/2,\,m_F=-7/2}$ and $\ket{\downarrow}= \ket{F=9/2,\,m_F=-9/2}$, the two lowest lying magnetic hyperfine states in the ground state hyperfine manifold. Details on the cooling scheme can be found in a previous publication~\cite{Bordia2016}. We then ramp up the lattices adiabatically in a series of linear ramps~\cite{Scherg2020}. Different initial states are prepared by varying the scattering length during the loading process, where attractive interactions benefit the formation of doublons. In contrast, pure singlon states are prepared by strong repulsion, weak confinement during the loading process and a $\SI{150}{\micro\second}$ long near-resonant light pulse to remove all residual doublons. Further details on this part are given in sec.~\ref{sec:Dlifetime}. The doublon fraction $n_D = N_d/N$ used in this paper is defined as the number of atoms on doubly-occupied sites $N_d$ divided by the total atom number $N$.

After the loading process, the dynamics is frozen in the deep 3D lattice with a depth of $18E_{r,s}$ for the primary (short) lattice and a superimposed long lattice with twice the primary wavelength and a depth of $20E_{r,l}$, creating a superlattice potential. The orthogonal lattices are kept at $55E_{r,\perp}$. $E_{r,j} = \hbar^2 k_j^2/(2m)$ is the respective recoil energy of the lattices with $j \in \{s,l,\perp\}$ and the wave vector $k_j = 2\pi/\lambda_j$. The atomic mass of ${}^{40}\mathrm{K}$ is denoted by $m$ and $\hbar$ is the Planck constant. Further, holding the gas in a strongly tilted superlattice for $\SI{30}{\milli\second}$ dephases residual coherences and the initial state is an incoherent mixture of product states with zero magnetization, represented by the density matrix $\rho_0 = \frac{1}{\mathcal{N}} \sum_\sigma \ket{\psi_0(\sigma)} \bra{\psi_0(\sigma)}$. The summation index runs over all spin permutations preserving the total magnetization and $\mathcal{N}$ is the normalization constant. The product state $\ket{\psi_0(\sigma)}$ can be expressed as $\ket{\psi_0(\sigma)} = \prod_{i\,\text{even}} \left(\hat{c}_{i,\uparrow}^\dagger \right)^{\hat{n}_{i,\uparrow}} \left(\hat{c}_{i,\downarrow}^\dagger \right)^{\hat{n}_{i,\downarrow}}$ with the fermionic creation operator $\hat{c}_{i,\uparrow(\downarrow)}^\dagger$ and $\langle \hat{n}_{i,\uparrow (\downarrow)} \rangle \in \{0,1\} \forall i$. During the freezing, the gradient is ramped to the final value, while the Feshbach field adjusts to compensate the field increase due to the homogeneous part of the gradient field. The dynamics is initiated by a sudden quench of the primary lattice to the desired depth of $12E_{r,p}$, while the long lattice is switched off. After the time evolution we suddenly increase both lattices back $18E_{r,s}$ and $20E_{r,l}$ respectively, ramp the Feshbach field to the non-interacting point and switch off the gradient field. After a hold time of $\SI{30}{\milli\second}$, which is required for the magnetic fields to settle to their final value, we perform a band transfer~\cite{Sebby2006,Foelling07} in combination with bandmapping and absorption imaging after $\SI{8}{\milli\second}$ time-of-flight to extract the imbalance from the quasi-momentum distribution. Most of the details on the tilt generation and calibration as well as a detailed sketch of the lattice loading sequence are given in~\cite{Scherg2020}. The main differences regarding this work are highlighted in the preceding paragraph. 

\subsection{Lattice loading, initial state distribution and doublon fraction}
\label{sec:doublons}
Here we characterize the global distribution of singlons and doublons in the initial state using occupation number resolved imaging with a resolution of 8.3~sites per pixel on the CCD camera chip~\cite{Scherg2018}. We simultaneously image singlons and doublons using a series of radio-frequency (RF) and microwave (MW) $\pi$-pulses and Landau-Zener sweeps at a magnetic field of $209.2\,\mathrm{G}$, corresponding to the non-interacting point of the Feshbach resonance between $\ket{\up}$- and $\ket{\dn}$-atoms. The overall scheme is illustrated in Fig.~\ref{fig:RFscheme} and can be summarized as follows:

\begin{figure*}
	\includegraphics[width=\textwidth]{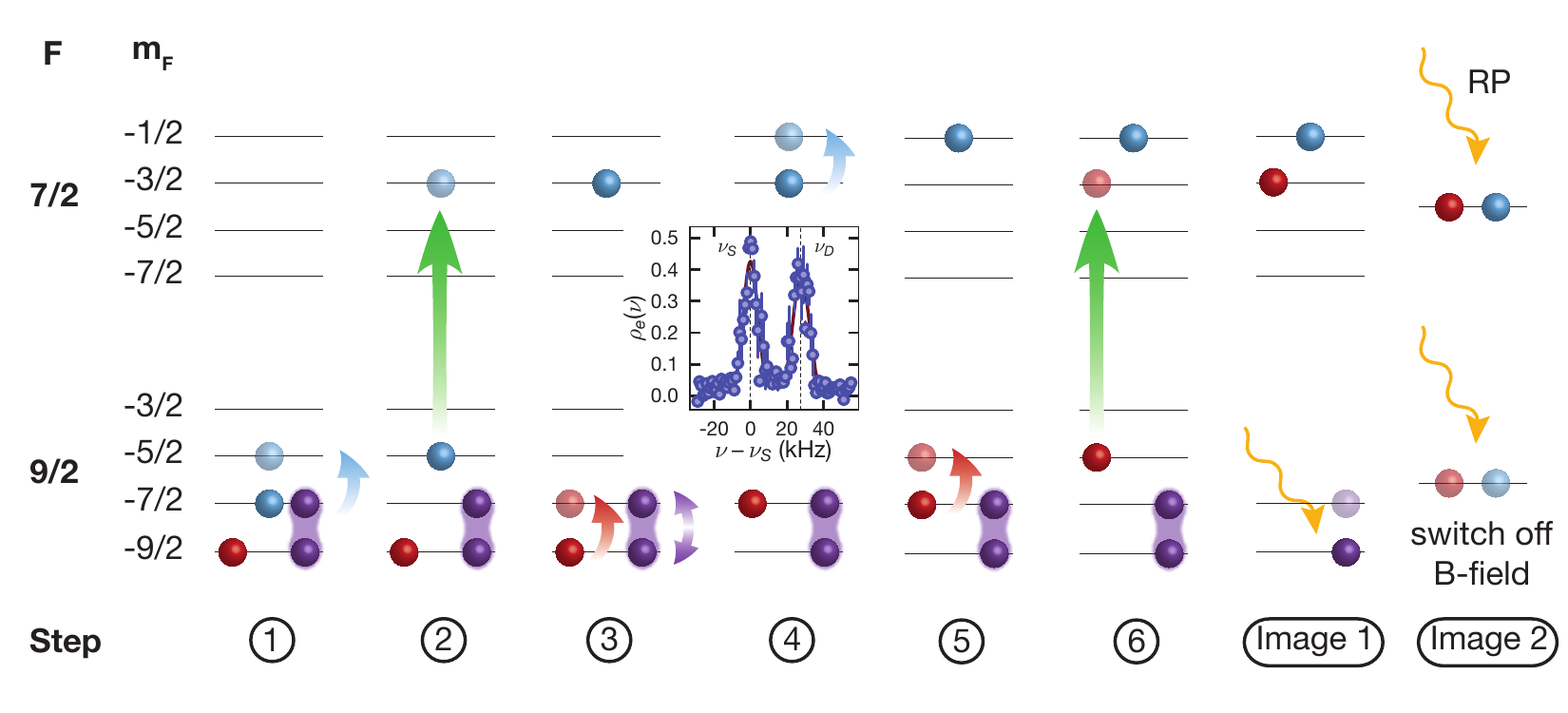}
	\caption{\textbf{Scheme for singlon- and doublon-resolved imaging:} This schematic figure illustrates the state operations with RF (red and blue arrows) and MW (green arrows) fields employed to image singlons and doublons separately. The color represents the initial spin state and the purple groups are opposite spins bound in doublons. RF sweeps and pulses operate in the same hyperfine manifold, while MW sweeps excite the atoms to the $F=7/2$-manifold. Apart from the second imaging step all steps are performed at the non-interacting point of the Feshbach resonance at $209.2(1)\,\mathrm{G}$. The inset shows an experimental RF spectrum to illustrate the interaction shift between singlons and doublons relevant for steps 1 and 5. Two distinct peaks at the singlon $\nu_S$ and doublon resonance $\nu_D$ emerge which can be clearly resolved. From the relative height of the peak we extract a doublon fraction that agrees well with the value from the main text [$n_D=0.47(4)$]. All data points are averaged thrice with the error bars showing the standard error of the mean and the fitting is done with a composite sinc-function (see text).}
	\label{fig:RFscheme}
\end{figure*}

\begin{enumerate}
	\item Separate $m_F=-7/2$ atoms on singly-occupied sites from those on doubly-occupied sites via the interaction shift as shown in the inset of Fig.~\ref{fig:RFscheme} and transfer them into $m_F=-5/2$ via an RF $\pi$-pulse. The interaction shift between initial and final states comes from the different scattering lengths before and after the pulse such that the interaction energy of the doublons changes from 0 to $\SI{27}{\kilo\hertz}$. This value is larger than the resolution of the pulse ($\SI{17}{\kilo\hertz}$) such that the separation works reliably.
	\item Transfer the singlons in $m_F=-5/2$ to the state $\ket{F=7/2,m_F=-3/2}$ in the $F=7/2$ ground-state hyperfine manifold with an MW sweep.
	\item Swap the occupations in the states $m_F=-7/2$ and $m_F=-9/2$ with an RF sweep. The width of the sweep is much larger than the interaction shift, such that no difference is made between singlons and doublons.
	\item Clear the $\ket{F=7/2,m_F=-3/2}$ state by sweeping the atoms to $\ket{F=7/2,m_F=-1/2}$, otherwise step 6 would sweep them back to the lower $F=9/2$ manifold.
	\item Repeat step 1.
	\item Repeat step 2.
\end{enumerate}
After this sequence the doublons end up in the states $m_F=-9/2$ and $m_F=-7/2$ in the $F=9/2$ ground-state manifold, while the singlons are transferred to the $F=7/2$ hyperfine manifold. The $m_F=-9/2$ doublon component is imaged in-situ in the presence of a magnetic field (the other component is off-resonant and hence invisible, but light-assisted collisions will remove all atoms on doubly-occupied sites from the lattice. Thus, this scheme is not suitable to extract the doublon fraction). After the first image, the magnetic field is switched off and the singlons are imaged by adding repumper light that pumps the atoms from the $F=7/2$ dark state back into the cycle to $F=9/2$. While the sweeps work with good fidelity, the overall efficiency of the scheme is limited to about $80\,\%$ by the magnetic field stability during the $\pi$-pulses. This results in an imperfect separation of singlons and doublons and adds noise to our detection signal. Therefore, we resort to an indirect technique for the measurement of the doublon imbalance.

\begin{figure}
	\includegraphics[width=3.3in]{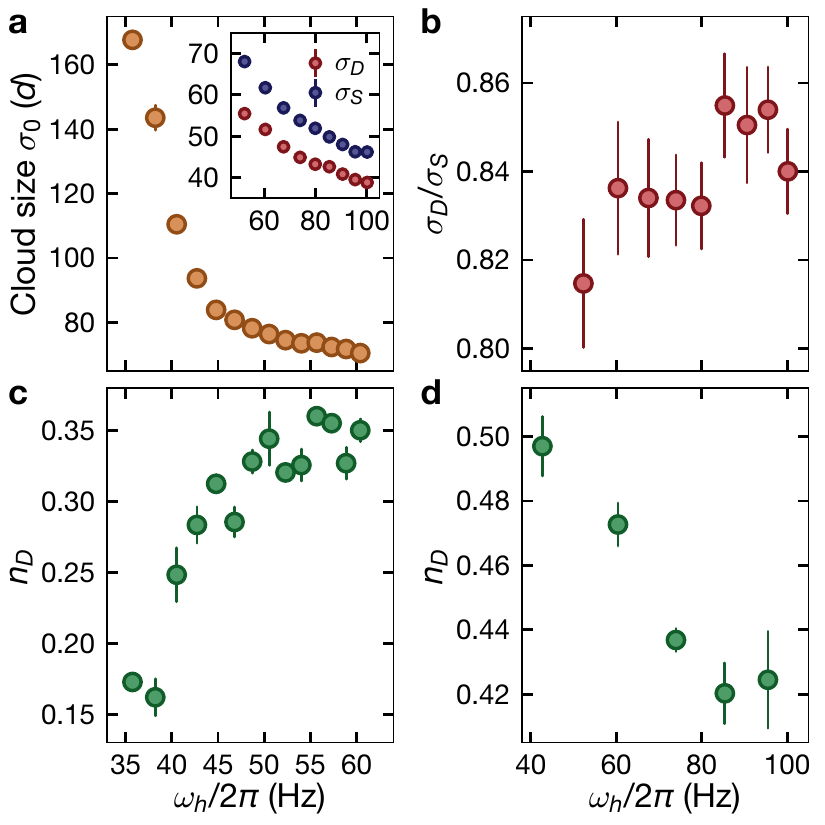}
	\caption{\textbf{Cloud sizes and doublon fraction:} \textbf{a} Cloud size $\sigma_0$ in units of the lattice constant $d$ as a function of the trap frequency $\omega_h$. The inset shows the same measurement for even larger confinement and  singlon- and doublon-resolved with cloud size $\sigma_S$ and $\sigma_D$ respectively. \textbf{b} Cloud size ratio of doublons and singlons depending on the external confinement. \textbf{c} Doublon fraction $n_D$ right after loading into the lattice versus the harmonic trapping frequency $\omega_h$. Up to this point the analysis was performed for a non-interacting loading process. \textbf{d} Doublon fraction as a function of trap frequency after $\SI{10}{\milli\second}$ in the lattice. This data is obtained at an attractive loading scattering length of $-20a_0$. The data points are averaged twice with the error bars representing the standard error of the mean.}
	\label{fig:latticeloading}
\end{figure}

\begin{figure}
	\includegraphics[width=3.3in]{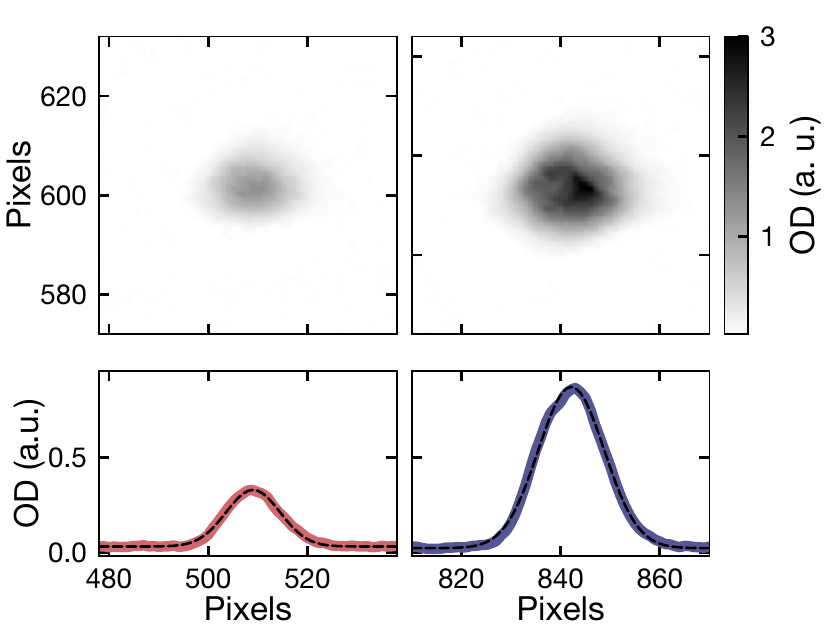}
	\caption{\textbf{In-situ distributions:} Top: Doublon (left) and singlon (right) cloud profile in-situ after lattice loading. The colorbar denotes the optical density in arbitrary units. Bottom: Integrated profiles along one vertical direction. The black dashed line is a Gaussian fit to the density. In this case $\sigma_D = 5.4(1)$ pixels and $\sigma_S = 6.7(1)$ pixels. Thereby 1 pixel corresponds to about $\SI{2.2}{\micro\meter}$.}
	\label{fig:insitu}
\end{figure}

However, we can separately image singlons and doublons in the same sequence with two images and use this information to characterize the initial state distributions as a function of the lattice loading parameters. These results are presented in Fig.~\ref{fig:latticeloading}. Our main parameter is the trap frequency $\omega_h = \sqrt{\alpha\hbar/(m d^2 \pi)}$ appearing in the typical harmonic potential $V_h = m \omega_h^2 id/2$ at site $i$ with the lattice spacing $d=\SI{266}{\nano\meter}$ and $\alpha \approx 10^{-3}J$ quantifies the strength of the quadratic potential~\cite{Scherg2020}. It characterizes the external harmonic confinement due to the interplay of the red-detuned dipole trap and the blue-detuned lattices. In Fig.~\ref{fig:latticeloading}a we observe that the final cloud size strongly depends on the confinement during loading and that the doublon cloud radius is smaller than the singlon cloud radius (Fig.~\ref{fig:latticeloading}a inset). This context is investigated closer in Fig.~\ref{fig:latticeloading}b, but we see no real dependence of the cloud size ratio on the confinement. We further measure the doublon fraction in the lattice as a function of the confinement in Fig.~\ref{fig:latticeloading}c and find that the doublon fraction increases for stronger confinement until it reaches a plateau that depends on other parameters such as the temperature. These results are in agreement with~\cite{Schneider2008}, where these relations were measured in our setup, although for a different lattice configuration. The reason why we do not choose to work at the strongest confinement is revealed in Fig.~\ref{fig:latticeloading}d. After a hold time of $\SI{10}{\milli\second}$ in the lattice the doublon fraction decreases for stronger confinement, owing to a reduced lifetime for strong compression. We thus decide to work around $\omega_h \approx 2\pi \times \SI{60}{\hertz}$. In Fig.~\ref{fig:insitu} we show an exemplary raw image of the singlon-doublon resolved measurement together with an integrated cloud profile. From these we extract the widths $\sigma_D$ and $\sigma_S$ of the doublon and singlon density distribution respectively via Gaussian fits.

\begin{figure}
	\includegraphics[width=3.3in]{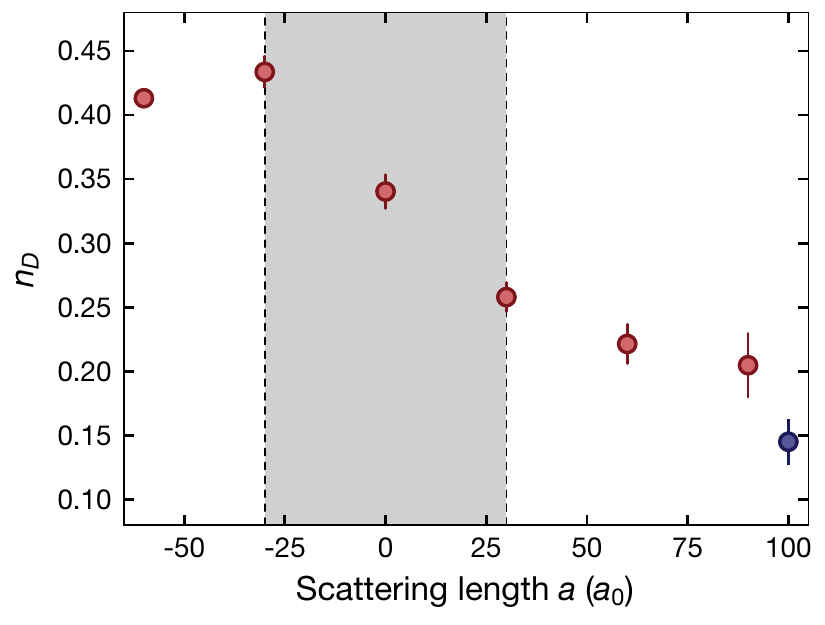}
	\caption{\textbf{Doublon fraction in the initial state as a function of the scattering length during lattice loading:} In the experiment we vary the scattering length between $-30a_0$ and $30a_0$. The blue point was measured at a different harmonic confinement to minimize the formation of doublons (see text). All data points are averaged three times and error bars are the standard error of the mean.}
	\label{fig:Doublonfraction}
\end{figure}

The doublon fraction of the initial state at a fixed harmonic confinement is controlled via the scattering length during the lattice loading process and the resulting fraction in the initial state is shown in Fig.~\ref{fig:Doublonfraction}. Attractive interactions benefit the formation of double occupations. We choose the range between $-30a_0$ and $30a_0$ such that our initial doublon fraction is roughly between $45\,\%$ and $25\,\%$. We do not employ even more attractive interactions as we see a reduction of the doublon fraction in this regime, probably due to a reduced lifetime. Moreover, in the limit of large attractive scattering length, the distribution of singlons and doublons becomes more uniform across the lattice and the hole fraction increases, while we rather thrive for a higher concentration near the trap center with a low hole fraction. For pure singlon initial states we load strongly repulsively ($100a_0$) and further reduce the confinement during the loading process in order reduce the density and hence suppress the doublon formation. Residual doublons are then removed with a near-resonant light pulse (see sec.~\ref{sec:Dlifetime}). A slight disadvantage of this method is the creation of additional holes in the singlon CDW initial state. Hole creation as an experimental imperfection is taken into account in the numerical simulations supporting our results (see main text).

\subsection{RF dressing}
\begin{figure*}[t]
	\includegraphics[width=\textwidth]{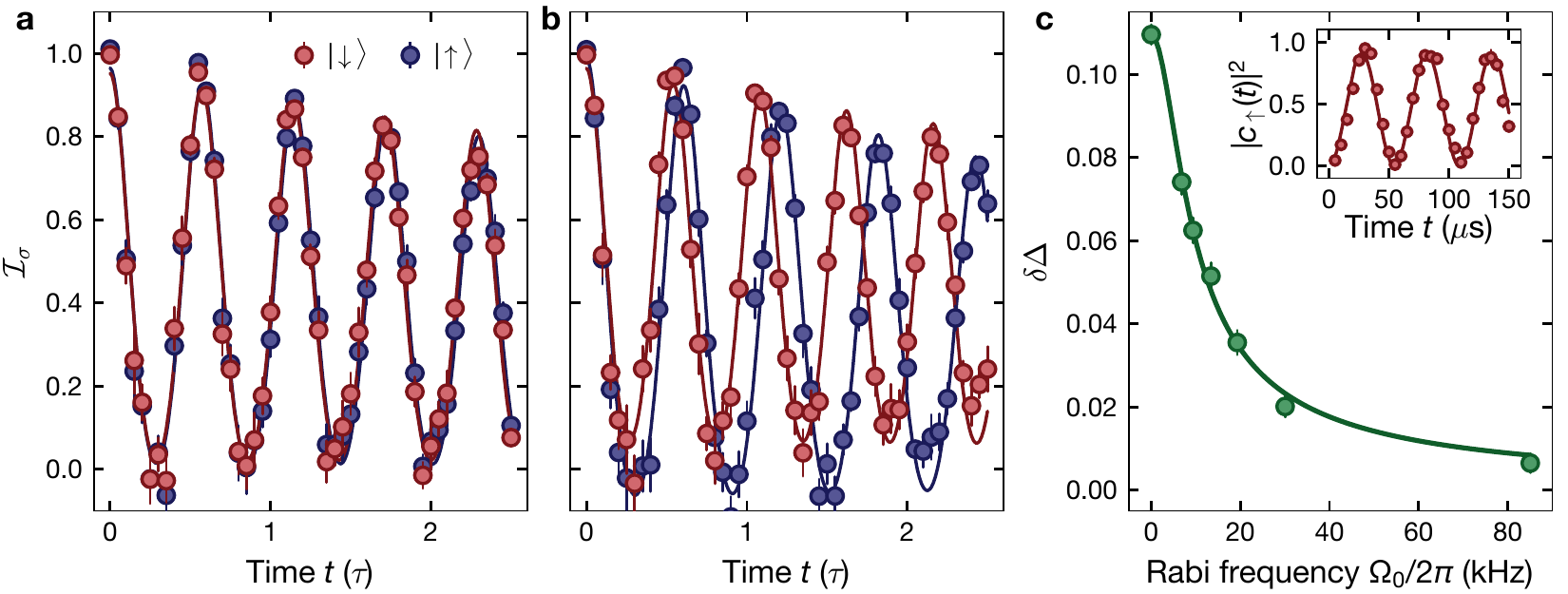}
	\caption{\textbf{RF dressing and tilt difference:} \textbf{a} Bloch oscillations with maximal dressing power ($\Omega_0 = \SI{85(1)}{\kilo\hertz}$) measured in a spin-resolved manner. \textbf{b} Bloch oscillations without RF dressing. The frequency difference is explained by the natural difference in the magnetic moment of the two spin states. The solid lines are analytical fits (see text) to extract the oscillation frequency and the error bars denote the standard error of the mean obtained from four averages. \textbf{c} Tilt difference $\delta\Delta$ as a function of the Rabi frequency $\Omega_0$. The solid line is an analytical fit to the data points with only one free parameter $\delta$ according to Eq.~(\ref{eqn:tilt_diff}). The inset shows a sample data set of resonant Rabi oscillations between the two spin states, which was used to calibrate the Rabi frequency with a sinusoidal fit for frequency and amplitude. The vertical axis is the excitation probability from $\ket{\dn}$ to $\ket{\up}$. Here, $|c_\uparrow(t)|^2$ is the population in $\ket{\up}$ at time $t$.}
	\label{fig:DressedOscillations}
\end{figure*}

In order to tune the naturally-given tilt difference between the spin components which is caused by the different $m_F$ quantum numbers, we employ the technique of RF dressing~\cite{Zwierlein2003,Skedrov2021}. During the time evolution we apply an RF field to drive the transition from $\ket{\downarrow}$ to $\ket{\uparrow}$. 
In the presence of the drive, the creation operators $\hat{c}_{i, \sigma}^\dagger$ are transformed according to a unitary operator:
\begin{align}
\begin{split}
\hat{c}_{i, \uparrow}^\dagger(t) =& e^{i \Omega t/2}\cos \left(\frac{\theta}{2}\right) e^{-i\phi/2} \hat{\tilde c}_{i, \uparrow}^\dagger \\&+e^{-i \Omega t/2} \sin \left(\frac{\theta}{2}\right) e^{i\phi/2} \hat{\tilde c}_{i, \downarrow}^\dagger \\
\hat{ c}_{i, \downarrow}^\dagger(t) &= -e^{i \Omega t/2}\sin \left(\frac{\theta}{2}\right) e^{-i\phi/2} \hat{\tilde c}_{i, \uparrow}^\dagger \\&+ e^{-i \Omega t/2} \cos \left(\frac{\theta}{2}\right) e^{i\phi/2} \hat{\tilde c}_{i, \downarrow}^\dagger .
\end{split}
\end{align}
Here, $\Omega$ is the generalized Rabi frequency of the RF driving. $\hat{\tilde c}_{\sigma, i}^\dagger$ are the dressed creation operators defined as
\begin{equation}
\begin{split}
\hat{\tilde c}_{i, \uparrow}^\dagger &=\cos \left(\frac{\theta}{2}\right) e^{i\phi/2} \hat{ c}_{i, \uparrow}^\dagger - \sin \left(\frac{\theta}{2}\right) e^{i\phi/2} \hat{ c}_{i, \downarrow}^\dagger \\
\hat{\tilde c}_{i, \downarrow}^\dagger &= \sin \left(\frac{\theta}{2}\right) e^{-i\phi/2} \hat{ c}_{i, \uparrow}^\dagger +  \cos \left(\frac{\theta}{2}\right) e^{-i\phi/2} \hat{ c}_{i, \downarrow}^\dagger
\end{split}
\end{equation}
where $\theta = \tan^{-1} \left(\Omega_0/ \delta \right)$, $\delta$  is the detuning, $\Omega_0$ is the resonant coupling strength of the driving with $\Omega=\sqrt{\delta^2+\Omega_0^2}$ and $\phi$ is the phase of the driving.  In this driven basis, the hopping term in the Hamiltonian is invariant.  
Moreover, it can be shown that the on-site interaction term in the 1D Fermi-Hubbard model is also invariant~\cite{Zwierlein2003} and therefore, we still have the regular Feshbach resonance to set the interaction strength between dressed states. The tilts, $\Delta_{\sigma}$, however transform to a weighted average under this unitary~\cite{Skedrov2021}:
\begin{equation}
\begin{split}
&\Delta_{\uparrow} \hat{c}_{i, \uparrow}^{\dagger}(t)\hat{c}_{i, \uparrow}(t)+\Delta_{\downarrow} \hat{c}_{i, \downarrow}^{\dagger}(t)\hat{c}_{i, \downarrow}(t) =\\ &(\Delta_{\uparrow}\cos^2 (\theta/2) + \Delta_{\downarrow}\sin^2 (\theta/2))\hat{\tilde c}_{i, \uparrow}^{\dagger}\hat{\tilde c}_{i, \uparrow}\\
&+(\Delta_{\downarrow}\cos^2 (\theta/2) + \Delta_{\uparrow}\sin^2 (\theta/2))\hat{\tilde c}_{i, \downarrow}^{\dagger}\hat{\tilde c}_{i, \downarrow}\\
& + (\Delta_{\uparrow} - \Delta_{\downarrow})\cos (\theta/2)\sin (\theta/2) e^{i (\Omega t-\phi)}\hat{\tilde c}_{i, \uparrow}^{\dagger}\hat{\tilde c}_{i, \downarrow}\\
& + (\Delta_{\uparrow} - \Delta_{\downarrow})\cos (\theta/2)\sin (\theta/2) e^{-i (\Omega t-\phi)}\hat{\tilde c}_{i, \downarrow}^{\dagger}\hat{\tilde c}_{i, \uparrow}
\end{split}
\label{eq:transTilts}
\end{equation}
The last two terms are dropped in a first-order Magnus expansion in the limit $\Omega \gg \Delta_{\downarrow}, \Delta_{\uparrow}$. Moreover, these time-dependent terms average to zero due to fluctuations in the magnetic field.  The effective tilts experienced by the dressed spin states can be read off directly from the first two terms in Eq.~(\ref{eq:transTilts}). They are $\tilde{\Delta}_{\uparrow}=\Delta_{\downarrow}\cos^2 (\theta/2) + \Delta_{\uparrow}\sin^2 (\theta/2)$ and $\tilde{\Delta}_{\downarrow}=\Delta_{\uparrow}\cos^2 (\theta/2) + \Delta_{\downarrow}\sin^2 (\theta/2)$. Thus, the tilt difference reduces to 
\begin{equation}\label{eqn:tilt_diff}
\tilde{\Delta}_\downarrow - \tilde{\Delta}_\uparrow = (\Delta_\downarrow - \Delta_\uparrow)\frac{|\delta|}{\sqrt{\delta^2 + \Omega_0^2}}
\end{equation}
and $\delta\Delta = (\tilde{\Delta}_\downarrow - \tilde{\Delta}_\uparrow)/ \Delta_\downarrow$ (main text and Fig.~\ref{fig:DressedOscillations}c).

The Rabi frequency $\Omega_0$ is determined by driving spin-polarized atoms without a tilt on resonance and extracting the frequency of the resulting Rabi oscillations by measuring the spin populations (Fig.~\ref{fig:DressedOscillations}c inset). In the lattice with a tilt, however, the resonance frequency is not constant across the lattice. This leads to a spatially-dependent detuning $\delta$. As a result, the relative tilt difference $\delta\Delta$ follows Eq.~(\ref{eqn:tilt_diff}) with a detuning that depends on the position in the lattice. We find that this simple analytical model with a single average detuning $\delta$ describes our measured data very well (Fig.~\ref{fig:DressedOscillations}c). The tilt difference shown here was measured using the spin-resolved imbalance $\mathcal{I}_\sigma$ in the absence of interactions and extracting their oscillation frequencies from analytical fits to the imbalance time traces for short times, where we see coherent Bloch oscillations. Harmonic confinement is known to dephase the dynamics and lead to an amplitude decay of the oscillations on our timescales~\cite{Scherg2020}. Here we imitate this decay with an exponential envelope and therefore use the fit function $\mathcal{I}_\sigma (t) = \exp(-t/\tau) \cdot \mathcal{J}_0 (8J/\tilde{\Delta}_\sigma \cdot \sin(\pi \tilde{\Delta}_\sigma t/h))$ with the decay time $\tau$, based on the results of~\cite{Scherg2020}. This is shown in Figs.~\ref{fig:DressedOscillations}a and~\ref{fig:DressedOscillations}b. As expected, the tilt difference depends on the RF power and we are able to almost compensate the tilt at the maximum available dressing power. From a fit of Eq.~(\ref{eqn:tilt_diff}) to the data points we extract an average detuning $\delta = \SI{6.5(2)}{\kilo\hertz}$.

We develop an intuitive picture to illustrate the physical origin of this global detuning $\delta$. Due to the magnetic field gradient the resonance frequency for the RF-dressing is spatially dependent across the atomic cloud. According to Eq.~\eqref{eqn:tilt_diff} this results in a spatially-dependent tilt seen by the dressed states $\ket{\tilde{\downarrow}}$ and $\ket{\tilde{\uparrow}}$ across the lattice according to
\begin{equation}
	\tilde{\Delta}_{\downarrow,\uparrow} = \frac{\Delta_\uparrow + \Delta_\downarrow}{2} \pm \frac{\Delta_\uparrow - \Delta_\downarrow}{2} \frac{|\delta|}{\sqrt{\delta^2+\Omega_0^2}}.
\end{equation}
In the experiment we automatically average across all tilts throughout the atomic cloud. This averaging effect is imitated by computing the resulting superposition of Bloch oscillations at the respective frequencies $\tilde{\Delta}_\downarrow$ and $\tilde{\Delta}_\downarrow$ and weighing their contribution with a Gaussian density distribution. For the limit of large tilt we approximate the analytical description of the Bloch oscillations (as shown in Fig.~\ref{fig:DressedOscillations}a and b) with a cosine-function:
\begin{equation}
	\mathcal{I}_{\tilde{\uparrow}}(t) = \frac{1}{L} \sum_{i=-L/2}^{L/2} \cos \left(\tilde{\Delta}_\uparrow(\delta_i)t/\hbar\right) \exp\left(-\delta_i^2/(2\sigma^2) \right)
	\label{eq:Bloch_superpos}
\end{equation}
and the analogous expression for $\mathcal{I}_{\tilde{\downarrow}}(t)$. Herein the sum runs over all lattice sites and $\sigma=\SI{14.5}{\kilo\hertz}$ denotes the standard deviation of the detuning distribution across the ensemble. We fit this superposition in Eq.~\eqref{eq:Bloch_superpos} with a single oscillation frequency $\tilde{\Delta}_{\uparrow}^{\mathrm{eff}}$ and extract the tilt difference in the dressed state basis $\tilde{\Delta}_{\downarrow}^{\mathrm{eff}} - \tilde{\Delta}_{\uparrow}^{\mathrm{eff}}$ for multiple values of the Rabi frequency $\Omega_0$. This results in a dressed-state tilt difference that depends on the resonant Rabi coupling similar to Fig.~\ref{fig:DressedOscillations}c. Fitting the same model to this data yields an average detuning of $\delta = \SI{9.8(1)}{\kilo\hertz}$, which is in reasonable agreement with the experimental value. The fact that larger coupling $\Omega_0$ leads to a better compensation of the tilt difference directly follows from Eq.~\ref{eqn:tilt_diff}.

In order to get a quantitative impression for the role of the tilt difference, it is insightful to compare it to the effective hopping that appears in the respective effective Hamiltonian. In the dipole-conserving regime ($\Delta \gg U,J$) the effective hopping is $J^{(3)}=J^2U/\Delta^2$ and thus the maximal tilt difference amounts to about $22J^{(3)}$ dominating the dynamical scale and leading to an effective inhibition of the dynamics as shown in Fig.~\ref{fig:fig4}a. The same argument holds in the regime $U \simeq 2\Delta$ with $J^{(2)} = J^2/\Delta$. Here, the maximal tilt difference corresponds to roughly $9J^{(2)}$ such that its effect is less pronounced, but still significant.

\subsection{Doublon lifetime}
\label{sec:Dlifetime}
In a lattice the lifetime of double occupations is typically limited by light-assisted collisions~\cite{Weiss1999}. This term describes a collision between two atoms, where one absorbs a photon during the collision process. Most of the lattices in this work are blue-detuned. This results in a gain of kinetic energy on the order of the photon detuning leading, which leads to rapid loss of both atoms from the trap. The characteristic lifetime depends on the compression in the lattice due to the extension of the wavefunction and the weight at the Condon point. Moreover, the lifetime depends on the interaction between the atoms, but since the absolute differences in the scattering length between the various regimes examined in this work are small ($-30a_0$ to $30a_0$, see Fig.~\ref{fig:Doublonfraction}), there is no large disparity in the doublon lifetimes.

\begin{figure}
	\includegraphics[width=3.3in]{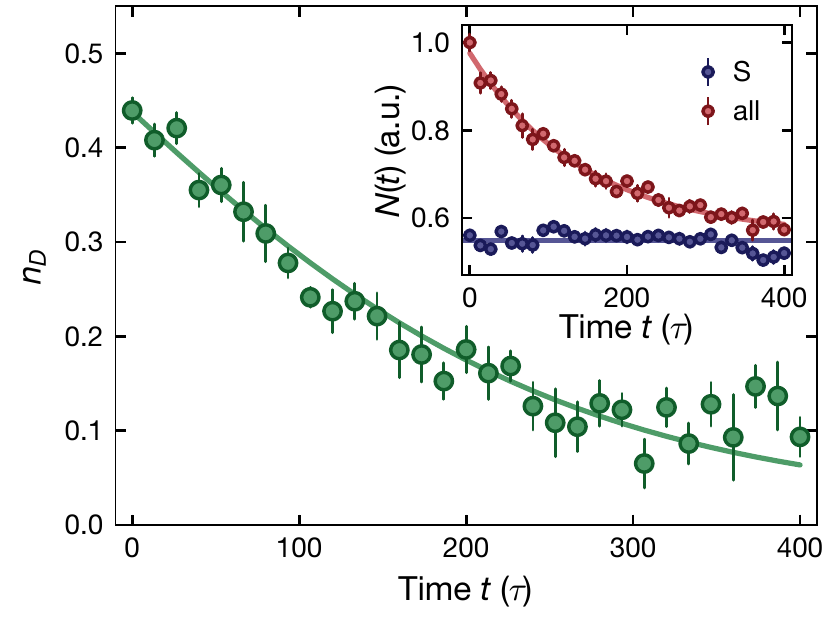}
	\caption{\textbf{Doublon lifetime:} Time evolution of the doublon fraction $n_D$ as a function of the hold time $t$ due to light-assisted collisions in the deep primary lattice at $18E_{r,s}$ and strongly repulsive interactions ($U=20J$). The green solid line represents the analytical function from the main text with the fit parameters obtained from the atom numbers in the inset. The inset shows the evolution of the raw atom numbers with (all atoms) and without doublons (only singlons), which was used to extract the doublon fraction $n_D$ shown in the main panel. Solid lines of the respective color are a constant (singlons) and exponential fit (all atoms), see main text. Error bars in inset and main panel denote the standard error of the mean obtained from ten averages.}
	\label{fig:Dlifetime}
\end{figure}

In this section, we analyze the doublon lifetime and its impact on the measured dynamics in more detail. In order to determine the lifetime independent of any dynamics we load the lattice as described in sec.~\ref{sec:general}, but remove the tilt and do not initiate the dynamics by performing a quench of the primary lattice and keep it at $18E_{r,s}$. Moreover, the interactions are constant and strongly repulsive with $U=20J$ such that doublons cannot dynamically decay into singlons. Using differential measurements of the total atom number with and without doublons, which are determined by removing doublons actively in every other experimental realization before imaging, we determine the singlon and doublon fraction as a function of time. The doublons are removed by applying an additional near-resonant light-pulse of $\SI{150}{\micro\second}$ duration which leads to a fast loss of doublons due to light-assisted collisions as explained above (see~\cite{Scherg2018} for details). The results are shown in Fig.~\ref{fig:Dlifetime}. We observe a constant singlon fraction, which is consistent with a negligible dynamical doublon decay. Hence, the observed doublon loss is mainly dominated by light-assisted collisions. The singlon lifetime is generally limited by off-resonant photon scattering and was determined in an independent measurement to be $2.6(2) \times 10^3 \tau$, which is much larger than any experimental timescale studied in this work. We fit an exponential function to the total atom numbers including doublons using $N(t) = N_0 \exp(-t/\tau)+N_\infty$, where $N_0$, $\tau$ and $N_\infty$ are free fit parameters. The doublon fraction is evaluated according to $n_D = 1-\bar{N}_S/N(t)$ where $\bar{N}_S$ is the average singlon number retrieved from a constant fit as shown in the inset of Fig.~\ref{fig:Dlifetime}. The result is shown in the plot and fits the data very well as expected. From this data we extract a doublon lifetime of $\tau = \SI{109(8)}{\milli\second} = 145(11)\tau$. The maximal measurement time for doublon initial states presented in the main text is chosen to approximately match the doublon lifetime ($\SI{100}{\milli\second} \approx 133\tau$) in order to minimize the impact of doublon loss on our experimental observations. We have also modelled the impact of doublons loss on the dynamics described in the main text using an approximate numerical description~\cite{Bharath2021} as described in sec.~\ref{sec:APX}.

\subsection{Interaction averaging}\label{sec:Uvariation}
\begin{figure}[t]
	\includegraphics[width=3.3in]{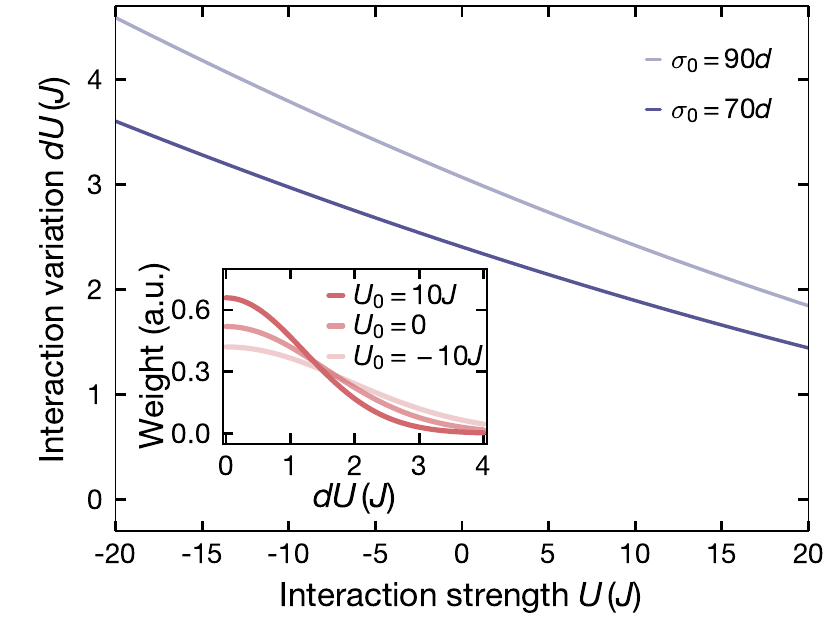}
	\caption{\textbf{Interaction averaging:} Effect of the magnetic field variation along a 1D tube for the lattice configuration used in the experiment and a constant tilt of $\SI{2}{\kilo\hertz} \approx 9J/h$, leading to a difference of the interaction strength $dU$ between the center and outermost lattice site. This effect depends on the cloud size and we plot the amplitude for a singlon and mixed initial state, which have different widths $\sigma_0$ in units of the lattice spacing $d$ due to the different dipole trap parameters. The inset depicts the relative weight of a certain variation $dU$ for three different values of the interactions $U_0$ and a cloud size of $90 d$, taking into account the Gaussian density distribution; here $U_0$ denotes the on-site interaction in the center of the 1D chain.}
	\label{fig:Uvariation}
\end{figure}

The magnetic field used to generate the linear external potential induces a spatial variation of the magnetic field across the system. This local variation of the magnetic field in turn results in a spatial dependence of the Hubbard interaction strength in the 1D tubes of the optical lattice. This variation is determined by the magnetic Feshbach resonance. The typical length of such a tube is evaluated from Fig.~\ref{fig:latticeloading}a and is between 290 and 360~lattice sites, dependent on the choice of initial state. From the equation for the magnetic Feshbach resonance we can compute the variation of the interaction strength between the center and outermost lattice site, denoted as $dU$. Fig.~\ref{fig:Uvariation} shows the result for a tilt of $\SI{2}{\kilo\hertz} \approx 9J/h$, representing an upper bound for the amplitude of the averaging effect. Note that the occupation of the lattice diminishes in the outer regions of the cloud such that the averaging effect has to be weighted by the respective atom number. This context is shown in the inset of Fig.~\ref{fig:Uvariation}. Using the Gaussian density distribution from Fig.~\ref{fig:latticeloading} we compute the distribution of interaction variation $dU$. We note that the biggest weight is near small values of $dU$ such that the value from the main panel is indeed only an upper bound. As illustrated, the distribution also depends on the set interaction $U_0$ due to the slope of the Feshbach resonance. We discuss the effect of interaction averaging in our numerical simulations in sec.~\ref{sec:APX}.

\begin{figure}[t]
	\includegraphics[width=3.3in]{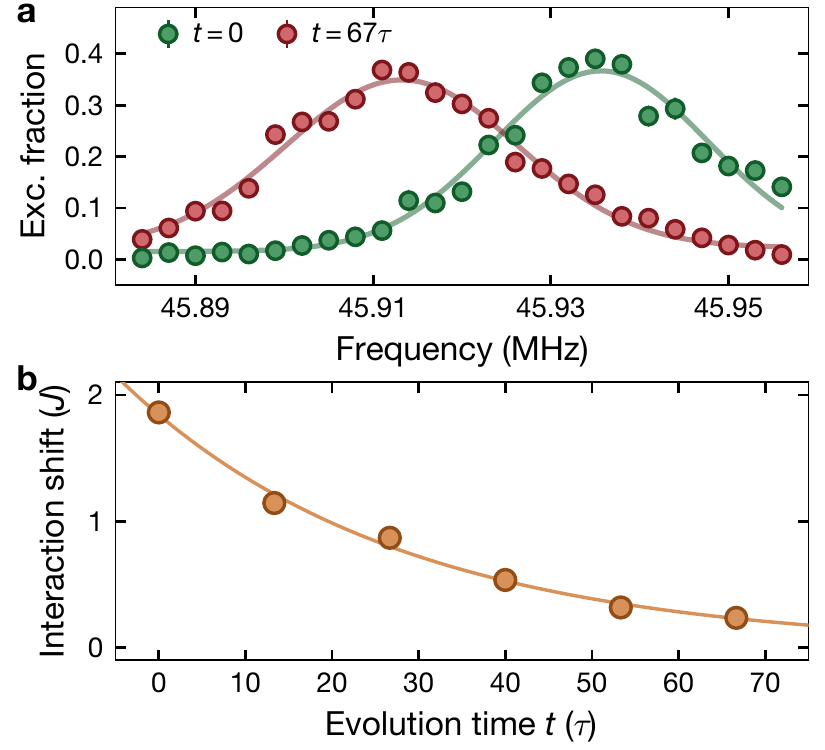}
	\caption{\textbf{Time-dependent interaction strength:} \textbf{a} Excitation spectrum of an RF $\pi$-pulse after two different evolution times in the lattice. The ramp time of the magnetic field was set to $\SI{30}{\milli\second}$. Solid lines are Gaussian fits for the extraction of the center frequency and the error bars represent the standard error of the mean from four averages. \textbf{b} Interaction shift $U(t)-U_0$ with the set interaction strength $U_0=0$ as a function of the evolution time extracted from the center frequencies in panel (a). The solid line is an exponential fit to the data points with a time constant of $32(6) \tau$.}
	\label{fig:int_evol}
\end{figure}

\subsection{Time-dependent interaction strength}
\label{sec:UofT}
The finite doublon lifetime in the optical lattice explained in sec.~\ref{sec:Dlifetime} sets additional constraints on the available wait times that can be introduced in the initial-state preparation, which is required in order to remove residual coherences and which is needed in order for the magnetic field to reach its final value. While the former happens fast, the latter requires about $\SI{140}{\milli\second}$ for the current in the coils to settle completely. Such a long wait time, however, would result in a considerable loss of doublons ($\approx 73\,\%$ according to the determined lifetime). As a compromise, we reduced the ramp duration and carefully characterized the residual time-dependence of the applied magnetic field. To this end we prepare a spin-polarized sample in the $\ket{\downarrow}$-state, send an RF $\pi$-pulse and scan its frequency across the $\ket{\downarrow} \leftrightarrow \ket{\uparrow}$ transition for various evolution times. We choose a ramp for the magnetic field time of $\SI{30}{\milli\second}$, the value used in the experiment for all doublon initial states, as zero point for the recorded excitation spectra in Fig.~\ref{fig:int_evol}a. The center frequency is extracted via Gaussian fits to the recorded line profile. From the center frequencies we can compute the magnetic field via the Breit-Rabi equation and based on that the Hubbard interaction strength in the lattice using the magnetic Feshbach resonance. This is shown in fig.~\ref{fig:int_evol}b. The shift is well-described by an exponential decay with a time constant of $\SI{24(4)}{\milli\second} = 32(6)\tau$. Thus, only the short-time dynamics is considerably affected by the time-dependence. This problem does not occur with the data for singlon initial states. There we choose a wait time of $\SI{140}{\milli\second}$ because we do not have to deal with the doublon loss and the singlon lifetime is about one order of magnitude larger. The singlon lifetime was determined in an independent measurement.

In order to understand the impact of the time-dependent interaction strength, we model the precise evolution using the approximate numerical description developed in Ref.~\cite{Bharath2021}. We study its impact on the experimental data in sec.~\ref{sec:APX} and compare it to the case of constant interaction strength.

\subsection{Singlon- and doublon-resolved imbalance}
The experimental data is presented in terms of the singlon, doublon and total imbalance so as to gain more insights into the dynamics governed by the effective Hamiltonians. In order to get access to these observables individually, we take successive images with and without doublon removal (sec.~\ref{sec:Dlifetime}). From these images we obtain information about the total imbalance $\mathcal{I}$ (no doublon removal) and the singlon imbalance $\mathcal{I}_S$ (with doublon removal). From the difference between both measurements we infer the doublon fraction $n_D$ and in particular the doublon imbalance $\mathcal{I}_D$ according to
\begin{equation}
	\mathcal{I}_D = \frac{\mathcal{I} - (1-n_D)\mathcal{I}_S}{n_D},
\end{equation}
where $n_D = N_d/N$ denotes the doublon fraction. The error bars are calculated via Gaussian error propagation.

In the numerics the different imbalances can be extracted more directly and are computed according to
\begin{eqnarray}
		\mathcal{I} &=& \langle \hat{\mathcal{I}} \rangle = \frac{1}{N} \sum_i (-1)^i \langle \hat{n}_i \rangle  \nonumber \\
		\mathcal{I}_S &=& \langle \hat{\mathcal{I}}_S \rangle = \frac{\sum_i (-1)^i \left(\langle \hat{n}_{i,\up}(1-\hat{n}_{i,\dn})\rangle +\langle \hat{n}_{i,\dn}(1-\hat{n}_{i,\up})\rangle\right)}{\sum_i \left(\langle \hat{n}_{i,\up}(1-\hat{n}_{i,\dn})\rangle+\langle \hat{n}_{i,\dn}(1-\hat{n}_{i,\up})\rangle \right)}  \nonumber\\
		\mathcal{I}_D &=& \langle \hat{\mathcal{I}}_D \rangle = \frac{\sum_i (-1)^i \langle \hat{n}_{i,\up} \hat{n}_{i,\dn})\rangle}{\sum_i \langle \hat{n}_{i,\up} \hat{n}_{i,\dn})\rangle}.
\end{eqnarray}
Note that the normalization factor in $\mathcal{I}_D$ ($\mathcal{I}_S$) depends on the doublon (singlon) number in the particular state such that it can diverge if no doublons (singlons) are present. Therefore, we only consider the charge imbalance in the construction of the analytical fragments as explained in sec.~\ref{sec:num_fragments}.

\subsection{Calibrations}
In this section we present detailed calibration measurements used to determine the microscopic parameters of the Fermi-Hubbard model [Eq.~(\ref{eq:Hamiltonian})] realized in the experiment.

\subsubsection{Tilt}
\label{sec:Tiltcalib}
A single-particle in a tilted lattice undergoes Bloch oscillations~\cite{Feldmann1992,Salomon1996} with the frequency set by the tilt. For the parity-projected Bloch oscillations measured by the imbalance $\mathcal{I}(t)$ it can be shown that~\cite{Scherg2020}
\begin{equation}
	\mathcal{I}(t) = \mathcal{J}_0 \left[ \frac{8J}{\Delta} \sin \left(\frac{\pi\Delta t}{h} \right) \right].
	\label{eq:I1}
\end{equation}
We measure the imbalance time evolution of a spin-polarized sample for short times (before the Bloch oscillations dephase) in the presence of a tilt and fit it with the analytical function [Eq.~(\ref{eq:I1})] to extract the tilt as fit parameter. This returns a very precise calibration of the tilt in our optical lattice. Exemplary time traces can be seen in the calibration of the RF dressing in Fig.~\ref{fig:DressedOscillations}a.

\subsubsection{Primary lattice depth and tunneling}
\label{sec:Jcalib}
\begin{figure}
	\includegraphics[width=3.3in]{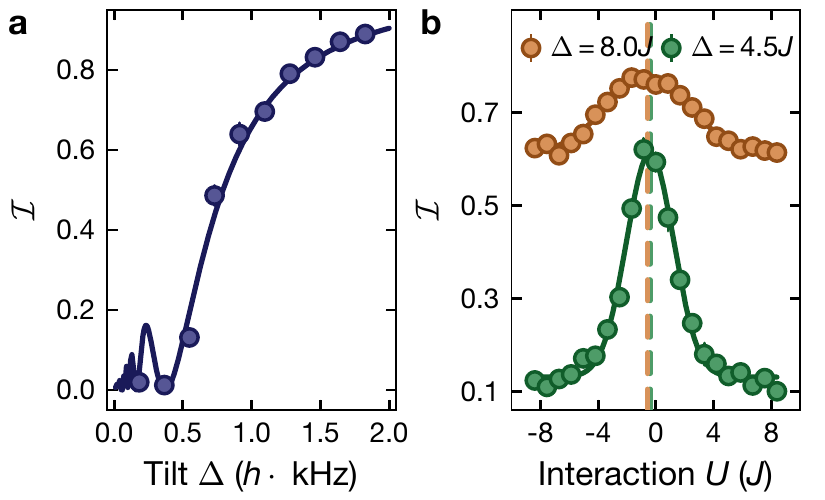}
	\caption{\textbf{Calibration of Hubbard parameters:} \textbf{a} Steady-state imbalance after $33\tau$ as a function of the tilt obtained from four averages. The solid line is an analytical fit according to Eq.~(\ref{eq:I2}) to extract the tunneling $J$. \textbf{b} Steady-state imbalance after $67\tau$ as a function of the set interaction in a 2D system with the tilt along the primary lattice axis for the case of large ($\Delta=8.0J$, 18 averages) and intermediate tilt ($\Delta=4.5J$, 10 averages). The data points are fitted with a Gaussian to extract the center, indicated by the vertical dashed lines. The data presented in this work is corrected for this systematic offset.}
	\label{fig:Calibration}
\end{figure}

At long times, as opposed to the situation in the previous section, a steady-state imbalance develops. For non-interaction atoms this is well described by a diagonal ensemble approach~\cite{Scherg2020}, which predicts
\begin{equation}
	\mathcal{I}_\infty(\Delta) = \mathcal{J}_0^2 \left(\frac{4J}{\Delta} \right).
	\label{eq:I2}
\end{equation}
We measure the imbalance of a spin-polarized sample after $33\tau$ as function of the tilt, calibrated by the method explained in the previous section. The data points are fitted with the analytical expression of Eq.~(\ref{eq:I2}) as depicted in Fig.~\ref{fig:Calibration}a. The extracted fit value is $J=h \cdot \SI{223(4)}{\hertz}$ in good agreement with the expectation for an ideal $12E_r$ primary lattice ($h \cdot \SI{216}{\hertz}$).

\begin{figure*}
	\includegraphics[width=\textwidth]{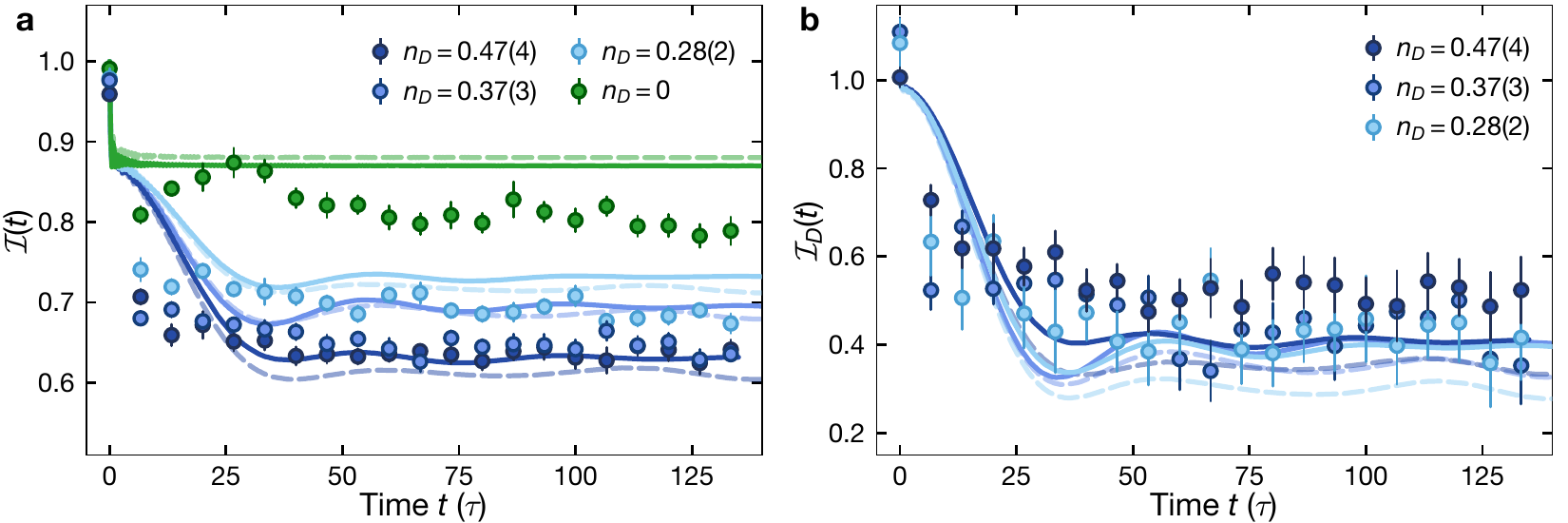}
	\caption{\textbf{Imbalance time traces in the dipole-conserving regime $\Delta \gg |U|,J$:} \textbf{a} Experimental and numerical charge imbalance traces for $\Delta/J = 8.0(2)$, $U/J = 2.7(2)$ and $\delta\Delta=0.6(2)\%$. Error bars denote the standard error of the mean obtained from ten averages. Dashed, transparent lines represent TEBD simulation results with the exact (Eq.~\ref{eq:Hamiltonian}), solid lines with the effective Hamiltonian (Eq.~\ref{eq:Heffdip}) including a hole fraction of $20\%$. Both Hamiltonians are simulated on a lattice with $L=101$ sites. \textbf{b} Time evolution of the doublon imbalance for three different initial doublon fractions. The experimental data is averaged ten times with the error bars representing the standard error of the mean. The lines represent numerical data with the same encoding and parameters as in panel (a).}
	\label{fig:Diptraces}
\end{figure*}

\subsubsection{Interaction offset}
We calibrate the non-interacting point of the Feshbach resonance by measuring the imbalance after $70\tau$ in a two-dimensional system, starting from a singlon initial state. For this purpose we lower the lattice along the orthogonal $y$-axis to $9E_{r,\perp}$ such that the tunneling elements $J$ along both axes are roughly equal. The tilt along the orthogonal axis was determined to be about $11\%$ of the value along the primary axis such that the dynamics is much less restricted. This 2D setting is more sensitive to interactions than the corresponding 1D system due to the larger configurational space (similar 2D systems were studied previously in a tilted lattice~\cite{Bakr2020} and with quasiperiodic disorder~\cite{Bordia2016}) and therefore a large signal amplitude is obtained around the non-interacting point. We fit a Gaussian to the data and extract the center, this is shown in Fig.~\ref{fig:Calibration}b. The obtained offset is taken into account for all the data presented in this work. We repeat this measurement for the weaker tilt ($\Delta=4.5J$) and find a smaller systematic offset compared to the strongly tilted case ($\Delta=8.0J$). This is expected since the correction of the Feshbach field required to compensate for the homogeneous part of the gradient field is smaller.

\section{Additional experimental data}
In this section we present additional data not shown in the main text. It includes complementary traces in the dipole-conserving regime ($\Delta \gg |U|,J$), experimental data for the regime of intermediate tilt and an investigation of the stability of imbalance time traces.

\subsection{Dipole-conserving regime}
In the main text we present central experimental time traces that show state-dependent dynamics in the dipole-conserving regime $\Delta \gg |U|,J$. Here, we show additional time traces and numerical simulations, which are used for the data points in Fig.~\ref{fig:fig2}c. In Fig.~\ref{fig:Diptraces}a we show charge imbalance time traces for $\Delta/J = 8.0(2)$ and $U/J = 2.7(2)$ for three types of initial states that differ in their initial doublon fraction between $n_D=0$ and $n_D=0.47(4)$. The experimental data is compared to time-averaged TEBD simulations of the full (Eq.~\ref{eq:Hamiltonian}) and effective Hamiltonian (Eq.~\ref{eq:Heffdip}). 

Besides the charge imbalance $\mathcal{I}$ we further show the doublon imbalance $\mathcal{I}_D$ in Fig.~\ref{fig:Diptraces}b. Unlike the total imbalance, it only shows a weak dependence on the initial doublon fraction, which can be directly explained by the action of the dominant hopping process $\hat{T}_3$. We find a larger disagreement with the simulations for higher doublon fractions. We associate this trend with an inhomogeneous charge distribution across the lattice that depends on the loading parameters and the averaging over the full cloud within one shot. 

\subsection{Doublon-number and spin-dependent behavior}
\begin{figure}
	\includegraphics[width=3.3in]{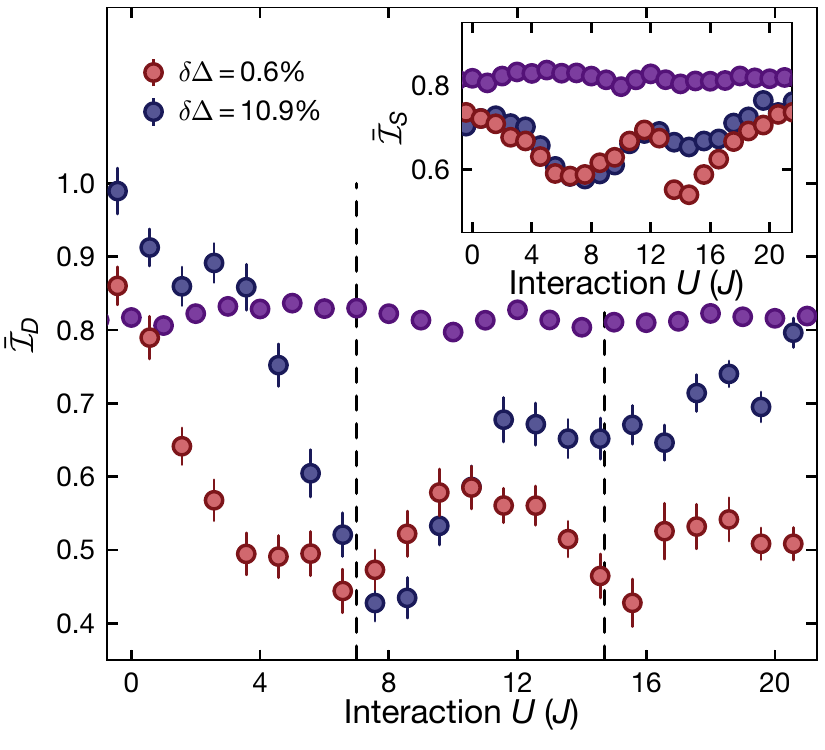}
	\caption{\textbf{Singlon- and doublon-resolved imbalance and spin-dependence as a function of the on-site interaction:} Doublon imbalance for minimal and maximal tilt
difference between both spin components versus on-site interaction
strength for $\Delta=8.0(2)J$ and $n_D=0.47(4)$, averaged between $67\tau$ and $80\tau$. For comparison we also show the singlon
initial state (purple) which shows no significant signal on this
timescale. The dashed vertical lines indicate the regimes of the tilt resonances $U = \Delta$ and $U = 2\Delta$. The inset shows the corresponding singlon imbalance
for the same setting. Data points contain eight averages
over five points in time, error bars are the standard error of
the mean. Each data point is averaged six times.}
	\label{fig:fig3suppUscan}
\end{figure}

We further show a full scan of the intermediate-time steady-state imbalance and their doublon-number and spin-dependent behavior for various values of the Hubbard interaction strength $U$ (Fig.~\ref{fig:fig3suppUscan}), which highlights the properties of the underlying microscopic processes in the emergent fragmented models and indicates the robustness of these features.

\subsection{Singlon-doublon resolved data for $|U|\simeq 2\Delta$}
\label{sec:Doubletilt_supp}
\begin{figure}
	\includegraphics[width=3.3in]{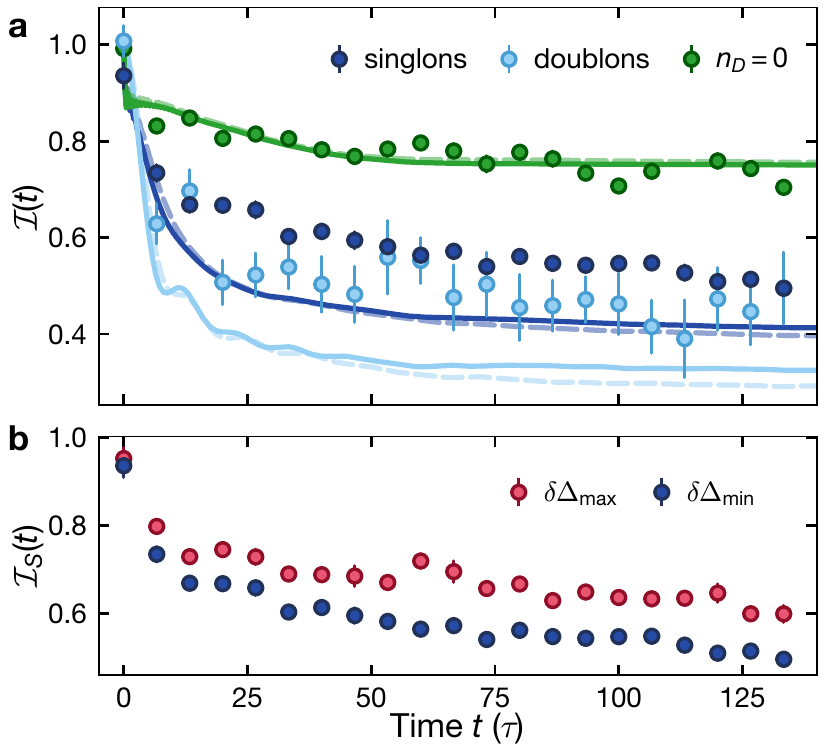}
	\caption{\textbf{Singlon-doublon-resolved imbalance time traces for $U\simeq 2\Delta$:} \textbf{a} Imbalance time traces at resonance $U\simeq 2\Delta$ [$\Delta=8.0(2)J$, $U = 14.7(2)J$] for a singlon initial state as well as singlon and doublon imbalance for a mixed initial state with $n_D=0.47(4)$. Lines are the results of TEBD simulations with the exact (dashed, transparent) and effective Hamiltonian (solid) on a system with $L=51$ sites, a doublon fraction of $n_D=0.46$ and a hole fraction of $n_h=0.2$. \textbf{b} Singlon imbalance $\mathcal{I}_S$ for the same mixed initial state as in (a) for minimal and maximal tilt difference $\delta\Delta$ as introduced in the main text ($\delta\Delta_\text{min}=0.6(2)\%$ and $\delta\Delta_\text{max}=11.0(2)\%$). Error bars denote the standard error of the mean from ten averages. See table~\ref{parameter_table} for more details.}
	\label{fig:fig3supp}
\end{figure}

In the main text we show the evolution of the imbalance $\mathcal{I}(t)$ in the regime $|U| \simeq 2\Delta$ (Fig.~\ref{fig:fig3}b), where we observe a significant dependence of the imbalance on the initial doublon density. However, the microscopic dynamics in this regime differs from the regime $\Delta \gg |U|,J$, as can be inferred from the effective Hamiltonian in Eq.~\eqref{eq:Heffdouble}. In order to resolve its microscopic processes we measure the singlon and doublon imbalance corresponding to the data shown in Fig.~\ref{fig:fig3}b in the main text. The results are depicted in Fig.~\ref{fig:fig3supp} and highlight characteristic properties of this regime in the tilted Fermi-Hubbard model, i.e. the doublon-dependence (Fig.~\ref{fig:fig3supp}a) and the impact of the spin-dependent tilt (Fig.~\ref{fig:fig3supp}b).

First, the difference between doublon and singlon dynamics is less pronounced than in the dipole-conserving regime (compare to Fig.~\ref{fig:fig2}b). This observation follows from the action of $\hat{T}_2$ [Eq.~\eqref{eq:Hdoub_terms}], which affects both singlons and doublons. Second, the imbalance of the singlon CDW is lower than in the dipole-conserving regime because the term $\hat{T}_1$ enables slow dynamics in this case via the dynamical formation of doublons. We qualitatively reproduce the experimental results with TEBD simulations of the exact and effective Hamiltonian. The residual offset between the traces is attributed to an inhomogeneous density distribution in the lattice.

\begin{figure}[t]
	\includegraphics[width=3.3in]{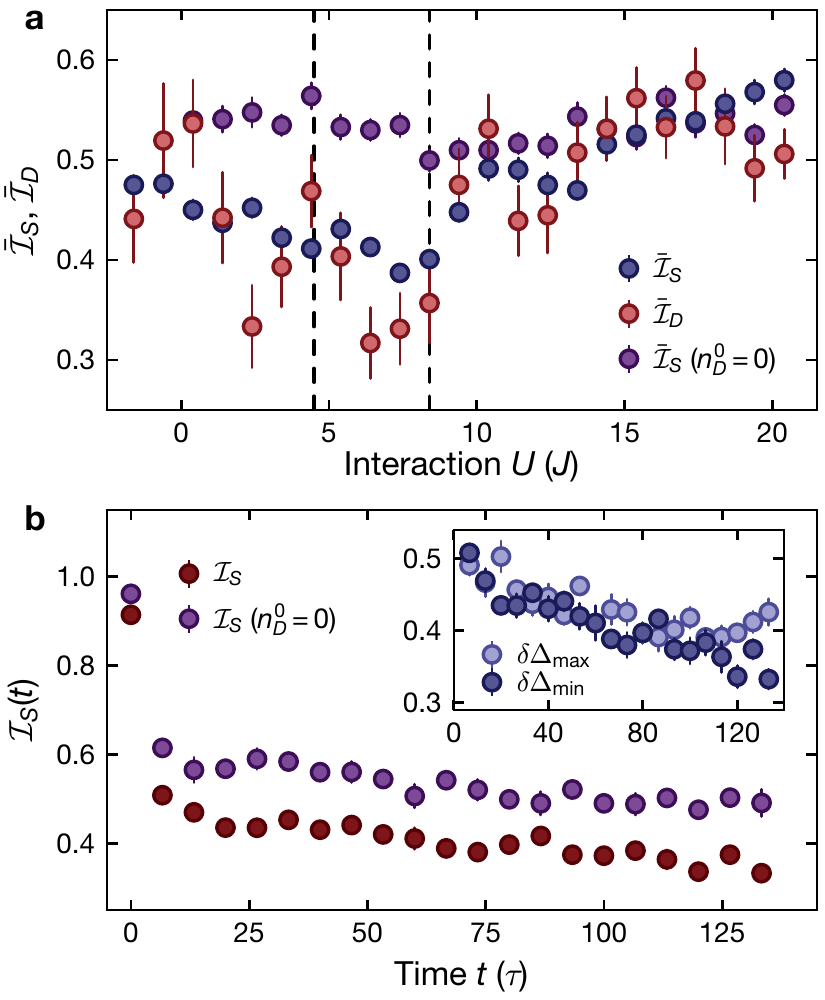}
	\caption{\textbf{Intermediate tilt regime at $\Delta=4.5J$:} \textbf{a} Singlon and doublon imbalance versus interaction strength averaged at five points in time between between $67\tau$ and $80\tau$ and averaged eight times in total. The imbalance rises for large interactions because the doublons become stable and immobile in the hard-core limit. The dashed vertical lines mark the regimes of the tilt resonance ($U=\Delta$) and second-order resonance ($U \simeq 2\Delta$). \textbf{b} Singlon imbalance time traces at the recorded second-order resonance ($U=7.7(2)J$) for an initial state with maximal doublon fraction $n_D=0.44(2)$ and for a singlon initial state for comparison. The inset shows the respective singlon imbalance evolution with and without the tilt difference. ($\delta\Delta_\text{min}=0.6(2)\%$ and $\delta\Delta_\text{max}=11.0(2)\%$). All data points are averaged ten times with the error bars representing the standard error of the mean.}
	\label{fig:fig5}
\end{figure}

\subsection{Regime of intermediate tilt}
While in the limit of large tilt the description via an effective Hamiltonian applies on exponentially long transient timecales, this is no longer an appropriate description for weak and intermediate tilts. In a previous work~\cite{Scherg2020} we explored the double-tilt resonance for intermediate tilts on the order of $\Delta \simeq 3J$ for the special case of a pure singlon CDW initial state. Here, we investigate to what extent our observations described in the main text, in particular the doublon-number dependent correlated tunneling processes and the sensitivity to a tilt difference, also apply in the intermediate tilt regime of $\Delta \simeq 4.5J$. We first measure the interaction dependence of the imbalance after an evolution time of $80\tau$. Fig.~\ref{fig:fig5}a presents the obtained results. We can identify the double-tilt resonance around $U/J \simeq 8.4$ as well as a third-order resonance at $U/J = 13$. For even stronger interactions we approach the hard-core limit, which is characterized by tightly-bound stable doublons. We further record time traces at the double-tile resonance $U \simeq 2\Delta$ like in Fig.~\ref{fig:fig3}b in the main text. The singlon imbalances are presented in Fig.~\ref{fig:fig5}b and only reflect a small difference between an initial singlon CDW and CDW with $n_D=0.44(4)$, a bit less pronounced than for the larger tilt presented in the main text (Fig.~\ref{fig:fig3}b). In the inset we further explore the effect of a tilt difference on the imbalance evolution and find that the traces are almost indistinguishable within our experimental uncertainty. This is also different compared to the regime of large tilt, where the tilt difference gives rise to significantly constrained dynamics. This context is highlighted in Fig.~\ref{fig:fig3supp}b showing a significant increase of the singlon imbalance in the presence of a tilt difference that can be explained by the additional constraints imposed by this energy mismatch. 

With this experimental data set we infer that even for intermediate tilts the central claims from the main text and the previous section~\ref{sec:Doubletilt_supp} remain valid. For the special case of the resonant regime $|U|\simeq 2\Delta$ we observed a pronounced dependence on the doublon number. Overall this effect may be slightly reduced, however, the overall value of the imbalance is also lower, which complicates a direct comparison. We further note that the sensitivity to a tilt difference between the spins is still present, but much less pronounced than in Fig.~\ref{fig:fig3supp}b. Finally, we observe a third-order resonance when $U \simeq 3\Delta$. These higher-order resonances are expected to be more prominent for weak tilt values, since the prefactors of the higher-order terms in the effective Hamiltonian are larger. For $\Delta/J=8.0$ we cannot investigate the appearance of higher-order resonances due to technical limitations in the accessible range of interaction strengths.

\subsection{Stability of imbalance time traces}
\begin{figure}
	\includegraphics[width=3.3in]{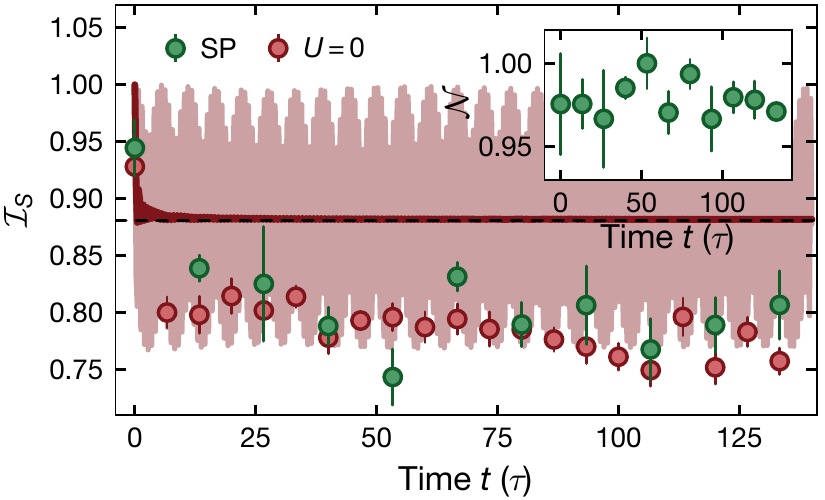}
	\caption{\textbf{Stability of imbalance traces:} Long-term stability of a spin-polarized (SP) and spin-mixed non-interacting sample ($U=0$) at fixed tilt $\Delta/J=8.0(2)J$. The red shaded area is a TEBD simulation on $L=101$ lattice sites exhibiting Bloch oscillations and the red solid line is the time averaged trace, while the black dashed line represents the analytical single-particle imbalance from Eq.~\eqref{eq:I2}. The inset displays the atom number of the spin-polarized sample and shows no sign of decay over the relevant time scales. Spin-polarized data was averaged thrice and non-interacting data ten times, error bars denote the standard error of the mean.}
	\label{fig:spinpol}
\end{figure}

As explained in the main text, we cannot reliably measure large imbalance values over long times. This is caused by heating due to noise on one of the lattices, which causes excitations and consequently a reduction in the detected imbalance. This effect is most pronounced for $\mathcal{I} \ge 0.8$ and most likely explains the discrepancy for the singlon time trace in Fig.~\ref{fig:fig2}b as well as in the tilt difference scan in Fig.~\ref{fig:fig4}a. At overall smaller imbalance values we can conciliate our experimental results with the numerics by considering the finite initial temperature in the lattice in terms of the hole fraction. In order to illustrate the decay effect over long times induced by the heating, we show an imbalance time trace of a spin-polarized sample in Fig.~\ref{fig:spinpol}. While the spin-polarized trace is stable within our resolution, the non-interacting trace (spinfull fermions with $U=0$) exhibits a weak decay on the observed timescale limiting our total experimentally-accessible observation times.

\section{Analytical and numerical techniques}
This section gives a detailed explanation of the theoretical and numerical concepts employed in this work. It comprises the effective Hamiltonians, simulations with exact diagonalization (ED), time-evolving block decimation (TEBD) and the approximate method recently developed in Ref.~\cite{Bharath2021}.

\subsection{Effective Hamiltonians}
The analytical expression of the effective Hamiltonian [Eq.~(\ref{eq:Heffdip}) in the main text] was derived previously in Ref.~\cite{Scherg2020}. We thus do not give the derivation again, but extend the discussion to the situation with a spin-dependent tilt. For the effective Hamiltonian in the resonance regime $|U| \simeq 2\Delta$ we also refer to the derivation performed in~\cite{Scherg2020}, but we provide the full analytical expression as used for the data in the main text (see Fig.~\ref{fig:fig3}b).

\subsubsection{Double-tilt resonance $|U|\simeq 2\Delta$}
When the Fermi-Hubbard Hamiltonian [Eq.~\eqref{eq:Hamiltonian} in the main text] is expanded up to leading order in perturbation theory for the special case of $U=2\Delta \gg J$, one obtains the following expression for the effective Hamiltonian (for simplicity we focus on $U\simeq 2\Delta$)
\begin{align}
	\begin{split}
		\hat{H}_{\mathrm{eff}}^{\mathrm{res}} = \hat{H}_0 &+ \frac{8J^2}{3\Delta} \sum_i \hat{n}_{i,\uparrow} \hat{n}_{i,\downarrow} - \frac{4J^2}{3\Delta} \hat{T}_{XY} \\
		&+ \frac{J^2}{\Delta} \hat{T}_1 - \frac{2J^2}{\Delta} \hat{T}_2 + \frac{2J^2}{3\Delta} \hat{T}_4 + \hat{H}_D.
	\end{split}
	\label{eq:Heffdouble}
\end{align}
Herein, the central energy scale is given by the effective hopping rate $J^{(2)} = J^2/\Delta$. The individual terms are off-diagonal hopping processes $\hat{T}_1$, $\hat{T}_2$ and $\hat{T}_4$, as well as diagonal contributions: 
\begin{widetext}
\begin{equation}
	\begin{split}
		\hat{H}_0 &= \Delta \sum_{i,\sigma} i \hat{n}_{i,\sigma} + 2\Delta \sum_i \hat{n}_{i,\up} \hat{n}_{i,\dn}, \\
		\hat{T}_{XY} &= \sum_{i,\sigma} (\hat{c}_{i,\bar{\sigma}}^\dagger \hat{c}_{i+1,\bar{\sigma}} \hat{c}_{i+1,\sigma}^\dagger \hat{c}_{i,\sigma} + \mathrm{h.c.}), \\
		\hat{T}_1 &= \sum_{i,\sigma} \left((1- \hat{n}_{i+2,\bar{\sigma}})(1- 2\hat{n}_{i+1,\bar{\sigma}}) \hat{n}_{i,\bar{\sigma}} \hat{c}_{i,\sigma}^\dagger \hat{c}_{i+2,\sigma} +\mathrm{h.c.} \right), \\
		\hat{T}_2 &= \sum_{i,\sigma} \left((1- \hat{n}_{i+2,\bar{\sigma}}) \hat{n}_{i,\sigma} \hat{c}_{i,\bar{\sigma}}^\dagger \hat{c}_{i+1,\bar{\sigma}} \hat{c}_{i+1,\sigma}^\dagger \hat{c}_{i+2,\sigma} +\mathrm{h.c.} \right), \\
		\hat{T}_4 &= \sum_{i,\sigma} \left((\hat{n}_{i,\sigma} - \hat{n}_{i+2,\bar{\sigma}})^2 (1-2(\hat{n}_{i+2,\bar{\sigma}} - \hat{n}_{i,\sigma}))  \times \hat{c}_{i,\bar{\sigma}} \hat{c}_{i+1,\bar{\sigma}}^\dagger \hat{c}_{i+1,\sigma}^\dagger \hat{c}_{i+2,\sigma} +\mathrm{h.c.} \right), \\
		\hat{H}_D &= -\frac{4J^2}{3\Delta} \left(2\sum_i \hat{n}_{i,\up} \hat{n}_{i,\dn} (\hat{n}_{i+1} + \hat{n}_{i-1}) + \sum_{i,\sigma} \hat{n}_{i,\sigma} \hat{n}_{i+1,\bar{\sigma}} \right).
	\end{split}
	\label{eq:Hdoub_terms}
\end{equation}

\end{widetext}

Due to renormalized interaction effects, the resonance does not exactly appear at $U=2\Delta$, but is shifted to $2\Delta - 8J^2/(3\Delta)$ up to second-order perturbation theory~\cite{Scherg2020}. In the provided effective Hamiltonian this detuning does only enter via the term $(U-2\Delta)\sum_{i}n_{i,\uparrow}n_{i,\downarrow}$ although in general it should also appear in every second-order process. In the following, we will show that it is sufficient to only keep this term at second order of perturbation theory and that it is not necessary to include additional corrections. To this end let us assume that $J \ll U,\Delta$ and at the same time $\abs{U-2\Delta}\ll J\ll \Delta $.  More precisely,  $\abs{U-2\Delta} =O\left( \frac{J^2}{\Delta}\right)$.  In this regime we can use degenerate perturbation theory, with well-defined degenerate subspaces of the Hamiltonian $H_0=U\sum_i n_{i,\uparrow}n_{i,\downarrow} + \Delta \sum_i i n_i$. We closely follow Ref.~\cite{LewensteinBook} in the following derivation. In particular let us focus on the off-diagonal second-order contributions in comparison to the detuning term. These are
\begin{align} \label{eq:Heff2ren} \nonumber
	& -J^2\left(\frac{1}{U+\Delta} + \frac{1}{U-\Delta} \right) \hat{T}_{XY} + \\
	\nonumber & \frac{J^2}{2}\left(\frac{1}{\Delta} + \frac{1}{U-\Delta} \right) \hat{T}_1 - \frac{2J^2}{U-\Delta} \hat{T}_2 \\
	\nonumber &+ J^2\left(\frac{1}{U-\Delta} + \frac{1}{\Delta} \right) \hat{T}^L_4  - J^2\left(\frac{1}{\Delta} - \frac{1}{U+\Delta} \right) \hat{T}^R_4 \\
	&+ (U-2\Delta)\sum_{i} \hat{n}_{i,\uparrow} \hat{n}_{i,\downarrow} + \hat{H}_0,
\end{align}
where we used the notation
\begin{align}
    \nonumber \hat{T}^L_4&= \sum_{i, \sigma} \left( \hat{n}_{i,{\sigma}}(1- \hat{n}_{i+2,\bar{\sigma}}) \hat{c}_{i,\bar{\sigma}} \hat{c}^{\dagger}_{i+1,\bar{\sigma}}  \hat{c}^{\dagger} _{i+1,\sigma} \hat{c}_{i+2,\sigma} \right. \\
	& \left. + \textrm{h.c.}\right),    \\
	\nonumber \hat{T}^R_4&= \sum_{i, \sigma}\left( (1- \hat{n}_{i,{\sigma}}) \hat{n}_{i+2,\bar{\sigma}} \hat{c}_{i,\bar{\sigma}} \hat{c}^{\dagger}_{i+1,\bar{\sigma}}  \hat{c}^{\dagger} _{i+1,\sigma} \hat{c}_{i+2,\sigma}  \right.
	\\ & \left.+ \textrm{h.c.}\right).
\end{align}

It is important to notice that all coefficients match with those in Eq.~(\ref{eq:Heffdouble}) in the limit $U \to 2\Delta$. In fact, let us write $U=2\Delta + \epsilon$ with $\abs{\epsilon}\ll J\ll \Delta$. Then any coefficient $J_{\textrm{eff}}$ in Eq.~\eqref{eq:Heff2ren} can be expressed as
\begin{align}
\nonumber&J_{\textrm{eff}}=J^2\frac{1}{c\Delta \pm \epsilon}=\frac{J^2}{c\Delta}\frac{1}{1\pm \epsilon/(c\Delta)}\\
 &=\frac{J^2}{c\Delta}\left(1\mp \frac{\epsilon}{c\Delta} + O\left(\frac{\epsilon^2}{\Delta^2}\right) \right)\approx \frac{J^2}{c\Delta},
\end{align}
for some $c\sim O(1)$ number.  For the terms in Eq.~\ref{eq:Heff2ren}, $c=1, 2$ or $3$. Therefore, we conclude that to second order, these corrections can be neglected, whenever $\abs{U-2\Delta}\ll J\ll \Delta$, though we should still keep the correction term $U\sum_i \hat{n}_{i,\uparrow} \hat{n}_{i,\downarrow} +\Delta\sum_i i \hat{n}_i - \hat{H}_0 = \epsilon \sum_i \hat{n}_{i,\uparrow} \hat{n}_{i,\downarrow}$. 

\subsubsection{Effective Hamiltonian with tilt difference}
The effective Hamiltonian in Eq.~(\ref{eq:Heffdip}) was obtained under the assumption of a spin-independent tilt $\Delta\equiv \Delta_\downarrow =\Delta_\uparrow$, together with $\Delta\gg J, \,\abs{U}$. Nevertheless, our experimental setup allows us to tune the tilts in the regime $|\tilde{\Delta}_\downarrow - \tilde{\Delta}_\uparrow|/\Delta_\downarrow \in [0.006, 0.11]$ using the technique of RF dressing. To simplify notations and because the experimental implementation does not play any role for the following discussion, we denote the spin-dependent tilt for the dressed states in this section simply as $\Delta_\sigma$. Now, as long as $\delta_\Delta\equiv \Delta_\downarrow - \Delta_\uparrow>0$, is small compared to the hopping rate $J$ ($\delta_\Delta \ll J$) our perturbative expansion also works up to some additional contributions. To see how this happens let us write Eq.~(\ref{eq:Hamiltonian}) as follows
\begin{equation}
\begin{split}
	\hat{H}=&-J\sum_{i,\sigma} \left( \hat{c}_{i,\sigma}^\dagger \hat{c}_{i+1,\sigma}+\textrm{h.c.} \right)+ U\sum_i \hat{n}_{i,\uparrow} \hat{n}_{i,\downarrow} \\
	&+ \Delta_\uparrow\sum_i i \hat{n}_i + \delta_\Delta \sum_{i}i \hat{n}_{i,\downarrow}.
\end{split}
\end{equation}

Since $\Delta_\uparrow\gg J \gg \delta_\Delta$, we just keep the small contribution in $\delta_\Delta$ and follow the same expansion in $J/\Delta_\uparrow$ as for a spin-independent tilt. Since $[\sum_{i}i \hat{n}_{i,\downarrow},\sum_i i \hat{n}_i]=0$, this contribution appears already at first order. Thus, the resulting effective Hamiltonian becomes
\begin{align} \label{eq:effdelta}
	\nonumber &\hat{H}_{\textrm{eff}}^{\textrm{dip}}=J^{(3)}(\hat{T}_3 + 2\hat{T}_{XY}) + 2J^{(3)} \hat{V} + \tilde{U}\sum_i \hat{n}_{i,\uparrow} \hat{n}_{i,\downarrow} \\
	&+ \Delta_\uparrow\sum_i i \hat{n}_i + \delta_\Delta\sum_i i \hat{n}_{i,\downarrow}.
\end{align}

To understand the effect of the additional term in $\delta_\Delta$, we consider the following family of states $\{\ket{n}\}=\{ \ket{ \dots \uparrow  \uparrow \uparrow \overset{n}{\updownarrow} 0 \uparrow  \uparrow \uparrow \dots}\}$ with a doublon $\updownarrow$ at lattice site $n$ and a hole at site $n+1$ in the chain surrounded by a polarized background. This background can be of whatever uniform polarization. This set defines a Krylov subspace left invariant under the action of $\hat{H}_{\textrm{eff}}^{\textrm{dip}}$. In particular, the term $\hat{T}_{XY}$ has a trivial action and every diagonal contribution, except the term in $\delta_\Delta$, is proportional to the identity in this subspace. Thus, we project Eq.~\eqref{eq:effdelta} into this subspace leading to the single-particle Hamiltonian
\begin{align}
	\nonumber  &\left.\hat{H}_{\textrm{eff}}^{\textrm{dip}} \right|_{\{\ket{n}\}}= - J^{(3)}\sum_n \ket{n}\bra{n+1} + \textrm{h.c.} 
	\\ & + \delta_{\Delta}\sum_n n \ket{n}\bra{n},
\end{align}
that describes a doublon-hole pair surrounded by a uniform polarized background and that propagates as a single-particle with hopping amplitude $J^{(3)}$ in the presence of a tilt $\delta_\Delta$. Therefore, quenching the system from an initial state $\ket{n}$ leads to a constant singlon imbalance $\mathcal{I}_S(t)=0$ and Bloch oscillations in the doublon imbalance with the analytical form $\mathcal{I}_D(t)=\mathcal{J}_0\left(\frac{8J^{(3)}}{\delta_\Delta}\sin\left(\frac{\pi \delta_\Delta t}{h} \right) \right)$, where $\mathcal{J}_0$ denotes the $0$th order Bessel function of the first kind and $h$ the Planck constant~\cite{Scherg2020}. This leads to a time-average value at infinite time equal to $\mathcal{J}_0^2\left( \frac{4J^{(3)}}{\delta_\Delta} \right)$, shown in Fig.~\ref{fig:fig2}d in the main text.

The relation with previous studies of dipole-conserving systems with spin-$1$ variables and 3-local Hamiltonians~\cite{Sala2020} can be made explicit by the following mapping. We can distinguish between charge and spin quantum numbers respectively defined on a site via $\hat{q}_i\equiv \hat{n}_{i}-1$ and $\hat{S}^z_i \equiv \hat{n}_{i,\uparrow}- \hat{n}_{i,\downarrow}$. Thus a state $\ket{n}$ can be interpreted as a dipole $\ket{+-}$ (in charge degrees of freedom) in a zero-charge background.

So far we have focused on the regime $J\gg \delta_\Delta$, where we could simply use the usual perturbative Schrieffer-Wolff technique to obtain an effective Hamiltonian (or equivalently a high-frequency expansion in the reference frame of the tilt~\cite{Scherg2020}). However, this is not the case when $J\geq \delta_\Delta$ and for generic tilts such that $\Delta_\uparrow/\Delta_\downarrow \not \in \mathbb{Q}$. This requires a different approach and it is in fact a direct application of the theory developed in~\cite{Else2017} (see in particular Sec.~VIII.B and Appendix A) for quasiperiodically driven systems. Let us write Eq.~(\ref{eq:Hamiltonian}) in the frame where it becomes a quasiperiodic Hamiltonian
\begin{equation}
\begin{split}
	&\hat{H}(t)= -J\sum_{i,\sigma} \hat{c}_{i,\sigma}^\dagger \hat{c}_{i+1,\sigma}e^{-i\Delta_\sigma t } + \textrm{h.c.} 
	\\ &+ U\sum_i \hat{n}_{i,\uparrow} \hat{n}_{i,\downarrow},
\end{split}
\end{equation}
with periods $T_\sigma=2\pi/\Delta_\sigma$ for $\sigma=\uparrow,\downarrow$. Then we can decompose $\hat{H}(t)=\sum_{n_1,n_2}e^{i(\Delta_\uparrow n_1 + \Delta_\downarrow n_2)t} \hat{H}_{(n_1,n_2)}$ in terms of a Fourier series with the only non-vanishing terms $\hat{H}_{(0,0)},\, \hat{H}_{(\pm 1,0)},\, \hat{H}_{(0,\pm 1)}$.

Following appendix A in~\cite{Else2017}, we can obtain an effective static Hamiltonian as a perturbative expansion in $J/\Delta_\sigma$. The $0$th order contribution is given by the diagonal term $\hat{H}_{(0,0)}=U\sum_i \hat{n}_{i,\uparrow} \hat{n}_{i,\downarrow}$  and the first order contribution vanishes up to boundary terms, in both cases agreeing with the result for a spin-independent tilt~\cite{Scherg2020}. Using~\cite{Else2017}, we explicitly obtain the second order contribution 
\begin{align}
	\nonumber &\sum_{\bm{n}\neq \bm{0}} \left( \frac{1}{ 2 (\bm{n}\cdot\bm{\omega})^2 }   [ \hat{H}_{\bm{n}}, [\hat{H}_{\bm{0}}, \hat{H}_{-\bm{n}}] ] \right.\\
	& \left.+\sum_{\bm{m}\neq \bm{0},\bm{n}}\frac{1}{ 3 (\bm{n}\cdot\bm{\omega})(\bm{m}\cdot\bm{\omega}) }   [ \hat{H}_{-\bm{n}}, [\hat{H}_{\bm{n}-\bm{m}}, \hat{H}_{\bm{m}}] ] \right)
\end{align}
where we have used the notation $\bm{n}=(n_1,n_2)$ and $\bm{\omega}=(\Delta_\uparrow,\Delta_\downarrow)$. Since $[\hat{H}_{(\pm 1,0)}, \hat{H}_{(0,\pm 1)}]=0$ as well as $[\hat{H}_{(+ 1,0)}, \hat{H}_{(-1,0)}]=0$ (up to boundary terms), the only non-trivial contribution is given by $\sum_i (\hat{n}_{i+1,\uparrow} - \hat{n}_{i,\uparrow})(\hat{n}_{i+1,\downarrow} - \hat{n}_{i,\downarrow})$, namely a diagonal contribution but no correlated hopping process. 
This is in line with the fact that dipole-conserving processes require the presence of interactions~\cite{Scherg2020, Taylor_Stark2020, Moudgalya2019}. For a spin-independent tilt the Hubbard interaction mediates these hoppings involving the two spin species and leading to the conservation of the dipole moment $\sum_i i \hat{n}_i$. On the other hand, when the tilts are different, we expect the dipole moment for each spin to be independently conserved. However, these processes would require longer-range interactions involving the same spin species (e.g., $\sum_{i,\sigma} \hat{n}_{i,\sigma} \hat{n}_{i+1,\sigma}$). Thus, we expect that the effective Hamiltonian for $J,U\ll \Delta_\sigma$ with $J\geq \delta$ and $\Delta_\uparrow/\Delta_\downarrow \not \in \mathbb{Q}$ obtained from Eq.~(\ref{eq:Hamiltonian}) to only contain diagonal contributions, thus leading to frozen dynamics.

\subsubsection{Role of the Schrieffer-Wolff transformation}
Given a general static Hamiltonian $\hat{H}=g \hat{H}_0 + \hat{V}$ with $[\hat{V},\hat{H}_0]\neq 0$, there are different approaches to obtain the effective Hamiltonian in the limit $g\to\infty$. The general idea is to find a unitary transformation that brings the Hamiltonian $\hat{H}$ close to a block-diagonal form in the eigenbasis of $\hat{H}_0$, though the choice of this rotation is not unique. One of those approaches is given by the Schrieffer-Wolff transformation, which allows us to obtain a local effective Hamiltonian in a perturbative manner~\cite{BRAVYI20112793}.  This is achieved through a unitary transformation $e^{i\hat{S}}$ (with $[\hat{S},\hat{H}_0]\neq 0$), which is chosen such that the off-diagonal terms $R$ can be made exponentially small $\|R\|\leq e^{-cg}$, when truncating the series to a finite optimal order $n^*$~\cite{Else2017, Abanin_theory}. This is
\begin{equation}
	e^{i\hat{S}}\hat{H}e^{-i\hat{S}}=\hat{H}_{\textrm{eff}} + R,
\end{equation}
with $[\hat{H}_0,\hat{H}_{\textrm{eff}}]=0$. Thus, it implies that $e^{-i\hat{S}} \hat{H}_0 e^{i\hat{S}}$ is an almost-conserved observable, in the sense of Theorem 3.2 in~\cite{Abanin_theory}. We may write 
\begin{equation}
\hat{H}_{\textrm{eff}} = \sum_k \hat{H}_{\textrm{eff}}^{(k)}
\end{equation}

Here,  $\hat{H}_{\textrm{eff}}^{(k)}$ is the contribution to the effective Hamiltonian at order $k$, which scales as $1/g^k$.  In addition to the emergent symmetry $[\hat{H}_0,\hat{H}_{\textrm{eff}}]=0$, the effective Hamiltonian, truncated to order $n$ is fragmented in the local number basis consisting of number states, $\ket{\bf{n}}$ for the regimes of interest.  Each fragment is a connected subset of the set of number states, where the connectivity is defined by the Hamiltonian treated as an adjacency matrix~\cite{Sala2020}.  Higher orders $\geq n$ will mix most of these fragments. Each of these fragments defines a subspace $\mathcal K=\{ \ket{\bf{n}}\}$, to which we can associate a projector $P_{\mathcal K}$ such that $[P_{\mathcal K}, \sum_{k=0}^n\hat{H}^{(k)}_{\textrm{eff}}]=0$. Thus, in the limit $g\to \infty$, we find that $[\hat{H},\tilde{P}_{\mathcal{K}}]\sim O\big(1/g^{n+1}\big)$ with $\tilde{P}_{\mathcal{K}} = e^{-i\hat{S}} P_{\mathcal{K}}e^{i\hat{S}}$. This means that fragmentation physics of the $n^{th}$-order effective Hamiltonian $\hat{H}^{(n)}_{\textrm{eff}} $ is expected to survive up to times $t\sim g^{n+1}$. This implies that this a transient phenomenon, when realized perturbatively as in the case of the tilted Fermi-Hubbard model. The fragment is spanned by the states $\tilde{\mathcal{K}}=\{ e^{-i\hat{S}}\ket{\bf{n}}\}$. This is not a fragment in the number basis but some locally dressed version of it since $\hat{S}$ is local and $1/g\ll 1$. However, since $\lim_{g \to \infty}e^{i\hat{S}} = 1$, this means $\tilde{\mathcal{K}}\to\mathcal{K}$.

\subsection{Details on numerical techniques}
In this section we discuss the numerical methods that were used to simulate the dynamics of the system as well as to investigate the fragmented structure. We can split these into two main approaches: (1) Tensor network (TN) algorithms and (2) Exact diagonalization (ED) for small system sizes.

\subsubsection{Tensor network algorithms and convergence of simulations}
We use time-evolving block decimation (TEBD)~\cite{Vidal2003} to simulate the real-time dynamics. This method is particularly amenable for nearest-neighbor Hamiltonians, as it trotterizes the time evolution into local gates, truncating after every step keeping the largest $\chi$ Schmidt values. In the following we fix the tilt $\Delta=8J$ and analyze the convergence of TEBD for the imbalance $\mathcal{I}(t), \mathcal{I}_S(t)$ and $\mathcal{I}_D(t)$ in the two regimes $U=2.7J$ (dipole-conserving regime) and $U=15.7J$ (resonant regime) for a given initial state (no averaging). Notice that requiring convergence for local observables does not imply convergence in other observables like the half-chain entanglement entropy. We use a second-order Trotter decomposition with time step $dt=0.01\tau$ and provide numerical data that shows that these parameters are sufficient to get a good convergence. Moreover, as we probe different families of initial states, we focus on the convergence of $n_D$ with the largest fractal dimension, which generates the largest amount of entanglement. In particular we fix $n_D=0.46$ as this is the highest doublon fraction in the experiment.

\begin{figure*}
	\includegraphics[width=6.6in]{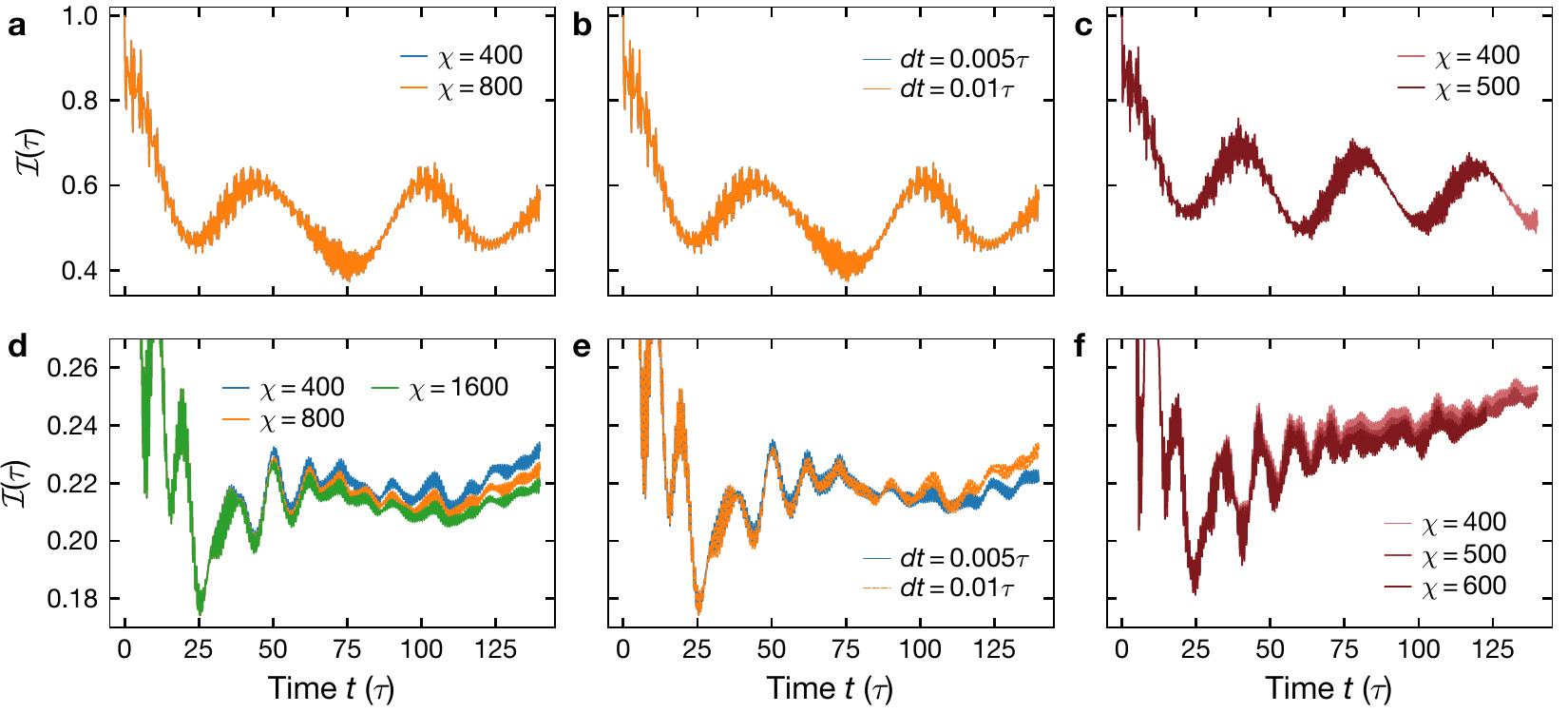}
	\caption{\textbf{Convergence analysis of TEBD simulations with exact Eq.~(\ref{eq:Hamiltonian}) and effective Hamiltonians Eq.~(\ref{eq:Heffdip}) and Eq.~(\ref{eq:Heffdouble}):} All traces are computed for one single initial state with $n_D=0.46$ without holes and on a system with 101 (panels a-c) or 51~lattice sites (panels d-f) \textbf{a} In the dipole-conserving regime ($\Delta = 8.0J$, $U=2.7J$) the charge imbalance time traces show perfect agreement upon doubling the bond dimension. \textbf{b} This observation equally applies for a reduced Trotter step size at constant $\chi=400$. \textbf{c} For the effective Hamiltonian the convergence with bond dimension is equally established. \textbf{d} In the resonant regime $U=15.7J \simeq 2\Delta$ the qualitative agreement is good throughout, but for evolution times $\ge 40\tau$ the traces start to diverge. Thus, in this regime the simulations are not yet fully converged even for bond dimension 1600. \textbf{e} The behavior is qualitatively the same upon a reduction of the Trotter step at $\chi=400$. \textbf{f} The convergence for the effective Hamiltonian shows a similar trend as the exact one in panel (d), i.e. at late times the traces start to diverge.}
	\label{fig:Convergence}
\end{figure*}

We first analyze the real-time evolution governed by the tilted Fermi-Hubbard chain in Eq.~(\ref{eq:Hamiltonian}). Fig.~\ref{fig:Convergence}a shows perfect convergence of the time traces when doubling the bond dimension for the dipole-conserving regime with system size $L=101$ and up to times $t=140\tau$.

Fig.~\ref{fig:Convergence}d shows results in the resonant regime for system size $L=51$ and various bond dimensions. Unlike for the previous case, the agreement is quantitative only for times $t\le 40\tau$ and only qualitative at later times. Although the quantitative disagreement between the traces is marginal, we observe that even at bond dimension 1600 the traces have not yet fully converged. Notice that this regime is less non-ergodic than the previous one due to the additional dynamical processes. This leads to faster spreading of correlations and thus stronger truncation errors at early times. Nevertheless, for doublon fraction  $n_D=0$, we find a perfect agreement at all times. We further analyze convergence in terms of the Trotter step $dt$ in Fig.~\ref{fig:Convergence}b for the dipole-conserving regime and Fig.~\ref{fig:Convergence}e for the resonant regime. Our conclusions are the same, i.e. perfect convergence is observed for the former upon halving the Trotter step, while deviations appear at resonance for $t\le 80\tau$.

Once we have understood the limitations of these methods for the exact evolution, we now address the implementation of the effective dynamics. The effective evolution of the system is given by~\cite{Scherg2020}
\begin{equation}
	\hat{U}(t,t_0)= e^{-i\lambda \hat{S}}e^{-i(t-t_0) \hat{H}_{\textrm{eff.}}} e^{i\lambda \hat{S}},
\end{equation}
with $\hat{S}$ a self-adjoint operator, which in these perturbative regimes is given by
\begin{equation}
\begin{split}
	&\hat{S}_{\textrm{dip}}= -i\sum_{i,\sigma}\left(1+\frac{U}{\Delta}\left(\hat{n}_{i+1,\bar{\sigma}}-\hat{n}_{i,\bar{\sigma}} \right)\right)\\
	& \times \hat{c}_{i,\sigma}^\dagger \hat{c}_{i+1,\sigma}+ \textrm{h.c},
\end{split}
\end{equation}
for the dipole-conserving regime and
\begin{align}
	\nonumber &\hat{S}_{\textrm{res}}= -i\sum_{i,\sigma}\left(1- 2\hat{n}_{i,\bar{\sigma}}-\frac{2}{3} \hat{n}_{i+1,\bar{\sigma}}\right.\\
	&\left.+\frac{8}{3} \hat{n}_{i,\bar{\sigma}} \hat{n}_{i+1,\bar{\sigma}} \right) \hat{c}_{i,\sigma}^\dagger \hat{c}_{i+1,\sigma}+ \textrm{h.c},
\end{align}
in the resonant regime, with $\lambda=J/\Delta\ll 1$ a small perturbative parameter. Thus, in order to implement the dynamics we apply three different time evolutions in the following way: after obtaining the MPS representation of the initial state $\ket{\psi(0)}$,
\begin{enumerate}
	\item In order to apply the operator $e^{i\lambda \hat{S}}$ we treat $\lambda$ as a ``time" and apply  a backward evolution with the operator $\hat{S}$ as the ``Hamiltonian" and duration $\lambda$. This will already generate a considerably amount of entanglement as $\hat{S}$ takes the form of a hopping Hamiltonian.
	\item For a given time duration $t$, we evolve with Hamiltonian $\hat{H}_{\textrm{eff}}$ thus obtaining $e^{-it \hat{H}_{\textrm{eff}}}e^{i\lambda \hat{S}}\ket{\psi(0)}$.
	\item Finally, to measure a given observable, we make a copy of the resulting MPS and apply the operator $e^{-i\lambda \hat{S}}$ again, by treating $\lambda$ as ``time" and applying a forward evolution with $\hat{S}$ and duration $\lambda$.
\end{enumerate}
Since the effective Hamiltonians are not of nearest-neighbor type, we group pairs of sites in order to apply the TEBD algorithm.
Finally, we investigate the convergence behavior of the effective Hamiltonians. We expect that the convergence in case of the effective Hamiltonians is faster than of the full Hamiltonian because they only contain the leading order terms that generate less entanglement. We show the result of the analysis in Fig.~\ref{fig:Convergence}c for the dipole-conserving regime. Here the agreement is perfect upon increasing the bond dimension. The traces from the resonant regime in Fig.~\ref{fig:Convergence}f show the same qualitative behavior as for the exact Hamiltonian with quantitative accordance for times $t \le 50\tau$.

\subsubsection{Time-averaged imbalance and comparison between exact and effective dynamics}
As mentioned in the main text, we reproduce our experimental time traces with TEBD simulations for system sizes $L=51$ and $L=101$. A better agreement with the experimentally prepared initial state is achieved by including $20\,\%$ holes on even sites. From an earlier experiment with similar conditions~\cite{Scherg2018} this number appears to be reasonable. Holes in the initial state effectively cut the chain and restrict the dynamics because the processes of the effective Hamiltonians require certain charge configurations for resonant hoppings to take place. Consequently, the observed imbalance is higher than in the ideal situation and this describes our experimental data much more accurately. In addition, the simulations typically show strong oscillations, especially in the dipole-conserving regime that are not observed experimentally. Since we average over the entire atomic ensemble, such coherent signatures quickly dephase and we measure an average value. Therefore, we compare the data in the main text to time-averaged imbalance traces defined as
\begin{equation}
	\mathcal{I}(t) = \frac{1}{t} \int_0^t \mathcal{I}(\tau) d\tau.
	\label{eq:timeaverage}
\end{equation}

\begin{figure}
	\includegraphics[width=3.3in]{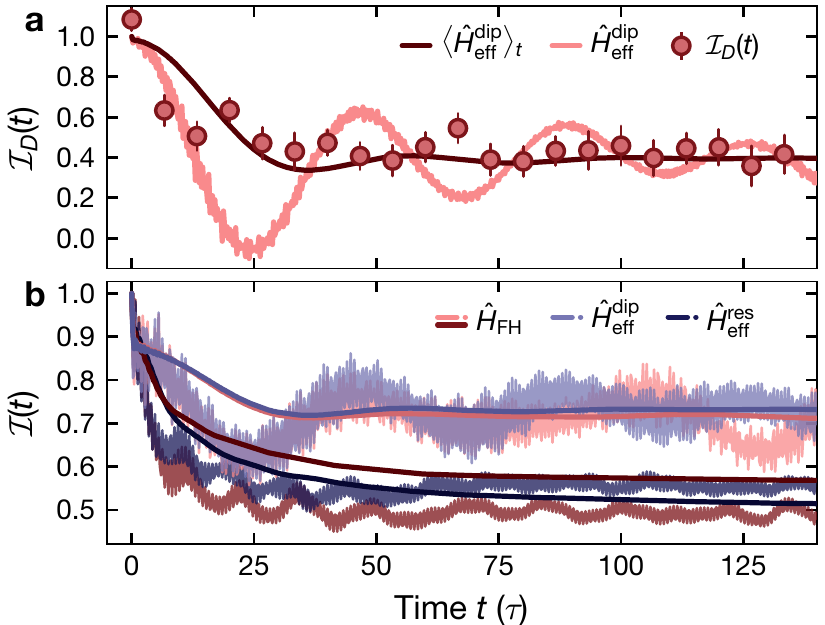}
	\caption{\textbf{Comparison between full and time-averaged traces:} \textbf{a} Experimental data in the dipole-conserving regime (see Fig.~\ref{fig:fig2}) with an initial doublon fraction of $n_D=0.28(2)$ together with TEBD simulations with $n_D=0.26$ of the effective Hamiltonian from Eq.~(\ref{eq:Heffdip}) on $L=101$ lattice sites [$U=2.7J$ and $\Delta=8.0J$]. The time-averaged trace captures the average value of the oscillating full trace and matches the experimental data. \textbf{b} Comparison of numerical time traces from the exact [Eq.~\eqref{eq:Hamiltonian}] and effective Hamiltonians [Eqns.~\eqref{eq:Heffdip} and~\eqref{eq:Heffdouble}] for both parameter regimes studied in the main text. Solid lines show the time-averaged traces.}
	\label{fig:Timeavg}
\end{figure}

The effect of the time average is illustrated in Fig.~\ref{fig:Timeavg}a for simulations with the effective Hamiltonian in the dipole-conserving regime for an initial state with $n_D=0.28(4)$ in the experiment and $n_D=0.26$ in the simulations. While the full simulation shows long-lived oscillations, these are erased in the time-averaged curve. We find that the mean value is very well captured and the shape of the experimental traces is well reproduced.

In Fig.~\ref{fig:Timeavg}b we provide an exemplary comparison between the exact and effective traces in both regimes studied. It shows that the traces agree very well throughout in the dipole-conserving regime and so do the time-averaged traces. In the resonance regime a systematic offset forms between the time traces after about ten tunneling times, a deviation that can be explained by higher-order contributions in the exact Hamiltonian. This offset is also captured by the time average. We notice the persistent oscillations in the former case, which are less pronounced on resonance.

\subsubsection{Doublon number conservation and infinite-temperature imbalance}
\label{sec:num_fragments}
While TEBD allow us to reach rather large system sizes on intermediate time scales, ED is more amenable to simulate the long-time behavior as well as exact properties of the system. Here we are limited to system sizes $L=17$ due to the spin degree of freedom and the large corresponding dimension of the Hilbert space. To simulate the system, we write Hamiltonian~(\ref{eq:Hamiltonian}) in the subspace with fixed particle numbers $N_{\uparrow}$ and $N_{\downarrow}$. In order to study the fragmented structure, we directly construct the effective Hamiltonians within the corresponding global symmetry sectors. Experimentally, the local $S^z$-basis is the relevant one to study the fragmented structure, as initial states are prepared as incoherent sums over product states in this basis. These are labeled by $N_{\uparrow}, N_{\downarrow}$ and the respective effective global conserved quantity. Focusing on the experimentally prepared CDW initial states for different $n_D$, we numerically construct their containing fragments, treating the effective Hamiltonian as an adjacency matrix~\cite{Scherg2020} in the local $S^z$ basis. This way, we identify all states dynamically connected to a given initial state and thus the full fragmented structure of the Hilbert space.

In the regime $\Delta \gg |U|,J$ the relevant conserved quantity is the dipole moment $\hat{P}=\sum_i i \hat{n}_i$. In fact, the fragmented structure in this regime can be partially understood using the analogous spin-$1$ Hamiltonian previously studied in~\cite{Sala2020} and an additional distinction between charge and spin degrees of freedom. For fixed $n_D$, different initial states may correspond to different dipole moments and thus different symmetry sectors, depending on the configuration of the additional doublons. Note, that in the experiment this is not relevant, since we always prepare initial states with the same dipole moment. As already mentioned in the main text, in this work the diagonal interaction energy $\tilde{U}$ is large compared to the effective hopping rate ($U \gg J^{(3)}$) such that processes, that do not conserve the doublon number, are suppressed. In order to model the short-time dynamics via the construction of the Krylov fragments appropriately, we can obtain the dynamically relevant sectors not only via the conserved dipole moment, but further impose the independent conservation of the doublon number under the action of the effective Hamiltonian in Eq.~(\ref{eq:Heffdip}). Since this is not a conserved quantity of $\hat{H}_{\mathrm{eff}}$ the term $\hat{T}_3$ in the original effective Hamiltonian thus needs to be replaced by the doublon number conserving term
\begin{equation}
	\hat{T}_3^\prime = \sum_{i,\sigma} \left( (\hat{n}_{i,\sigma}-\hat{n}_{i+2,\bar{\sigma}})^2 \hat{c}_{i,\bar{\sigma}} \hat{c}_{i+1,\bar{\sigma}}^\dagger \hat{c}_{i+1,\sigma}^\dagger \hat{c}_{i+2,\sigma} + \mathrm{h.c.} \right).
	\label{eq:T3_doublon}
\end{equation}

In Fig.~\ref{fig:fig4} in the main text we show the infinite-temperature imbalance as a function of the doublon fraction $n_D$. From the construction of the dynamical fragments of an effective Hamiltonian as explained above we obtain all states connected within the respective fragment. By computing their respective imbalance and weighing all states equally we finally retrieve the infinite-temperature imbalance $\mathcal{I}(T=\infty)$. At the same doublon fraction $n_D$ not all states have the same dipole moment and hence, they do not live within the same symmetry sector. Consequently, we average over all fragments that we identify at a given doublon fraction, resulting in the numerical error bars shown in Fig.~\ref{fig:fig4} in the main text.

\begin{figure*}[ht]
	\includegraphics[width=\textwidth]{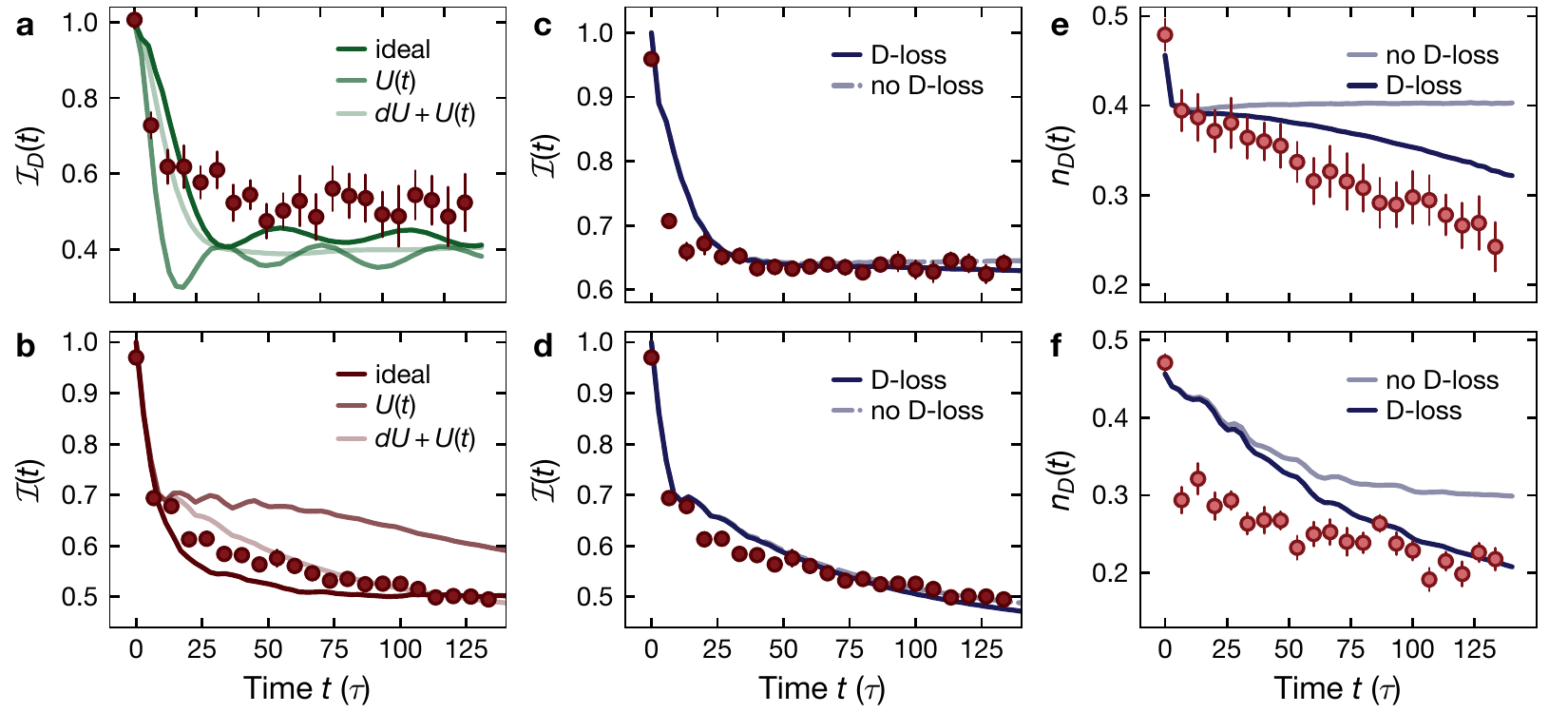}
	\caption{\textbf{Impact of experimental imperfections:} Calculations based on the approximate method in Ref.~\cite{Bharath2021} for the investigation of experimental imperfections in the dipole-conserving regime $U=2.7J$ (top row) and the resonance regime $U\simeq 2\Delta$ (bottom row). \textbf{a} Doublon imbalance time evolution for the ideal case as well as with time-dependent interaction strength $U(t)$ as well as the interaction averaging $dU$. These effects basically average out oscillations without altering the steady-state value. The experimental data not only shows excellent agreement, but further shows that we cannot resolve the effect of imperfections. \textbf{b} Charge imbalance on resonance for a doublon fraction of $45\,\%$. The opacity encodes the same effects as in (a) and is explained in the legend. \textbf{c} and \textbf{d} Imbalance time trace emphasizing the effect of the doublon loss, the other imperfections are included. The effect is only marginal on the observation time. \textbf{e} Time evolution of the doublon fraction for the same parameters as in \textbf{a}-\textbf{d} and numerical simulations with the doublon loss included. Our recorded doublon fraction is very well explained by this model. \textbf{f} The same comparison for the resonance regime.}
	\label{fig:APX}
\end{figure*}

\subsection{Simulations with the approximate method}
\label{sec:APX}
As explained in the previous sections of the supplementary material, we have several experimental imperfections, which possibly influence the dynamics. While these effects are not included in the computations with the effective Hamiltonian and respective ED simulations, we tackle them with a newly-developed approximate method~\cite{Bharath2021} for the full Hamiltonian of the system in Eq.~(\ref{eq:Hamiltonian}). We show that the imperfections in our system do not have a significant quantitative influence on the experimental observations presented in the main text.

We consider three experimental imperfections that are described in detail in the sections above:
\begin{itemize}

\item[1.]\textbf{Time-dependent interaction strength:}  As presented in sec.~\ref{sec:UofT}, the time-dependent interaction strength can be modeled with an exponential defined as $U(t)/J = U_0/J + 1.8 \cdot\exp(-t/32.0)$, where $U_0$ is the set interaction strength and $t$ denotes the evolution time in tunneling times.

\item[2.] \textbf{Interaction averaging: } The interaction strength is not uniform along the lattice due to a gradient in the total magnetic field as described in sec.~\ref{sec:Uvariation}. We can model this effect using a spatially varying interaction, $U(i)$, where $i$ is the lattice site number.

\item[3.]\textbf{ Doublon loss:} Due to light-assisted collisions we have a finite doublon lifetime in the experiment, as described in sec.~\ref{sec:Dlifetime}. These losses are not included in the ED calculations in the main text, since this cannot be accomplished without a master equation approach~\cite{Dalibard1992,Carmichael1992,Fischer2016}, which would exceed the scope of this work. 
\end{itemize}
We use a Lindblad master equation,
\begin{equation}
\dot{\rho} = -\frac{i}{\hbar}[\hat{H}, \rho] + \gamma \sum_{i=1}^L \hat{L}_i \rho \hat{L}_i^{\dagger} -\frac{1}{2}\{\hat{L}_i^{\dagger}\hat{L}_i, \rho\}
\end{equation}
with the jump operator $\hat{L}_i = \hat{c}_{i, \downarrow} \hat{c}_{i, \uparrow}$ representing the loss of a doublon at site $i$. Here, $\rho$ is the many-body mixed state and $\hat{H}$ is the Hamiltonian from Eq.~(\ref{eq:Hamiltonian}) with a spatially and temporally varying interaction strength, $U(i, t)$.  We use the method described in Ref.~\cite{Bharath2021} to solve the above master equation.

We use parameters $\ell = 5, k_{\uparrow}=3$ and $k_{\downarrow}=3$. We introduce a new parameter $k_{\uparrow\downarrow}$, analogous to $ k_{\uparrow}$ and $k_{\downarrow}$ that represents the diameter of a shell around a given atom, the doublons inside which are considered for the time dynamics of the given atom (see Ref.~\cite{Bharath2021} for more details). We use $k_{\uparrow\downarrow}=3$.  The values of these parameters were chosen following a convergence test.  In order to estimate the error, we perform some of the simulations for a set of larger values, i.e., $\ell = 6, k_{\uparrow}=k_{\downarrow}=k_{\uparrow\downarrow}=4$ and compute the change in the mean value of the imbalance.  We find that this change is $\sim 0.006$ for doublon imbalance and $\sim 0.001$ for singlon imbalance.  We approximate the loss of doublons, described by the jump operators, by introducing an exponentially small decay of the projection of the quantum state to the subspace consisting of at-least one doublon. More precisely,  if $\hat{P}$ is the projector to the singlon subspace, we apply $e^{-\hat{P}\gamma \delta t}$ at each time step, followed by a normalization.  Here, $\gamma$ is the rate of the loss and and $\delta t$ is the timestep.  The value of $\gamma$ was calibrated to be $1/145 \times 2\pi J$ using data shown in Fig.~\ref{fig:Dlifetime}. In Fig.~\ref{fig:APX}, we show the time dynamics computed with and without the above mentioned imperfections. As we can see, the imperfections don't have a significant effect on the time dynamics of the charge imbalance and the doublon imbalance alike. We attribute the slight disagreement between the computation and the data in the doublon number dynamics (Fig.~\ref{fig:APX}c and Fig.~\ref{fig:APX}f) to the errors in the computations due to the approximation made in modeling the time evolution under the jump operators, mentioned above.

\section{Numerical parameters}
Here we provide the parameters used for the simulations presented in the main text and this supplementary information.

\begin{table*}[t]
	\centering
	\begin{tabular}{|m{1cm}|m{0.8cm}|m{2cm}|m{1cm}|m{2cm}|m{1.8cm}|m{2cm}|m{6cm}|}
	\hline
		\textbf{Panel} & $U/J$ & $L$ & $n_D$ & $n_h$ & $\chi$ & $e^{i\lambda\hat{S}}$? & \textbf{Comments} \\ \hline
		1c & 2.7 & 101 \newline 52 & 0.0 \newline 0.27 & 0 & 400 (eff.) & Yes & \\ \hline
		2a & 2.7 & 101 (ex.) \newline 101 (eff.) & 0.0 \newline 0.26 & 0.2 & 700 (ex.) \newline 400 (eff.) & Yes & If not mentioned otherwise, time traces are averaged over 10 random initial states \\ \hline
		2b & 2.7 & 101 (ex.) \newline 101 (eff.) & 0.26 & 0.2 & 700 (ex.) \newline 400 (eff.) & Yes & \\ \hline
		2d & 2.7 & 101 & 0.46 & 0.2 & 700 (ex.) & - & Only exact Hamiltonian\\ \hline
		3b & 15.7 & 51 (ex.) \newline 51 (eff.) & 0.46 & 0.2 & 400 (ex.) \newline 500 (eff.) & Yes & \\ \hline
		4a & 2.7 & ED: 13, 15, 17 \newline TEBD: 101 ($n_D=0.62$: 51) & see figure & 0.2 (TEBD) \newline 0.0 (ED) & 500 ($n_D=0.62$) \newline 400 else & $\hat{H}_{\mathrm{eff}}^{\mathrm{dip}}$: Yes \newline $T=\infty$: No & In ED we combined multiple system sizes to realize more values for $n_D$ \newline $L=13$: $n_D=0$, 0.29, 0.5, 0.67 \newline $L=15$: $n_D=0.25$, 0.44, 0.60 \newline $L=17$: $n_D=0.22$, 0.40, 0.55 \\ \hline
		4b & 15.7 & ED: 13 \newline TEBD: 51 & see figure & 0.2 (TEBD) \newline 0.0 (ED) & 500 ($n_D=0.26$, $0.33$, $0.46$) \newline 400 else & $\hat{H}_{\mathrm{eff}}^{\mathrm{dip}}$: Yes \newline $T=\infty$: No \newline $T_{n_D}=\infty$: No & \\ \hline
		S10a & 2.7 & 101 (ex.) \newline 101 (eff.) & 0.0, 0.26, 0.33, 0.46 & 0.2 & 700 (ex.) \newline 400 (eff.) & Yes & \\ \hline
		S10b & 2.7 & 101 (ex.) \newline 101 (eff.) & 0.0, 0.26, 0.33, 0.46 & 0.2 & 700 (ex.) \newline 400 (eff.) & Yes & Same data as in S10a, but showing the doublon imbalance $\mathcal{I}_D$ \\ \hline
		S12a & 15.7 & 51 (ex.) \newline 51 (eff.) & 0 \newline 0.46 & 0.2 & 400 (ex.) \newline 500 (eff.) & Yes & Same data set as in Fig. 3b \\ \hline
		S14 & 0 & 101 & 0 & 0.2 & 700 & Yes & Only exact Hamiltonian \\ \hline
		S15a & 2.7 & 101 & 0.46 & 0 & see legend & Yes & All traces of the convergence analysis are obtained for the same single initial state. \newline Exact Hamiltonian \\ \hline
		S15b & 2.7 & 101 & 0.46 & 0 & 400 & Yes & Exact Hamiltonian \\ \hline
		S15c & 2.7 & 101 & 0.46 & 0 & see legend & Yes & Effective Hamiltonian \\ \hline
		S15d & 15.7 & 51 & 0.46 & 0 & see legend & Yes & Exact Hamiltonian \\ \hline
		S15e & 15.7 & 51 & 0.46 & 0 & 400 & Yes & Exact Hamiltonian \\ \hline
		S15f & 15.7 & 51 & 0.46 & 0 & see legend & Yes & Effective Hamiltonian \\ \hline
		S16a & 2.7 & 101 & 0.26 & 0.2 & 400 & Yes & Only effective Hamiltonian \newline Same data as in 2b \\ \hline
		S16b & 2.7 \newline 15.7 & 101 ($U=2.7J$) \newline 51 ($U=15.7J$) & 0.26 & 0.2 & \underline{$U/J=2.7$}: \newline 700 (ex.) \newline 400 (eff.) \newline \underline{$U/J=15.7$}: \newline 700 (ex.) \newline 500 (eff.) & Yes & \\ \hline \hline
		\textbf{Panel} & $U/J$ & $L$ & $n_D$, $n_h$ & $(\ell, k_\uparrow, k_\downarrow, k_d)$ & $\gamma$ & \textbf{Effects} & \textbf{Comments} \\ \hline
		S17a & 2.7 & 100 & 0.45, 0.2 & (5,3,3,3) & 0 & $U(t)$, $dU$ & Simulations with the approximate method developed in~\cite{Bharath2021} \newline $U$ varies in space and time \\ \hline
		S17b & 15.7 & 100 & 0.45, 0.2 & (5,3,3,3) & 0 & $U(t)$, $dU$ & \\ \hline
		S17c & 2.7 & 100 & 0.45, 0.2 & (5,3,3,3) & $>0$ & $U(t)$, $dU$, $\gamma$ & Including the doublon loss at rate $\gamma = 1/145\tau$\\ \hline
		S17d & 15.7 & 100 & 0.45, 0.2 & (5,3,3,3) & $>0$ & $U(t)$, $dU$, $\gamma$ & \\ \hline
		S17e,f & 2.7 \newline 15.7 & 100 & 0.45, 0.2 & (5,3,3,3) & $>0$ & $U(t)$, $dU$, $\gamma$ & Same simulations as in Panels 15c,d show the dynamical doublon fraction\\ \hline
	\end{tabular}
	\caption{\textbf{Numerical parameters:} The table summarizes the numerical parameters used in the ED and TEBD simulations employed in this work ($n_h$: hole fraction, $\chi$: bond dimension, $e^{i\lambda\hat{S}}$: Schrieffer-Wolff transformation respected?). The standard Trotter step of TEBD simulations (if not mentioned otherwise) is $0.01\tau$.}\label{parameter_table}
\end{table*}

\end{document}